\documentclass[aps,prl,preprint,groupedaddress]{revtex4-2}

\usepackage{amsmath}
\usepackage{graphicx}
\begin{document}

%\preprint{}

\title{Ground- and excited-state properties of LiNb$_{1-x}$Ta$_x$O$_3$ solid solutions}



\author{Felix Bernhardt}
%\thanks{}                                                                                                     
\affiliation{Institut f\"ur Theoretische Physik, Justus-Liebig-Universit\"at Gie{\ss}en, 
Heinrich-Buff-Ring 16, 35392 Gie{\ss}en, Germany}
\affiliation{Center for Materials Research (ZfM/LaMa), Justus-Liebig-Universit\"at Gie{\ss}en, 
Heinrich-Buff-Ring 16, 35392 Gie{\ss}en, Germany}

\author{Florian A. Pfeiffer}
%\thanks{}                                                                                                     
\affiliation{Institut f\"ur Theoretische Physik, Justus-Liebig-Universit\"at Gie{\ss}en, 
Heinrich-Buff-Ring 16, 35392 Gie{\ss}en, Germany}
\affiliation{Center for Materials Research (ZfM/LaMa), Justus-Liebig-Universit\"at Gie{\ss}en, 
Heinrich-Buff-Ring 16, 35392 Gie{\ss}en, Germany}

\author{Felix Schug}
%\thanks{}                                                                                                     
\affiliation{Institut f\"ur Theoretische Physik, Justus-Liebig-Universit\"at Gie{\ss}en, 
Heinrich-Buff-Ring 16, 35392 Gie{\ss}en, Germany}
\affiliation{Center for Materials Research (ZfM/LaMa), Justus-Liebig-Universit\"at Gie{\ss}en, 
Heinrich-Buff-Ring 16, 35392 Gie{\ss}en, Germany}

\author{Anton Pfannstiel}
%\thanks{}
\affiliation{School of Physics, University of Osnabrueck, Barbarastrasse 7, 49076 Osnabrueck, Germany}

\author{Tobias Hehemann}
%\thanks{}
\affiliation{School of Physics, University of Osnabrueck, Barbarastrasse 7, 49076 Osnabrueck, Germany}

\author{Steffen Ganschow}
%\thanks{}
\affiliation{Leibniz-Institut f\"ur Kristallz\"uchtung, Max-Born-Str. 2, 12489 Berlin, Germany}

\author{Mirco Imlau}
%\thanks{}
\affiliation{School of Physics, University of Osnabrueck, Barbarastrasse 7, 49076 Osnabrueck, Germany}

\author{Simone Sanna}
\email[]{simone.sanna@theo.physik.uni-giessen.de}
%\thanks{}
\affiliation{Institut f\"ur Theoretische Physik, Justus-Liebig-Universit\"at Gie{\ss}en, 
Heinrich-Buff-Ring 16, 35392 Gie{\ss}en, Germany}
\affiliation{Center for Materials Research (ZfM/LaMa), Justus-Liebig-Universit\"at Gie{\ss}en, 
Heinrich-Buff-Ring 16, 35392 Gie{\ss}en, Germany}


\date{\today}

\begin{abstract}
LiNb$_{1-x}$Ta$_x$O$_3$ solid solutions are investigated from first principles and 
by optical spectroscopy. The ground- and excited-state
properties of the solid solutions are modelled within density functional theory as 
a function of the Ta concentration using special quasirandom structures spanning the entire 
composition range between LiNbO$_3$ and LiTaO$_3$. Deviations from a Vegard behavior are predicted 
for the lattice parameters, the heat capacity, the electronic bandgap, and consequently 
the absorption edge. The latter is measured for crystals of different composition by 
low temperature optical spectroscopy, qualitatively confirming the theoretical 
predictions. 
The LiNb$_{0.11}$Ta$_{0.89}$O$_3$ composition is found to be a highly unusual 
crystal with a permanent macroscopic electric polarization and nonetheless zero 
birefringence.  
\end{abstract}

%\keywords{}

\maketitle

\section{\label{sec:intro}Introduction}
%\subsection{}
%\subsubsection{}

Solid solutions are, according to the IUPAC definition \cite{IUPAC}, solids in 
which two or more components are compatible and form a unique phase. In this 
definition \textit{compatible} means, that the components (also referred to as 
parent compounds, end compounds, or solute and solvent) obey the Hume-Rothery 
rules \cite{Rules1,Rules2}. They state that if the two end compounds have the 
same crystal structure, similar electronegativity, valency, and comparable atomic 
radii (15\% or less difference), they may form a solid solution. This can occur 
if the particles of one component are accommodated into the space between the 
particles of the other (interstitial solid solutions), or if the solute substitutes 
the particles in the lattice of the solvent (substitutional solid solutions). In 
general, interstitial solid solutions form over a limited concentration range, 
while substitutional solid solutions may from with any composition between the 
two end compounds.

In both cases, the properties of the solvent are modified by the presence of 
foreign particles, which destroy, strictly speaking, the translational symmetry 
of the crystal, and thus the physical and electronic homogeneity on
the microscopic scale.
As a general rule of thumb, the material properties of the solid solutions such 
as the lattice parameters \cite{Keloglu1972}, the effective mass \cite{Polimeni2004}, 
the heat capacity, and the electronic bandgap \cite{Sanna2004} vary almost 
linearly with the composition. This behavior is described by the Vegard’s law 
\cite{Vegard1,Vegard2}. Yet, many exceptions exist, for which some property 
varies non linearly between those of the parent compounds \cite{Bombardi03,Baidya16}.
In this case, the considered property of the solid solutions might exceed that of both the end
compounds for some concentration. 

Solid solutions are extremely appealing for technological and industrial 
applications, as the mixed crystals often have superior properties with 
respect to the parent compounds. Moreover, the magnitude of the material 
properties can be tuned by composition and tailored for specific applications.

A prime example of solid solutions with a huge potential for technological 
applications are the ferroelectric LiNb$_{1-x}$Ta$_x$O$_3$ mixed crystals 
\cite{Wood08}. They combine the excellent 
piezoelectric, electro-optical, and electro-acoustic properties of lithium 
niobate (LiNbO$_3$, LN) with the thermal stability of lithium tantalate 
(LiTaO$_3$, LT) \cite{UlianaDez23,Gureva23}. 
Thus, they are very appealing for the realization of different sensors that 
might be operated in harsh environments, especially at high temperatures. 

Below their ferroelectric Curie temperature, both LN and LT crystallize in 
the same rhombohedral structure belonging to the $R3c$ space group \cite{Inbar96,Weis85}. 
In this structure, Nb and Ta ions are pentavalent. Due to the similarity of 
their ionic radii (both 0.78\,{\AA} for pentavalent, octahedrally coordinated 
ions \cite{webelements}), and of their electronegativity (1.6 and 1.5 in 
Pauling units for Nb and Ta, respectively \cite{webelements}), Nb and Ta can 
be readily exchanged. Accordingly, the substitutional LiNb$_{1-x}$Ta$_x$O$_3$ (LNT)
solid solutions can be grown over the whole composition range \cite{Bartasyte12}. 
According to the actual understanding of the solid solutions, Ta and Nb ions
are expected to be randomly distributed on the Ta/Nb lattice sites to 
maximize the entropic contributions to the free energy \cite{Manzoor18}.

Some properties of the LNT solid solutions, such as the birefringence
\cite{Wood08}, the lattice dynamics \cite{Micha16}, 
the linear optical response \cite{XUE2000581}, and the
phase transition \cite{FatimaDez23}, have been investigated in the past. However,
the overall knowledge of LNT mixed crystals is still scarce. 
Moreover, a theoretical characterization of the mixed crystals is 
missing up to date. Simplified approaches modelling a few compositions of 
the solid solutions as ordered crystals are available \cite{Riefer13SS,Sanna13SS}, 
which, however, do not model the correct cation distribution in random alloys. 
In this work, we model LiNb$_{1-x}$Ta$_x$O$_3$ solid solutions from first 
principles using special quasirandom structures (SQS) \cite{Wei90,Petzold10}, 
which allow to model random alloys within periodic boundary conditions. 
We predict the dependence of ground and excited state properties such as lattice 
parameters, heat capacity, and electronic structure on the composition. 
Deviations from the Vegard behavior are predicted for the fundamental electronic 
bandgap, which is verified by specifically performed low temperature optical 
measurements. 
Moreover, the solid solution with LiNb$_{0.11}$Ta$_{0.89}$O$_3$ composition 
is found to be a peculiar crystal with an unusual combination of permanent 
polarization and optical isotropy.  

% If in two-column mode, this environment will change to single-column
% format so that long equations can be displayed. Use
%\begin{widetext}
%\end{widetext}

\section{\label{sec:Methods}Methodology}
\subsection{Special Quasirandom Structures}
From the point of view of theorists, the lack of studies on LNT solid solutions 
is not surprising, as the atomistic modelling of non-ordered crystals is a conceptually 
challenging task. 
Ordered crystals and their properties are efficiently calculated within density 
functional theory (DFT), at least at 0\,K, in conjunction with 
periodic boundary conditions. On the contrary, mixed crystals with random 
site occupation are not easily modeled within a unit cell that is periodically 
repeated in the three cartesian directions, as it models by definition a periodic 
crystal. The brute-force method, consisting in modelling many randomly generated 
distributions for the same composition and then averaging the calculated quantities, 
is not feasible due to the huge number of possible configurations. To cope with this 
problem, many approaches have been developed. 

Within the Virtual Crystal Approximation, the ions are described by a virtual 
hybrid atom consisting of two or more species weighted according to the concentration 
\cite{VCA}. Thus, a mixed crystal can be modeled at the same computational cost of 
the end compounds. Within the Coherent Potential Approximation, the potential within 
the solid solution is modelled as an ordered, composition dependent array of effective 
potentials \cite{CPA1}. Unfortunately, both methods are non-structural approaches, 
which neglect finer details of disordered structures such as local distortions, 
which, however, are crucial for the description, e.g., of polarons. 

A further way to model non ordered crystals is the Cluster Expansion Method. Within 
this approach, it is assumed that scalar properties (e.g., the total energy) can be 
expanded in a series of structural motifs (clusters) \cite{Sanchez84,Laks92}. 
An effective Hamiltonian is then constructed by the sum of the interactions 
of given clusters. The parameters for 
the expansion are typically calculated within DFT, and the overall computational 
demand is quite high. Extensions of the method to tensor properties are available 
as well \cite{CExp}.

A pragmatic alternative, which we employ in the present work, is given by the special 
quasirandom structures (SQS). They are supercells that are optimized to provide 
the best representation of non-ordered systems within the supercell approach. Thereby, 
the site occupation is chosen such that the pair correlation functions are as close as 
possible to that of an ideal, infinite random solid solution. SQS are able to accurately 
predict thermodynamical properties at moderate computational cost \cite{Ghosh08,Petzold10}.

According to the original idea of Zunger and co-workers \cite{Wei90}, the atomic lattice 
is discretized into figures labeled by $f=(k,m)$, where $k$ is the number of vertices in 
the figure and $m$ is the maximum distance spanned by the figure edges. The latter is 
given in units of nearest-neighbor cells.
In this formalism, an analytic expression for the lattice averaged correlation function 
of an ideal, truly random alloy is known:

\begin{equation}
\label{eq:analytic}
\langle \overline{\Pi}_f \rangle_R = \left( 2x-1 \right)^k,
\end{equation}

which holds for finite figures ($k>0$) and impurity concentration $x\in [0,1]$.
The label $R$ denotes randomness, true disorder.

The products of a spin variable $\sigma_i=\pm 1$ which
assumes the values $+1$ and $-1$ depending on which ion occupies a given lattice 
site:

\begin{equation}
\label{eq:spinprod}
\Pi_l = \prod_{l=1}^{k}\sigma_i,
\end{equation}

can be used to define the lattice averaged correlation function of the 
cluster $f$ in the structure $S$, averaging over all appearances of that 
cluster within the structure $S$:

\begin{equation}
\label{eq:true}
\overline{\Pi}_f=\frac{1}{N D_f}\sum_l \Pi_l.
\end{equation}

In this equation, which represents the correlation function of the cluster $f$ 
in the structure $S$, $D_f$ denotes the multiplicity of the figure $f$ defined 
by its symmetry, and $N$ is the number of lattice points that can be occupied
either by Nb or Ta. 
Finally, introducing the mean distance between vertices $i,j$ in a figure:

\begin{equation}
\label{eq:dist}
\overline{d}_f=\frac{1}{N} \sum^k_{\substack{i,j \\ i<j}}d_{i,j},
\end{equation}

we are able to calculate the deviation $\varepsilon(S)$ of the averaged correlation function of a 
particular structure $\overline{\Pi}_f(S)$ from the corresponding correlation function 
of a truly random, infinite alloy of the same composition $\langle \overline{\Pi}_f \rangle_R$:

\begin{equation}
\label{eq:errf}
\varepsilon(S)=\sum_f\frac{D_f}{(k\overline{d}_f)^n}\left| \overline{\Pi}_f(S)-\langle \overline{\Pi}_f \rangle_R\right|.
\end{equation}

$\varepsilon(S)$ as defined by equation \ref{eq:errf} is referred to as \textit{error function} 
in Ref. \cite{Petzold10}.
Larger clusters with more vertices $k$ and larger mean distance $\overline{d}_f$ have a
minor contribution to $\varepsilon(S)$. This should mirror the fact that most physical 
properties stem mainly from short–ranged interactions, making correlation functions of small 
clusters more important for the choice of SQS. The exponent $n$ further determines the weight 
distribution between small and large clusters, as well as the unit of $\varepsilon(S)$. It can 
be shown that the order of the best SQS does not vary for $1 < n < 5$. So, we chose 
$n = 2$ in this work, yielding $\varepsilon(S)$ in units of inverse square meters. 
 We furthermore chose a cutoff distance of 11\,{\AA}, 9\,{\AA}, 7\,{\AA} and 
7\,{\AA} for $k = 2$, $k = 3$, $k = 4$ and $k = 5$, respectively. Where the configuration space
allows it, about 10$^5$ randomly generated candidate structures were considered. 

\begin{figure}
\includegraphics[width=0.6\linewidth]{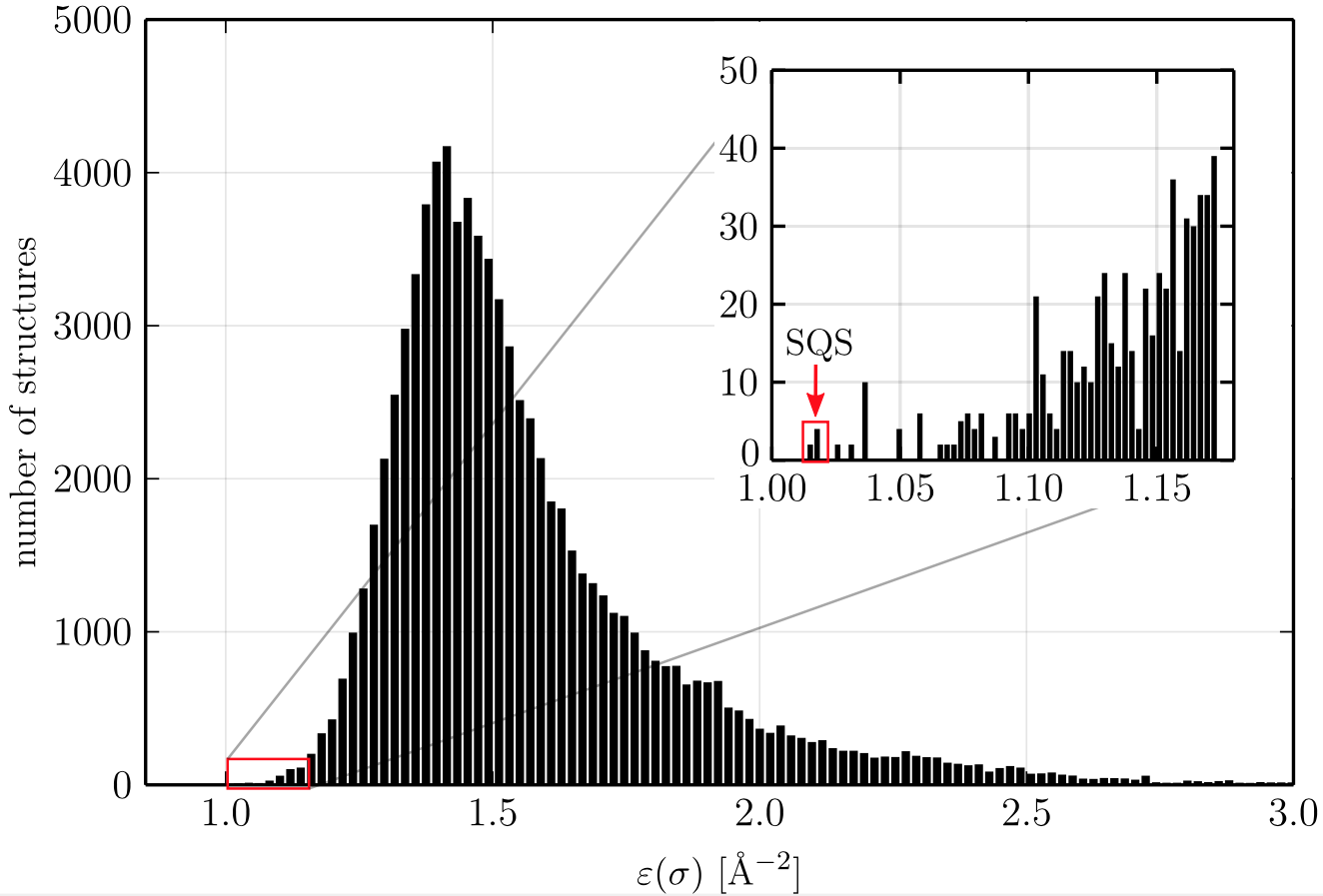}
\caption{\label{fig:sqs} Distribution of the error function $\varepsilon(S)$ for all the considered
randomly generated atomic configurations modelling a LiNb$_{0.58}$Ta$_{0.42}$O$_3$
solid solution. The red box in the inset indicates the best SQS candidates.}
\end{figure}

The best five SQS for each concentration (or at least three, where the configuration space
is restricted) were selected for further investigation by means of DFT.
This procedure is illustrated exemplarily for a LNT crystal with $x=0.42$ in figure
\ref{fig:sqs}.

\subsection{Computational Details}

Atomistic calculations are performed within the DFT as implemented in VASP
\cite{Kresse1993,Kresse1996,Kresse1996_2}. PAW potentials \cite{Bloechl94,Joubert1999}
with exchange-correlation functional in the PBEsol formulation \cite{Perdew2008} and
electronic configurations 1s$^2$2s$^1$, 4p$^6$4d$^3$5s$^2$, 5s$^2$5p$^6$5d$^3$6s$^2$
and 2s$^2$2p$^4$ are employed for Li, Nb, Ta and O, respectively. Plane waves up to a
cutoff energy of 450\,eV are used as a basis for the expansion of the electronic wave
functions. A Gaussian smearing with width 0.02\,eV is applied to the occupancies of
the electronic states.
For the calculations, three sets of orthorhombic supercells are employed, corresponding
to $2\times 1\times 1$, $1\times 2\times 1$ and $1\times 1\times 2$ repetitions of the	
orthorhombic unit cell and containing 120 atoms in all cases.
These allow to model LiNb$_{1-x}$Ta$_x$O$_3$ solid solutions with concentration
steps $\Delta x=0.041\overline{6}$. Equilibrium volumina are estimated fitting different
calculated energy-volume pairs by the Murnaghan state equation \cite{Murnaghan44}:

\begin{equation}
\label{eq:murnaghan}
F(V)=\frac{BV}{B^\prime}\left[ 
\frac{\left(\frac{V_0}{V}\right)^{B^\prime}}{B^\prime -1} 
+1\right] + \text{const.},
\end{equation}

where $B$ and $B^\prime$ are the bulk modulus and its first pressure derivative,
respectively, both evaluated at the equilibrium volume $V_0$.

Corresponding to the supercell symmetry, $2\times2\times2$, $4\times2\times2$ and
$4\times4\times1$ $\Gamma$-centered  Monkhorst-Pack K-point meshes \cite{Pack1977} are used.
For the calculation of the density of states (DOS) the K-point mesh was expanded to
include 3240 K-points.
The ionic positions are optimized, such that the Hellmann-Feynman forces acting on
the ions \cite{Forces} are lower than 0.005\,eV/{\AA}.

Harmonic phonon frequencies are obtained by the finite-displacement method perfomed 
within phonopy \cite{phonopy-phono3py-JPCM,phonopy-phono3py-JPSJ}. The orthorhombic 
SQS are doubled in one of the cartesian axes, such that the resulting supercells used 
for these calculations have roughly equal lattice constants in all directions. The 
K-point meshes are adjusted accordingly. Due to numerical issues, small imaginary 
frequencies in a restricted region around $\Gamma$ are present. These frequencies 
are assumed to be real for the thermodynamic calculations. Moreover, for the calculation 
of the thermodynamic properties, the phonon frequencies are interpolated onto a 
20$\times$20$\times$20 K-point mesh using the Parlinski-Li-Kawazoe method \cite{Parlinski1997}. 
This yields an error of only about 3\% for the considered property compared to a 50\% 
denser K-point mesh.

Linear optical properties are calculated at different levels of approximation
using the QuantumEspresso package
\cite{QE-2017}, an energy cutoff of 100\,Ryd, the previously introduced K-point meshes 
and optimized norm-conserving Vanderbilt pseudopotentials 
\cite{Hamann13}.
Within the independent particle approximation 
(IPA), the imaginary part of the dielectric function $\varepsilon$ 
reads as

\begin{equation}\label{eq:IPA}
  \Im[\varepsilon_{ab}(\omega)] \sim
  \lim_{q\to 0}\frac{1}{q^2}\sum_{c,v,\vec{k}}2\delta(E_{c\vec{k}}-E_{v\vec{k}}-\omega) \times
  \langle u_{c\vec{k}+\vec{e}_aq}|u_{v\vec{k}}\rangle \langle u_{c\vec{k}+\vec{e}_bq}|u_{v\vec{k}}\rangle^{*},
\end{equation}

with cartesian indices $a$ and $b$, conduction and valence band indices $c$ and $v$, 
respectively, band energies $E$ at K-point $\vec{k}$, and the Bloch functions $u$. 
In total, 800 states are considered for the summation, for all the considered LNT crystals. Furthermore, 
the K-point meshes are increased to twice their previous sizes in each direction, in 
order to achieve convergence of the dielectric function up to around 10\,eV.
The real part of $\varepsilon$ is calculated via Kramers-Kronig relations. 
With the knowledge of the dielectric function, basic optical properties such as the
frequency dependent refractive index can be calculated:

\begin{equation}
\label{eq:brechzahl}
n(\omega)=\sqrt{\frac{1}{2}\sqrt{\Im[\varepsilon(\omega)]^2+\Re[\varepsilon(\omega)]^2}+Re[\varepsilon(\omega)]}.
\end{equation}

Quasiparticle energies are calculated within G$_0$W$_0$ approximation using the 
generalized plasmon pole model as implemented in the software package BerkeleyGW 
\cite{Deslippe2012,Hybertson1986}. We calculate the energy corrections for about 
130 electronic bands centered around the highest valency band, using the same 
K-Point meshes as for the optical calculations in DFT. As these calculations are 
computationally quite demanding, we only consider one SQS for the selected Ta 
concentrations.

\subsection{Experimental setup}

LNT crystals in the composition range $0<x<1$ are measured experimentally 
by means of absorption spectroscopy in the ultraviolet (UV) and visible 
(VIS) spectral range, i.e. in the energy range from 1.0\,eV to 5.0\,eV, to  
verify the theoretical predictions. 
We use crystal samples polished on both sides to optical quality 
($\lambda$/10) with thicknesses in the range of approx. 1\,mm and 
dimensions of approx. 10\,mm $\times$ 10\,mm. The samples were 
prepared from commercially available wafers (LN, LT from the Precision 
Microoptics USA Inc.) or from crystal boules which are grown as single crystals 
using the Czochralski method (LNT, IKZ Berlin). A commercial UV/VIS 
two-beam photospectrometer (UV-3600, Shimadzu Deutschland GmbH) is used 
for the measurement series, which was extended with a closed-cycle helium 
cryostat (RDK 10-320, Leybold GmbH) with optical quartz windows. 
Here, the samples are placed against a copper-pinhole on the cooling 
finger of the cryostat and are temperature modulated in the range of 30\,K 
to 200\,K in steps of 10\,K. 
The combined use of a deuterium lamp ($<$280\,nm) and halogen lamp ($>$280\,nm) 
for probe light generation together with a multialkali photomultiplier tube 
for detection makes it possible to capture the spectral range in which the 
position of the optical absorption edge is expected for all crystal samples 
(spectral resolution of $<$5\,nm FWHM). All spectra for each temperature point 
are corrected for optical losses induced by the cryostat setup via the 
difference spectra with and without a sample in the optical beam path. 

\begin{figure}
\includegraphics[width=0.62\linewidth]{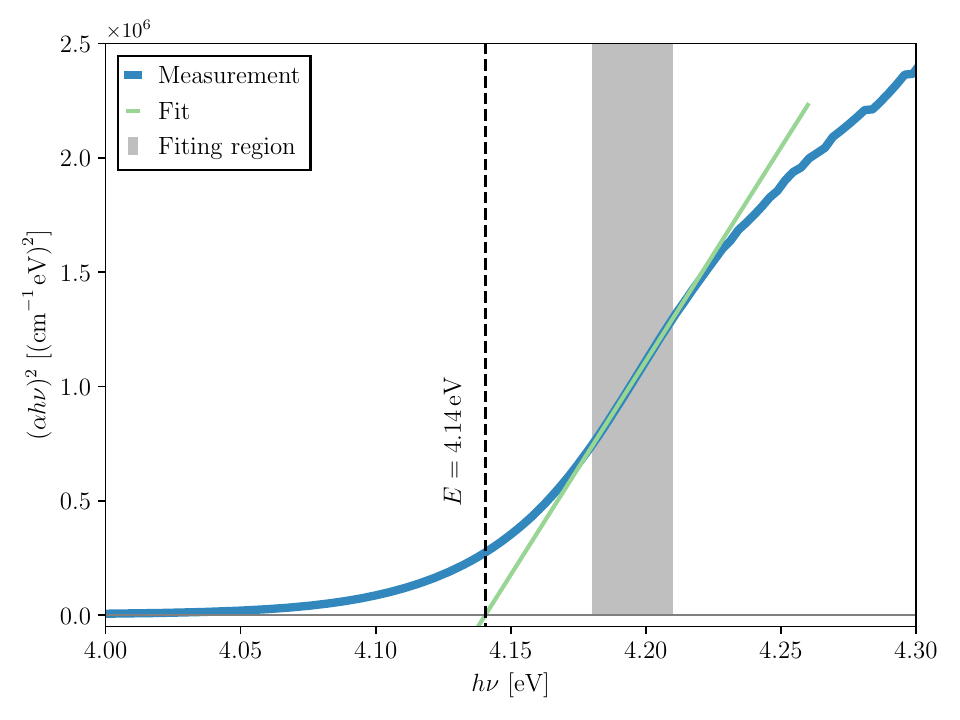}%
\caption{\label{fig:tauc} Exemplary Tauc fit for the direct transition of 
lithium niobate at 50\,K. The shaded area represents the region where the 
linear function (green line) is fitted to the measurement (blue line). The 
vertical dashed indicator shows the intercept with the abscissa which defines 
the band gap.}
\end{figure}

For analysis of the optical absorption gap, the spectra are linearized using 
the Tauc method \cite{Tauc66,Zanatta19,Bock19,Bhatt12}

\begin{equation}
\label{eq:tauc}
\alpha h\nu = A\left( h\nu-E_g \right)^{\frac{1}{n}}
\end{equation}

Where $n=2$ is chosen for the direct, moment conserving band to band 
transition and $n=\frac{1}{2}$ for transitions under participation of 
lattice phonons. The linearized region is fitted according to equation \ref{eq:tauc}. 

\begin{figure}
\includegraphics[width=0.62\linewidth]{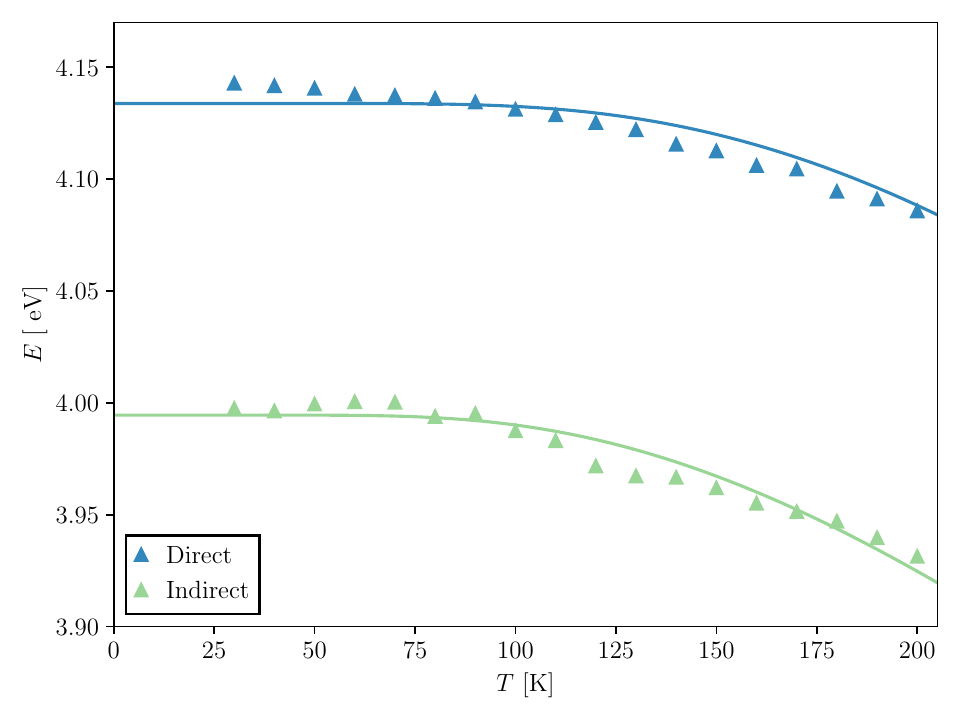}%
\caption{\label{fig:temp} Temperature dependence of the measured direct (blue) 
and indirect (green) optical transitions (exemplarily shown for LiNbO$_3$) as 
obtained by the Tauc procedure. The solid lines are fits performed according to 
the model of O'Donnel and Chen \cite{ODonnel91}.}
\end{figure}

The procedure is exemplary depicted for the direct transition of lithium 
niobate in figure \ref{fig:tauc}. Band gap energies are determined by 
extrapolation of the linear fit to the intercept with the abscissa, which 
thus defines the band gap. This analysis is performed for both transition 
types (direct and indirect) over all investigated temperatures and samples. 
The temperature dependence of the optical band gap is empirically modelled 
as described in Ref. \cite{ODonnel91} via:

\begin{equation}
\label{eq:kevinpeter}
E\left(T\right)=E_0-S\left\langle\hbar\omega\right\rangle\left[\frac{2}{\exp{\left(\frac{\left\langle\hbar\omega\right\rangle}{k_\mathrm{B}T}\right)}-1}\right],
\end{equation}

where $S$ denotes a dimensionless electron-phonon coupling strength and 
$\langle\hbar\omega\rangle$ a system average effective coupling phonon energy 
according to Huang and Rhys \cite{Huang50}. Optical band gap values for 
absolute zero $E_0$ are found by fitting of equation \ref{eq:kevinpeter} 
to the experimental data as shown in figure \ref{fig:temp}. 
The procedure is equivalently performed for all investigated samples.

\section{\label{sec:res}Results}

\subsection{Structural properties}

Crystals belonging to the trigonal group are usually described either by a hexagonal or by a 
rhombohedral unit cell. Moreover, a supercell of orthorhombic symmetry such as the 
SQS employed in this work is used for the specification of the materials properties 
that are described by tensors such as the electro-optic or piezoelectric coefficients
\cite{Sanna17surf}. In this work, we discuss the lattice paramters of the hexagonal 
structure, which can be readily converted into the rhombohedral lattice parameters as 
described in Ref. \cite{Weis85}.

\begin{figure}
\includegraphics[width=0.62\linewidth]{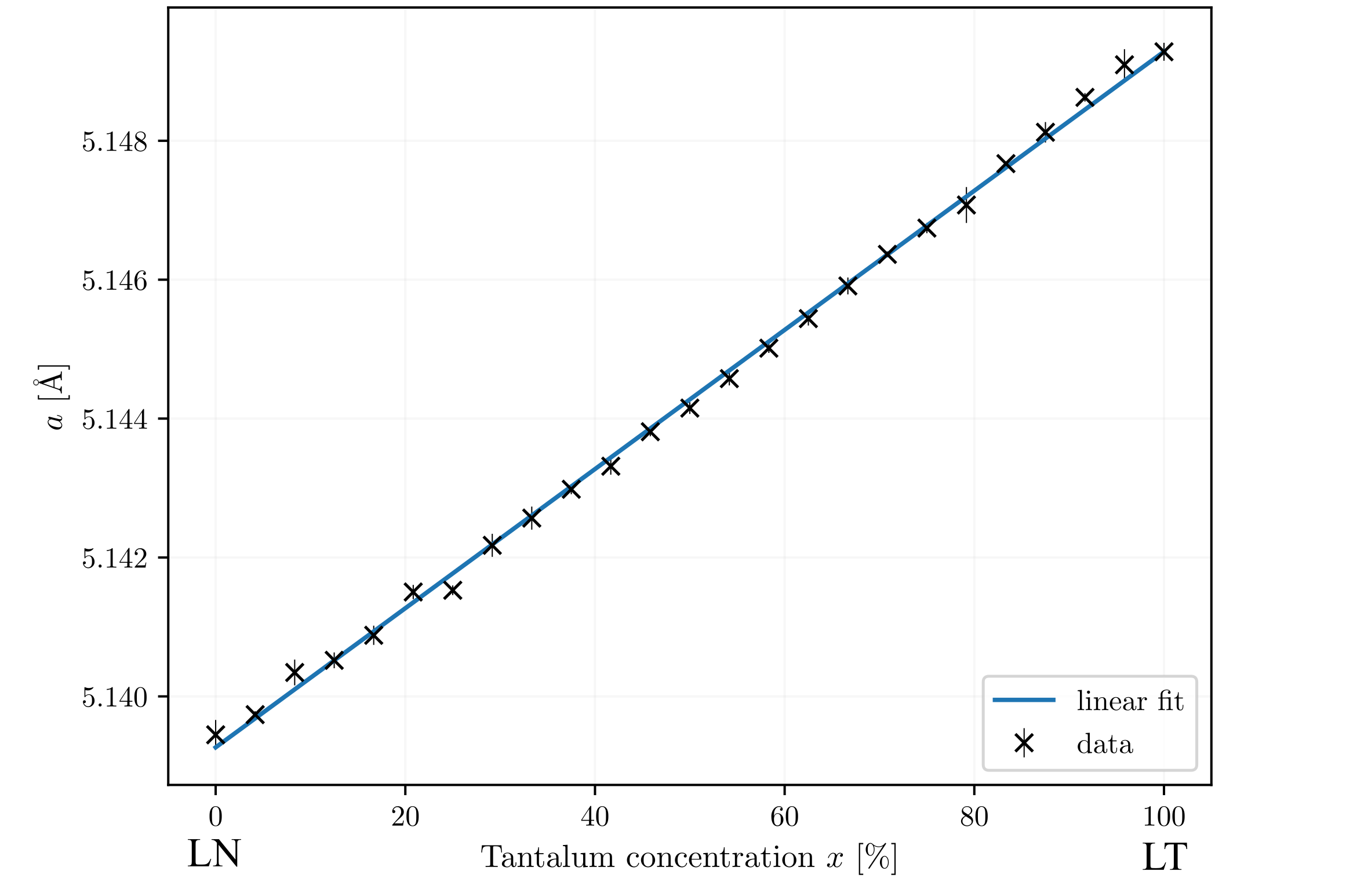}%

\includegraphics[width=0.62\linewidth]{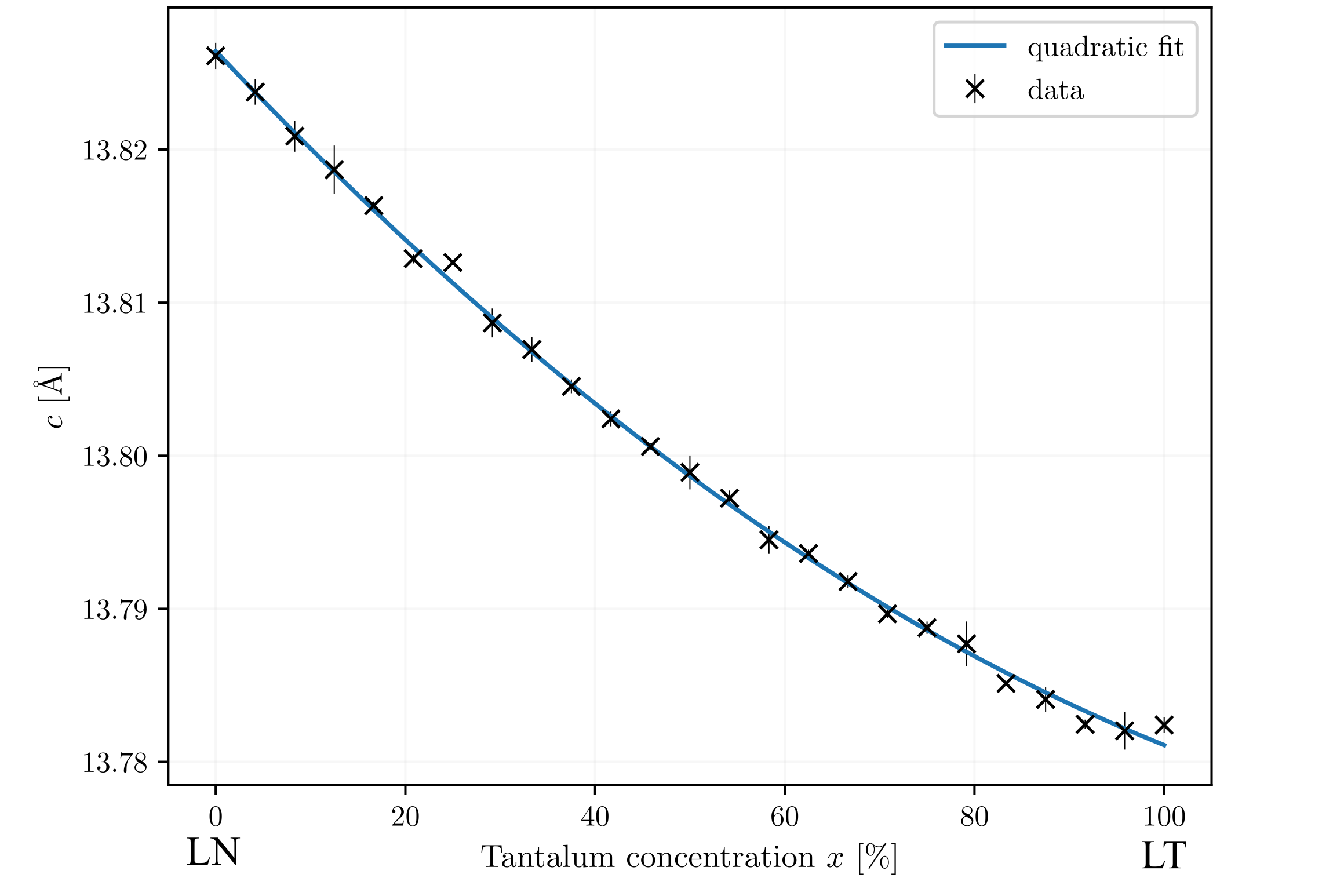}%

\includegraphics[width=0.62\linewidth]{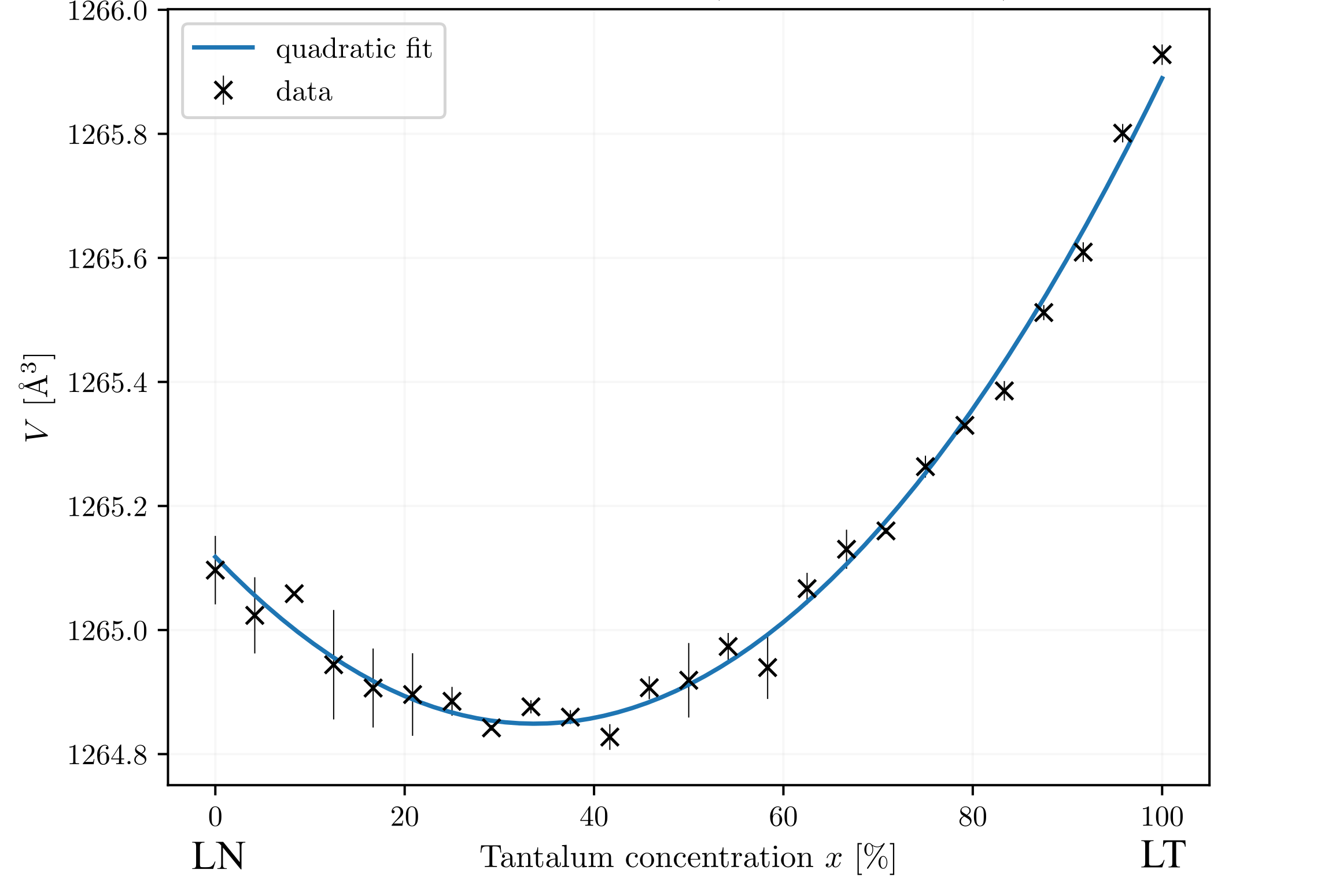}%
\caption{\label{fig:lattice} DFT calculated hexagonal lattice parameters $a$ and $c$
as well as volume $V$ of the LiNb$_{1-x}$Ta$_x$O$_3$ solid solutions as a function of 
the composition. In the plot, $V$ is twice the volume of the orthorhombic unit cell.}
\end{figure}

The structural properties of the LiNb$_{1-x}$Ta$_x$O$_3$ solid solutions as a function 
of the composition are shown in table \ref{tab:structure} and figure \ref{fig:lattice}.
The values calculated with different SQS are slightly different, in particular if
SQS of different supercell symmetry are employed. Therefore, we show in figure \ref{fig:lattice}
the values averaged over all employed supercells of a given composition. The error
bars show the largest deviations from the mean value obtained using different SQS.

The lattice parameter $a$ (figure \ref{fig:lattice}, upper panel), which describes the 
basal plane of the hexagonal structure, varies only in a small range (about 0.2\,\%) and 
grows rather linearly with the composition, with a dependence which is nicely described 
by the Vegard law:
 
\begin{equation}
\label{eq:vegard}
a_{\text{LiNb}_{1-x}\text{Ta}_x\text{O}_3}=(1-x)\cdot a_{\text{LiNbO}_3 }+ x\cdot a_{\text{LiTaO}_3}.
\end{equation}

Thus, the composition can be exploited to a certain extent to optimize the in-plane lattice
matching of LNT thin films on different substrates.
The lattice parameter $c$, describing the height of the hexagonal cell, 
decreases sub-linearly with the Ta content, instaed. The crystal volume $V$ 
(figure \ref{fig:lattice}, lower panel) has a pronounced non-linear dependence on the 
composition. A minimum volume of ca. 316.19\, \AA$^3$ per rhombohedral unit cell is 
calculated for solid solutions with a Ta content of about 40\,\%.

\begin{figure}
\includegraphics[width=0.62\linewidth]{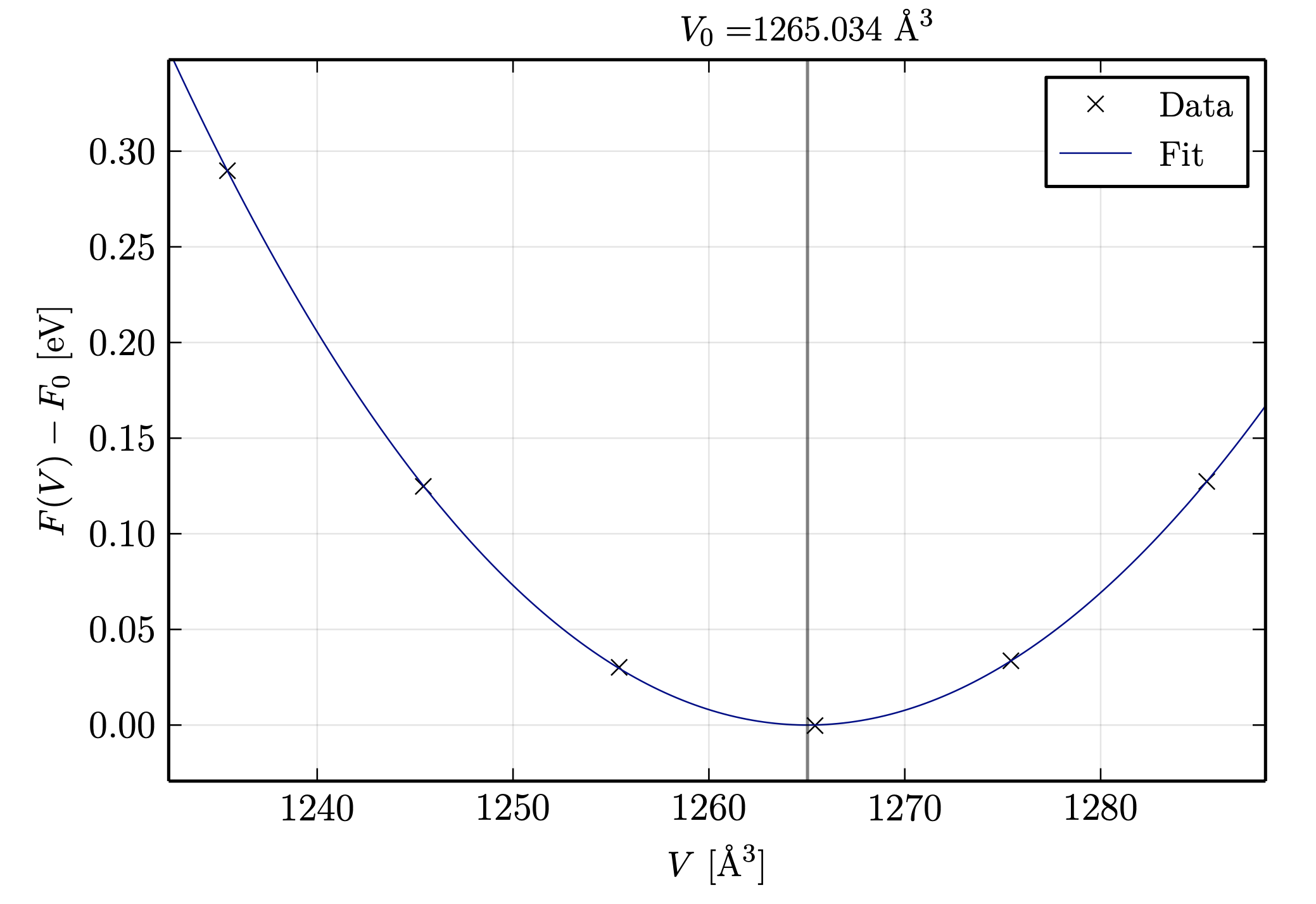}%
\caption{\label{fig:murnaghanfit}Total energy of the orthorhombic, 120 atoms supercell modelling
LiTaO$_3$ for various volumes and fit by equation \ref{eq:murnaghan}. The predicted 
DFT equilibrium volume is indicated.}
\end{figure}

\begin{figure}
\includegraphics[width=0.62\linewidth]{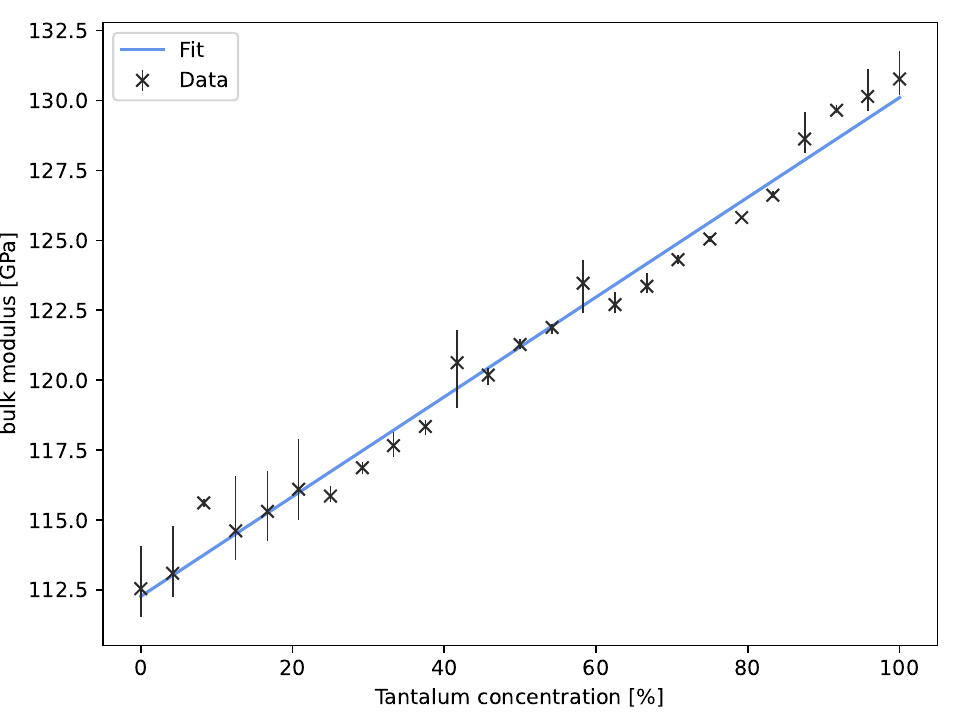}%
\caption{\label{fig:bulkmodulus} DFT calculated bulk modulus of the LiNb$_{1-x}$Ta$_x$O$_3$ 
solid solutions as a function of the composition. The blue line is a linear fit of the
calculated values (black crosses).}
\end{figure}

The bulk modulus $B$ and its first derivative with respect to the pressure $B^\prime$
are calculated by fitting the energy-volume curves (calculated at $T = 0$\,K) to 
the Murnaghan equation (see equation \ref{eq:murnaghan}). This is shown exemplarily 
for $x=1$  (pure LiTaO$_3$) in figure \ref{fig:murnaghanfit}. 
The values calculated for the parent compounds ($B_{\text{LT}}$=130.8\,GPa, 
$B^\prime_{\text{LT}}$=3.51 and $B_{\text{LN}}$=112.6\,GPa, $B^\prime_{\text{LT}}$=5.9) 
are in good agreement with available experimental values \cite{Gaillac16} as well
as with other theoretical results ($B^{\text{theo}}_{\text{LT}}$
=124\,GPa and $B^{\text{theo}}_{\text{LN}}$=102\,GPa \cite{Jain2013}). 
Between the end compounds, the bulk modulus varies almost linearly with the
composition, as shown in figure \ref{fig:bulkmodulus}. 

\begin{table}
\caption{\label{tab:structure}Hexagonal lattice parameters $a$ and $c$ (both in {\AA}), 
crystal volume $V$ (in {\AA}$^3$), bulk modulus $B$ (in GPa) and 
its first pressure derivative $B^\prime$ of the LiNb$_{1-x}$Ta$_x$O$_3$                                          
solid solutions as a function of the composition.}
\begin{ruledtabular}
\begin{tabular}{r|ccccc}
$x$ & $a$ & $c$ & $V$ & $B$ & $B^\prime$ \\
\hline 
  0.0 & 5.139   & 13.826  & 316.279 &   112.553  &  5.859 \\
  4.2 & 5.140   & 13.824  & 316.257 &   113.102  &  6.081 \\
  8.3 & 5.140   & 13.821  & 316.271 &   115.619  &  3.958 \\
 12.5 & 5.140   & 13.819  & 316.222 &   114.620  &  5.862 \\
 16.7 & 5.141   & 13.816  & 316.241 &   115.320  &  5.797 \\
 20.8 & 5.141   & 13.813  & 316.194 &   116.110  &  5.671 \\
 25.0 & 5.141   & 13.812  & 316.206 &   115.865  &  6.552 \\
 29.2 & 5.142   & 13.809  & 316.176 &   116.873  &  6.040 \\
 33.3 & 5.142   & 13.807  & 316.208 &   117.668  &  5.909 \\
 37.5 & 5.142   & 13.805  & 316.141 &   118.352  &  5.922 \\
 41.7 & 5.143   & 13.802  & 316.182 &   120.630  &  4.761 \\
 45.8 & 5.143   & 13.801  & 316.185 &   120.188  &  5.671 \\
 50.0 & 5.144   & 13.799  & 316.194 &   121.282  &  5.428 \\
 54.2 & 5.144   & 13.797  & 316.200 &   121.890  &  5.431 \\
 58.3 & 5.145   & 13.795  & 316.207 &   123.472  &  4.999 \\
 62.5 & 5.145   & 13.794  & 316.191 &   122.707  &  5.648 \\
 66.7 & 5.146   & 13.792  & 316.277 &   123.363  &  5.693 \\
 70.8 & 5.146   & 13.790  & 316.259 &   124.315  &  5.589 \\
 75.0 & 5.147   & 13.789  & 316.301 &   125.053  &  3.315 \\
 79.2 & 5.147   & 13.788  & 316.310 &   125.821  &  5.655 \\
 83.3 & 5.148   & 13.785  & 316.365 &   126.618  &  5.694 \\
 87.5 & 5.148   & 13.784  & 316.367 &   128.627  &  4.528 \\
 91.7 & 5.149   & 13.782  & 316.405 &   129.654  &  4.906 \\
 95.8 & 5.149   & 13.782  & 316.451 &   130.147  &  3.499 \\
100.0 & 5.149   & 13.782  & 316.485 &   130.773  &  3.518 \\
\end{tabular}
\end{ruledtabular}
\end{table}

\begin{figure}
\includegraphics[width=0.68\linewidth]{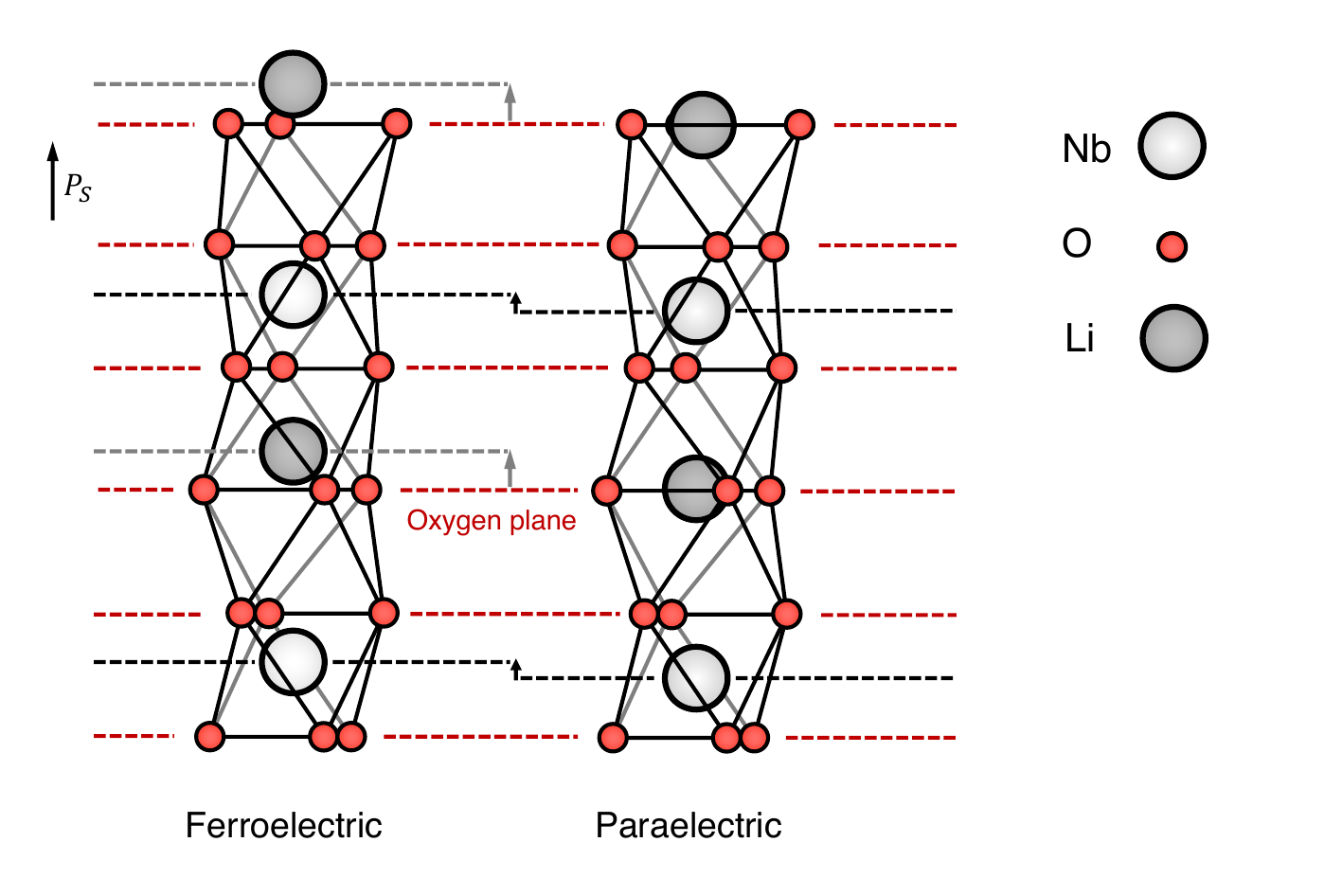}%
\caption{\label{fig:phasen}Schematic representation of the ferroelectric
(lhs) and paraelectric phase of LiNbO$_3$. The oxygen layers represent the 
average position of the Li atoms in the paraelectric phase. The Nb atoms
are in the paraelectric phase exactly between two oxygen layers.}
\end{figure}

Strictly connected with the composition dependent change of the structural 
parameters is the spontaneous polarization $P_S$ of LiNbO$_3$ and LiTaO$_3$.
$P_S$ originates from a separation of the centers of mass of the positive and
negative charges, which results in an electric dipole moment in each unit cell
(see figure \ref{fig:phasen}).
Although all the ions in the two oxides are nominally isovalent, the spontaneous 
polarization decreases from 71\,$\mu$C/cm$^2$ for 
LiNbO$_3$ \cite{Chen01} to 60\,$\mu$C/cm$^2$ for LiTaO$_3$ \cite{Kitamura98}, respectively.
This effect can be mainly traced back to the different relative position of 
anions and cations in the two materials. The latter is discussed in the
following as a function of the composition.

\begin{figure}
\includegraphics[width=0.48\linewidth]{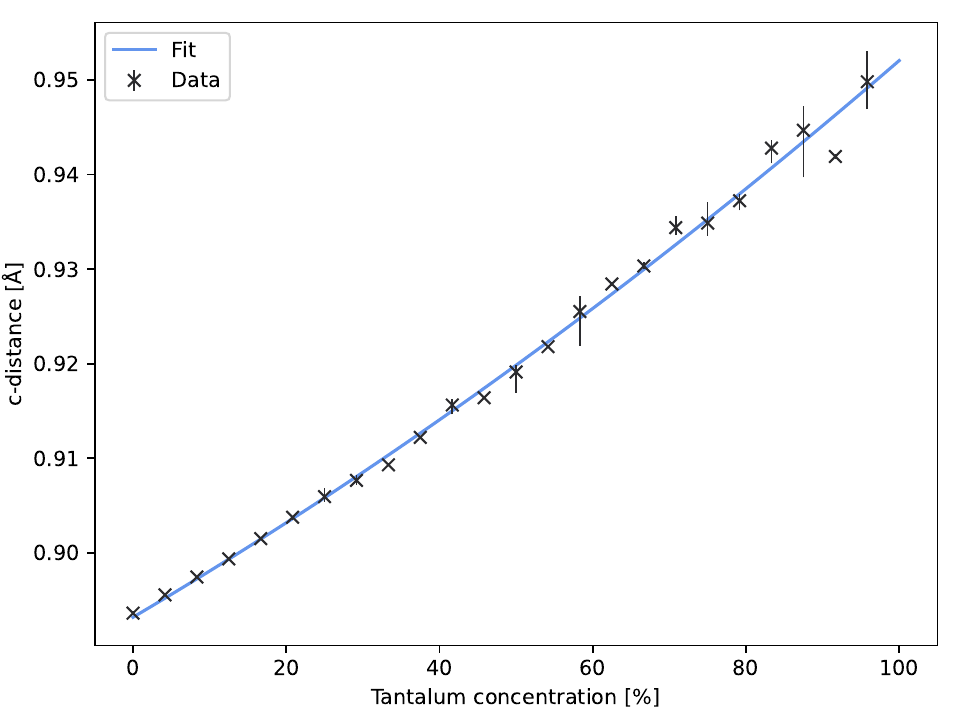}%
\includegraphics[width=0.48\linewidth]{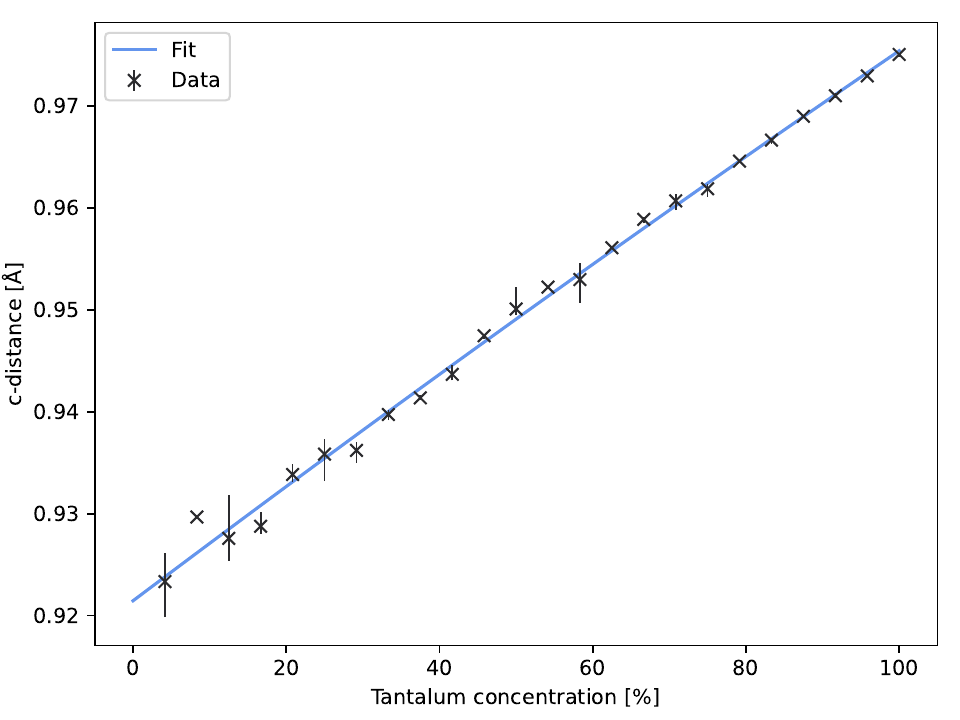}%
\caption{\label{fig:NbTa_dist}Average distance of the Nb ions (left panel) and of the
Ta ions (right panel) from the closest oxygen layer perpendicular to the $c$-axis in
LiNb$_{1-x}$Ta$_x$O$_3$ solid solutions as calculated within DFT 
as a function of the composition. The oxygen layer represents the top of 
the oxygen octahedra containing the Nb or Ta ions. The increasing distances thus denote
a movement towards the cage center.}
\end{figure}

The relative position of the ions within the solid solutions depends,
as revealed by our models, on the 
composition. Figure \ref{fig:NbTa_dist} shows the average distance of the Nb ions 
(left panel) and of the Ta ions (right panel) from the closest oxygen layer 
perpendicular to the $c$-axis as a function of the composition. The oxygen layer 
represents the top of the oxygen octahedra containing the Nb or Ta ions and higher
distances from this layer correspond to points closer to the center of the octahedra. 
The latter represents the Ta and Nb positions in the paraelectric phase.
For a given concentration, Ta ions are in general closer to the cage center 
than the Nb ions (see different scales of the plots in figure \ref{fig:NbTa_dist}). 
This reflects the higher spontaneous 
polarization in LiNbO$_3$ than in LiTaO$_3$. The distances of both the Nb and Ta 
cations from the oxygen layer increase roughly linearly with the Ta concentration.
This denotes a movement of both species towards the cage center and corresponds, 
again, to a decrease of the spontaneous polarization from LiNbO$_3$ to LiTaO$_3$. 

\begin{figure}
\includegraphics[width=0.48\linewidth]{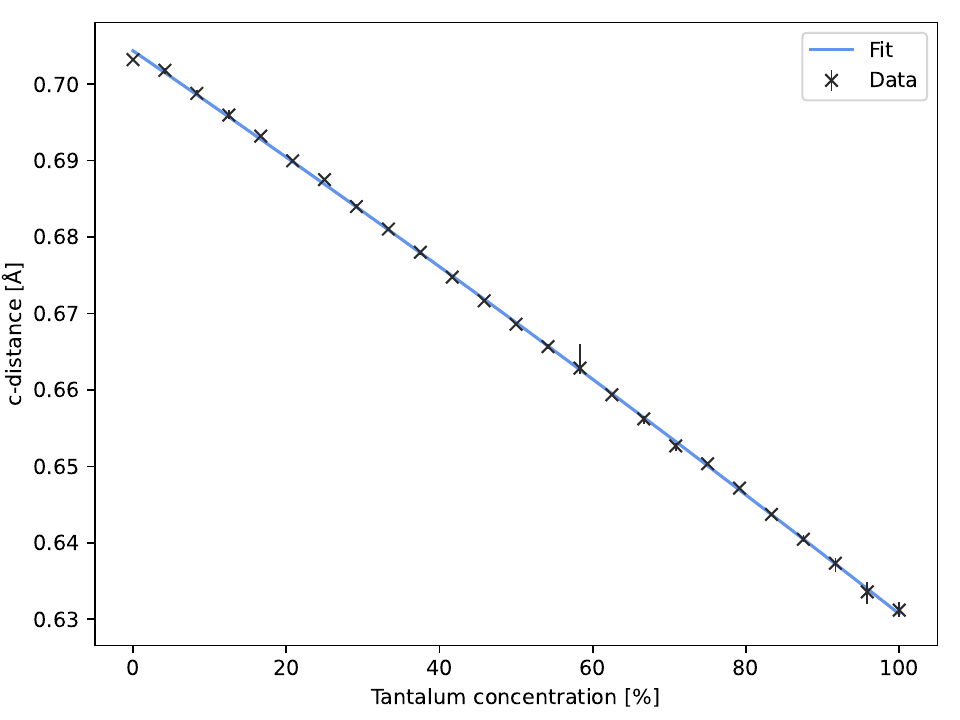}%
\caption{\label{fig:Li_dist}Average distance of the Li ions from the closest oxygen 
layer perpendicular to the $c$-axis in LiNb$_{1-x}$Ta$_x$O$_3$ solid solutions as 
calculated within DFT as a function of the 
composition. The decreasing distances thus denote a movement towards the oxygen layer. 
The latter represents the average position of the Li atoms in the paraelectric phase.}
\end{figure}

Figure \ref{fig:Li_dist} shows the average distance of the Li ions from the closest 
oxygen layer perpendicular to the $c$-axis as a function of the Ta content. The 
roughly linearly decreasing distances thus denote a movement towards the oxygen 
layer, representing the average position of the Li atoms in the paraelectric phase. 
Also in this case the composition dependence of the average position points out a
decreasing spontaneous polarization from LiNbO$_3$ to LiTaO$_3$.

\begin{figure}
\includegraphics[width=0.48\linewidth]{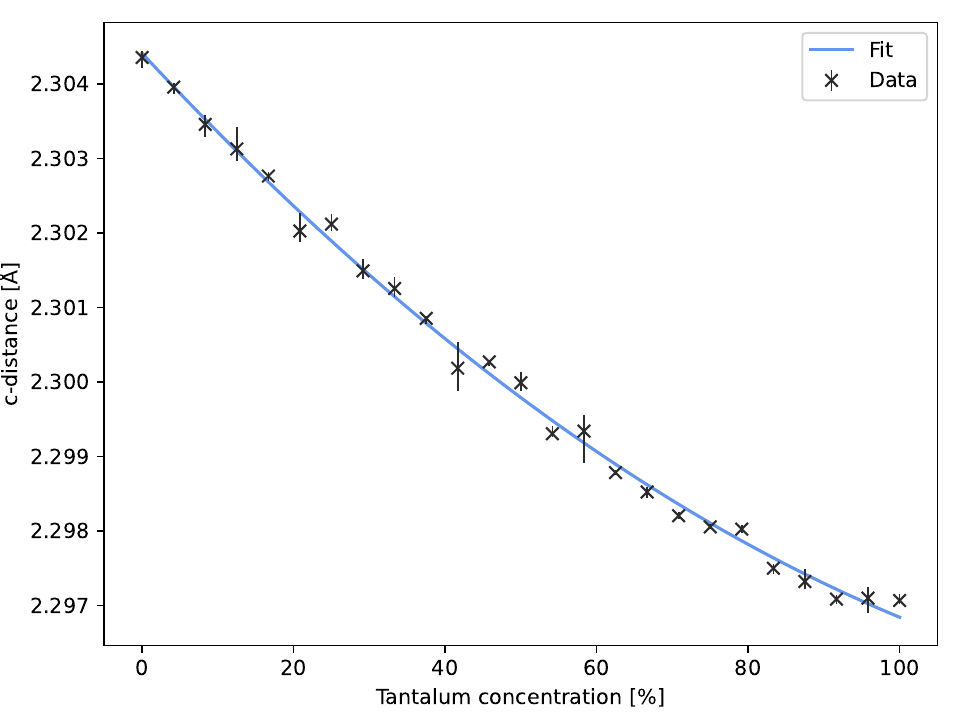}%
\caption{\label{fig:OO_dist}Average distance between the oxygen 
layers perpendicular to the $c$-axis in LiNb$_{1-x}$Ta$_x$O$_3$ solid solutions as 
calculated within DFT as a function of the 
composition. The interlayer distances are only slightly affected by the composition,
although a minor decrease can be discriminated, suggesting that the oxygen octahedra 
are slightly more compressed for increasing Ta concentration.}
\end{figure}

Finally, figure \ref{fig:OO_dist} shows the average distance between the oxygen
layers perpendicular to the $c$-axis as a function of the composition. 
The interlayer distances are only slightly affected by the composition,
although a very minor, sublinear decrease can be discriminated. This 
suggests that the oxygen octahedra are slightly more compressed for increasing 
Ta concentration.

Summarizing, the relative position of anions and cations in LiNb$_{1-x}$Ta$_x$O$_3$ 
solid solutions varies roughly linearly from LiNbO$_3$ to LiTaO$_3$ and explains,
at least qualitatively, the corresponding decrease of the spontaneous polarization.

\subsection{Electronic properties}

Vegard’s law holds not only for the lattice parameters, but also for the fundamental 
electronic bandgap of many binary semiconducting systems at thermal equilibrium:

\begin{equation}
\label{eq:vegard_gap}
E^g_{\text{LiNb}_{1-x}\text{Ta}_x\text{O}_3}=(1-x)\cdot E^g_{\text{LiNbO}_3 }+ x\cdot E^g_{\text{LiTaO}_3}.
\end{equation}

Previous calculations modelling a few compositions of LNT solid solutions as small, 
ordered crystals have predicted deviations from the behavior described by equation  
\ref{eq:vegard_gap} \cite{Riefer13SS}. Deviations from the Vegard's law can be quantified 
by a bowing parameter $b$, which can be negative (sub-Vegard behavior) or positive 
(super-Vegard behavior):

\begin{equation}
\label{eq:nonvegard_gap}
E^g_{\text{LiNb}_{1-x}\text{Ta}_x\text{O}_3}=(1-x)\cdot E^g_{\text{LiNbO}_3 }+ x\cdot E^g_{\text{LiTaO}_3} +bx(1-x).
\end{equation}

In Ref. \cite{Riefer13SS}, bowing factors of -0.6\,eV and -0.3\,eV are calculated 
within the independent particle approximation (IPA) and within the independent 
quasiparticle approximation (IQA), respectively. The SQS employed in this work allow
to investigate this aspect more precisely.

\begin{figure}
\includegraphics[width=0.65\linewidth]{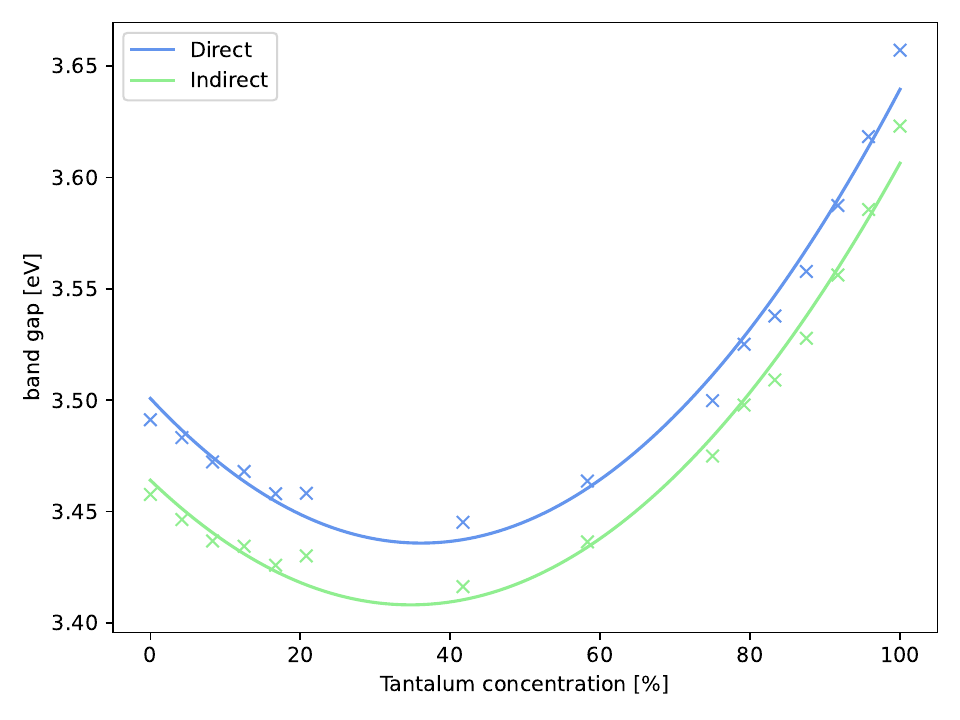}%
\caption{\label{fig:bandgap}Direct as well as indirect electronic bandgap of the 
LiNb$_{1-x}$Ta$_x$O$_3$ solid solutions calculated within DFT in the independent particle
approximation as a function of the composition. The direct bandgap occurs always at the 
$\Gamma$ point. For the sake of clarity, only data points calculated within SQS of the
same supercell symmetry are shown.} 
\end{figure}

The electronic band structures of the solid solutions (not shown in this work), 
feature the same flat dispersion known for the end compounds LiNbO$_3$ and LiTaO$_3$ 
\cite{Krampf_2021}. The direct bandgap (which is always located at the $\Gamma$-point, 
independently on the concentration and on the particular SQS) and the indirect bandgap 
as calculated within DFT in the independent particle approximation are shown in 
figure \ref{fig:bandgap}. 
Deviations from Vegard’s behavior are evident, with a bowing factor quantified in
-0.5\,eV (both for the direct and indirect bandgap), in substantial agreement with Ref. \cite{Riefer13SS}.
The (indirect) bandgap minimum is predicted for solid LNT solutions with a Ta
content of 34.3\%.

\begin{figure}
\includegraphics[width=0.48\linewidth]{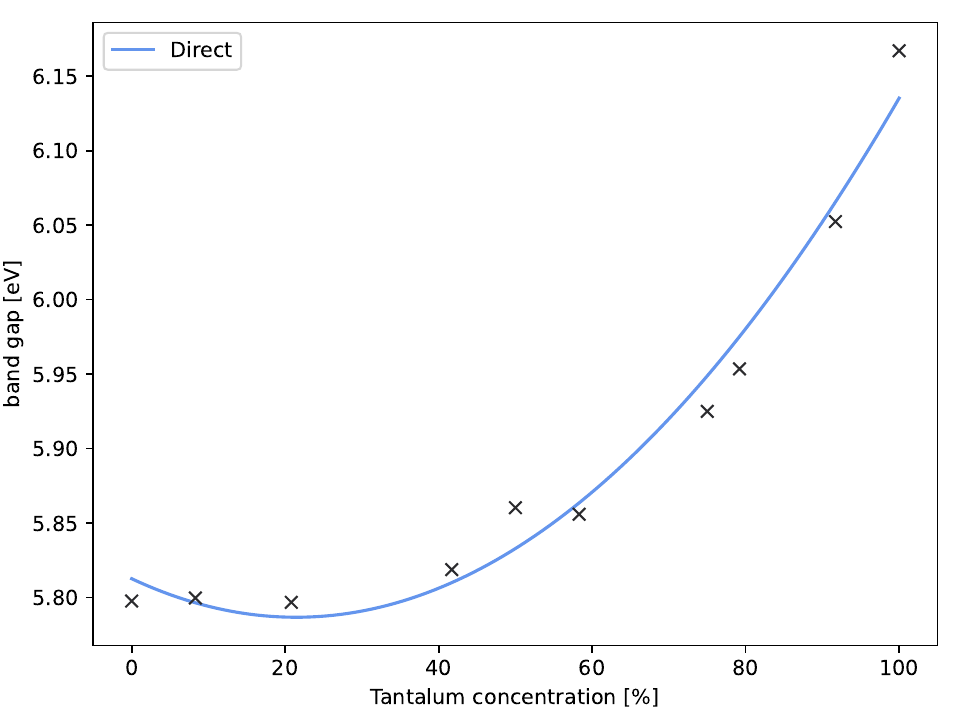}%
\includegraphics[width=0.48\linewidth]{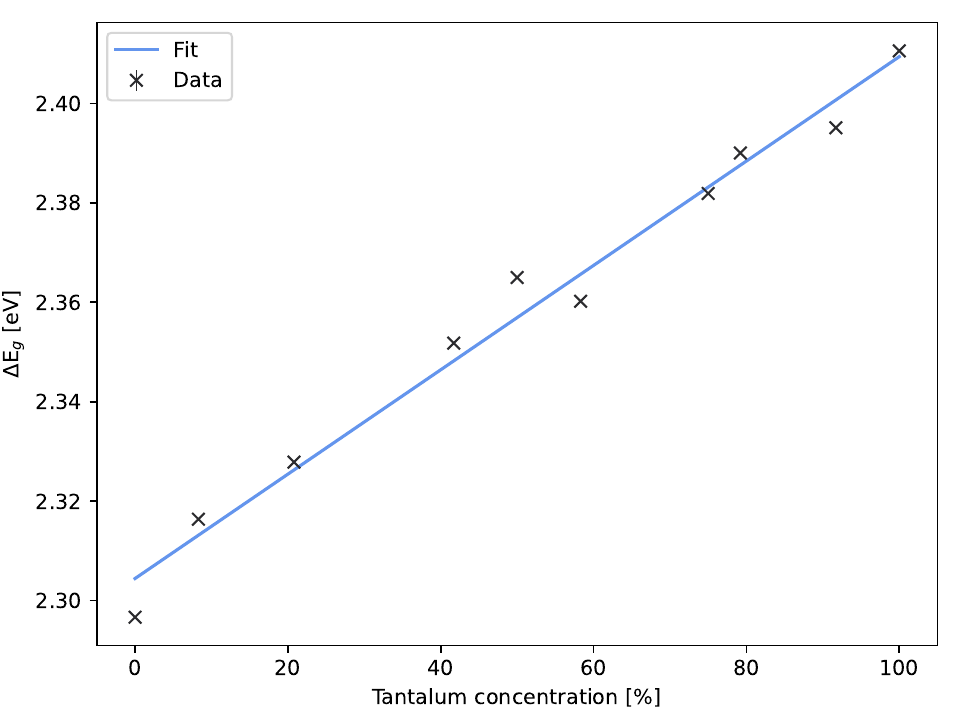}%
\caption{\label{fig:bandgap_GW}(lhs) Direct electronic bandgap of the
LiNb$_{1-x}$Ta$_x$O$_3$ solid solutions calculated within DFT in 
the independent quasiparticle approximation (G$_0$W$_0$) as a 
function of the composition. (rhs) Difference of the IPA and IQA bandgap as 
a function of the composition. Many-body effects affect Ta rich compositions 
more than Nb rich compositions.}
\end{figure}

Figure \ref{fig:bandgap_GW} (lhs) shows the composition dependence of the fundamental 
gap if many-body effects are considered. As expected, the GWA yields a larger direct 
electronic band gap compared to the DFT-IPA calculations. 
The broadening of the gap depends linearly on the composition and is more pronounced 
for higher Ta concentrations, as shown in figure \ref{fig:bandgap_GW} (rhs). The linear
dependence of the quasiparticle shifts on the concentration does not substantially
modify the magnitude of the bowing parameter (-0.56\,eV), however the 
minimum of the electronic bandgap is shifted to lower concentrations (21.2\%).

The predicted nonlinear dependence of the direct and indirect bandgap on the Ta 
content is highly relevant, as it non trivially affects the optical response of 
LNT based electro-optic devices and waveguides. Although a similar dependence is known at
high \cite{Becker24} or room temperatures from nanoparticles \cite{MircoNLO24}, 
we verify experimentally the theoretical predictions for single domain crystals 
in the limit of 0\,K. 
Thereby we perform reflectance spectroscopy measurements at low temperatures and
extrapolate the 0\,K behavior as described in the experimental section. 
 
\begin{figure}
\includegraphics[width=0.65\linewidth]{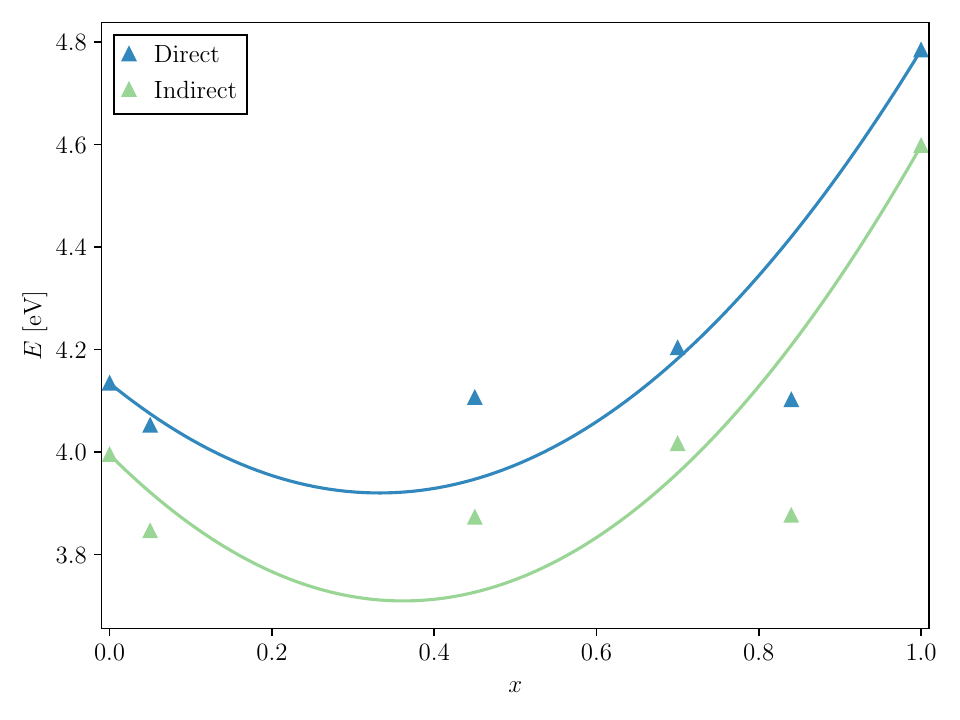}%
\caption{\label{fig:bandgap_exp} Composition dependence of the direct (blue) 
and indirect (green) optical transitions at absolute zero according to the model 
of O'Donnel and Chen \cite{ODonnel91}. The solid lines are fits of the bowing 
modified Vegard law to the respective data points with a strong sub-Vegard 
behavior.}
\end{figure}

The band gap values $E_0$ depicted in blue and green figure \ref{fig:bandgap_exp} show the values 
for the found direct and indirect electronic transitions, respectively. Although some uncertainty 
may derive from the chosen fitting procedure, the measurements clearly demonstrate 
a deviation from the Vegard behavior, confirming the DFT calulations. In 
further agreement with the theoretical results, the composition dependence of the
direct and indirect bandgap is nearly identical, and the direct gap can be obtained
in first approximation by a rigid vertical shift of about 0.3\,eV of the indirect bandgap, 
almost irrespective of the concentration.

Differently from the theoretical predictions, which can be nicely fitted by a second
order polynomial, the experimental data display a certain scattering. We trace the
scattering to a inherently different crystal quality. Due to the highly challenging 
growth procedure, different samples may indeed feature, e.g., homogeneity differences.

When comparing theoretical and experimental values, a word of caution is in 
order. We remark that we are comparing the calculated electronic bandgap
with the measured optical bandgap. The latter includes excitonic effects
that are not considered in our theoretical models.
Although the agreement between experimental and theoretical data mut be 
considered of 
qualitative nature, the measurements put the nonlinear dependence of the 
fundamental bandgap on the Ta content beyond any doubt. Unfortunately,
they do not reveal the origin of the nonlinear dependence.

\begin{figure}
\includegraphics[width=0.73\linewidth]{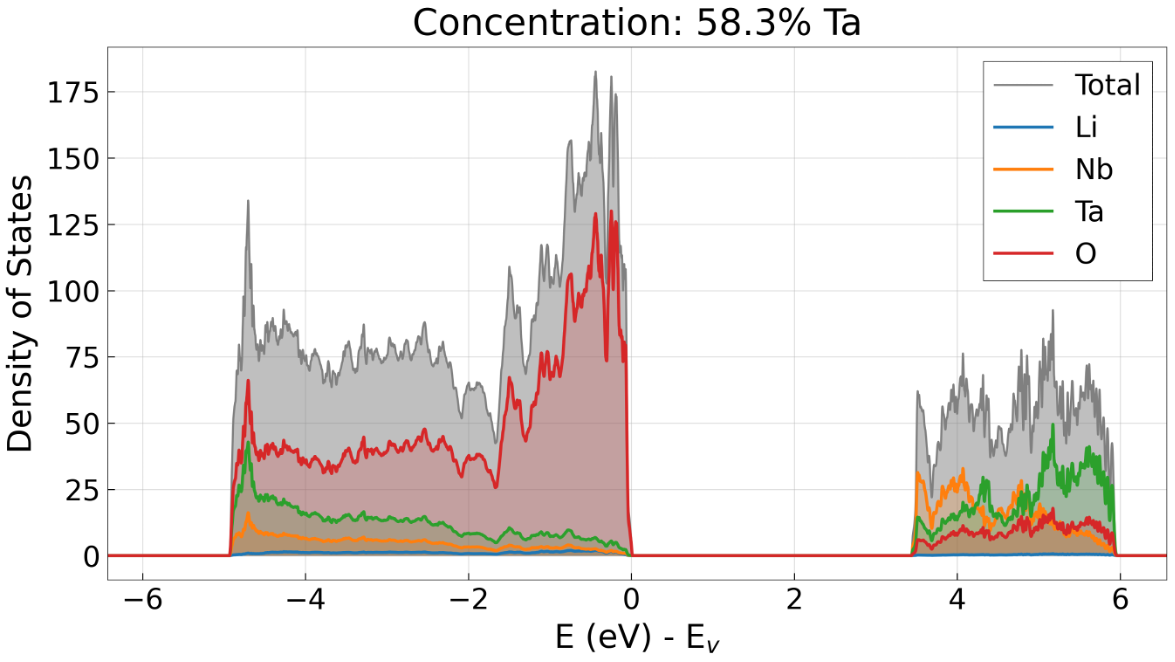}%
\caption{\label{fig:dos} Total (gray) and site projected DOS calculated
within DFT-PBEsol for the LiNb$_{41.7}$Ta$_{58.3}$O$_3$ solid solution.}
\end{figure}

\begin{figure}
\includegraphics[width=\linewidth]{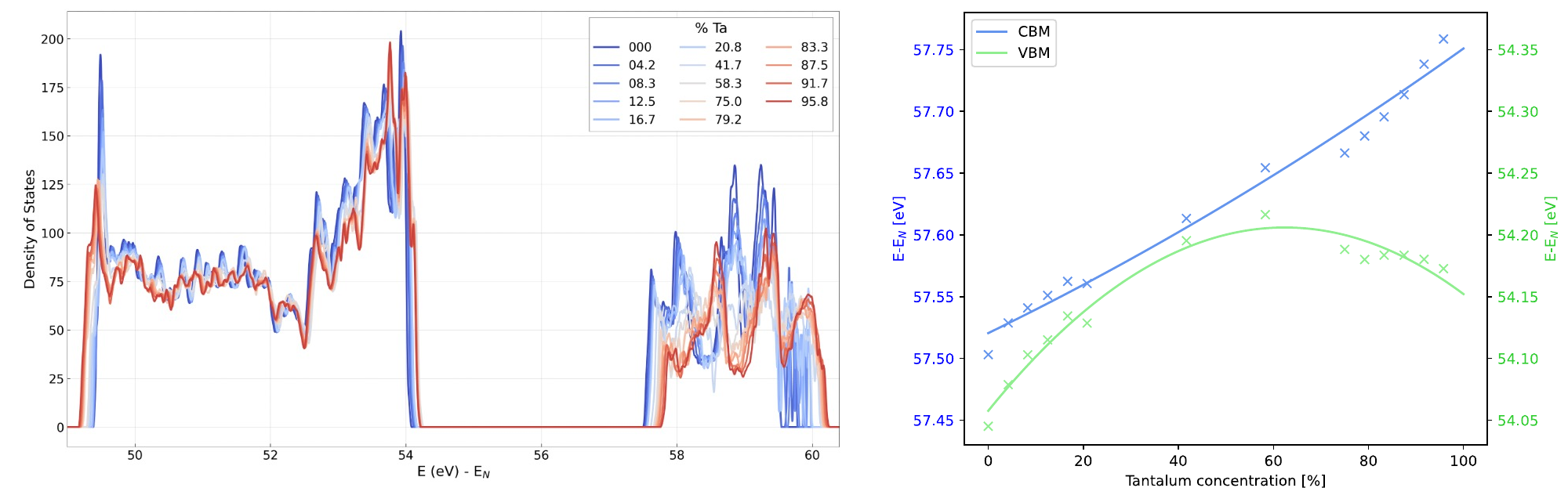}%
\caption{\label{fig:dosall} Left panel:
Total DOS of the LiNb$_{1-x}$Ta$_x$O$_3$ solid solutions as a function of the composition.
Right panel: Composition dependent position of the valence band maximum (in green) and 
conduction band minimum (in blue). The solid lines are polynomial fits of the calculated 
data and serve only as a guide to the eye.}
\end{figure}

The density of states provides an interpretation for this behavior. The upper panel 
of Figure \ref{fig:dos} shows the DOS exemplarily for the LiNb$_{0.42}$Ta$_{0.58}$O$_3$ 
solid solution (grey line) as well as the site projected DOS, demonstrating the contribution
of the single species in an energy region around the fundamental gap. The partial DOS 
is calculated by projection of the electronic wavefunction $\phi_{n,\vec{k}}$ on
spherical harmonics $Y_{lm}^{\alpha}$ centered at ion $\alpha$. The DOS reveals that
the valence band top has mainly O character, while the conduction band bottom is a O-Nb/Ta 
hybrid, as known for the parent compounds LN and LT \cite{SimoMD23}, 

The lhs panel of Figure \ref{fig:dosall} shows the total DOS of LNT solid solutions as
a function of the composition. Interestingly, the band edges shift toward higher energies 
with the Ta concentration. However, while the conduction band minimum shifts rather 
linearly, the position of the valence band maximum is a nonlinear function of the 
concentration. In particular, for low Ta content the valence band edge rapidly moves 
upwards, thus reducing the fundamental gap. For higher Ta content, the valence bad 
maximum hardly moves, while the conduction band minimum still linearly rises with the 
Ta concentration, so that the electronic bandgap becomes larger. This dependence is 
explicitly shown in the rhs panel of Figure \ref{fig:dosall}.

\subsection{Thermodynamical properties}

Further thermodynamical properties of the solid solutions are derived from the
lattice dynamics calculated as described in the methodological section. The phonon
dispersion of the end compounds calculated within the described approach can be 
found elsewhere \cite{SimoMD23}.
The heat capacity, e.g., can be computed in harmonic approximation from the dynamical properties
(i.e., with the knowledge of the phonon dispersion) \cite{Friedrich15} according to equation 
\ref{eq:cv}:

\begin{equation}
\label{eq:cv}
c_V=\left(\frac{\partial U}{\partial T}\right)_V=\sum_{\vec{q},\nu} k_B \left(\frac{\hbar\omega(\vec{q},\nu)}{k_BT}\right)^2
\frac{\exp\left(\frac{\hbar\omega(\vec{q},\nu)}{k_BT}\right)}{\left[\exp\left(\frac{\hbar\omega(\vec{q},\nu)}{k_BT}\right)-1\right]^2}.
\end{equation}

where $\omega_{\vec{q},\nu}$ denotes the frequency of phonon mode $\nu$ at q-point 
$\vec{q}$, and $k_\text{B}$ and $T$ denote the Boltzmann constant and the temperature, 
respectively.

\begin{figure}
\includegraphics[width=0.48\linewidth]{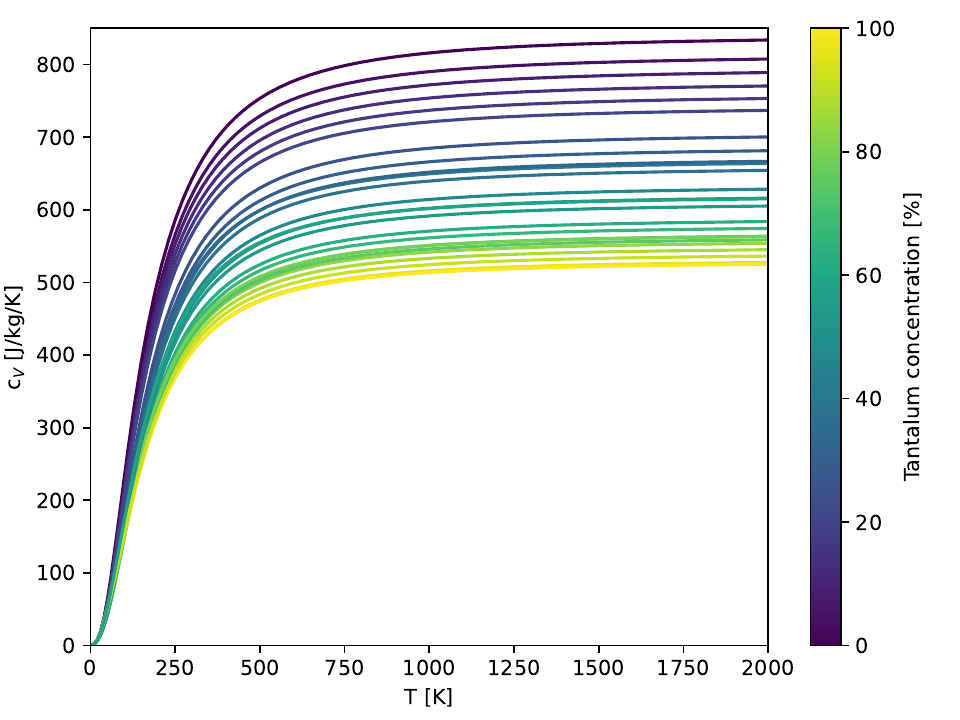}%
\includegraphics[width=0.48\linewidth]{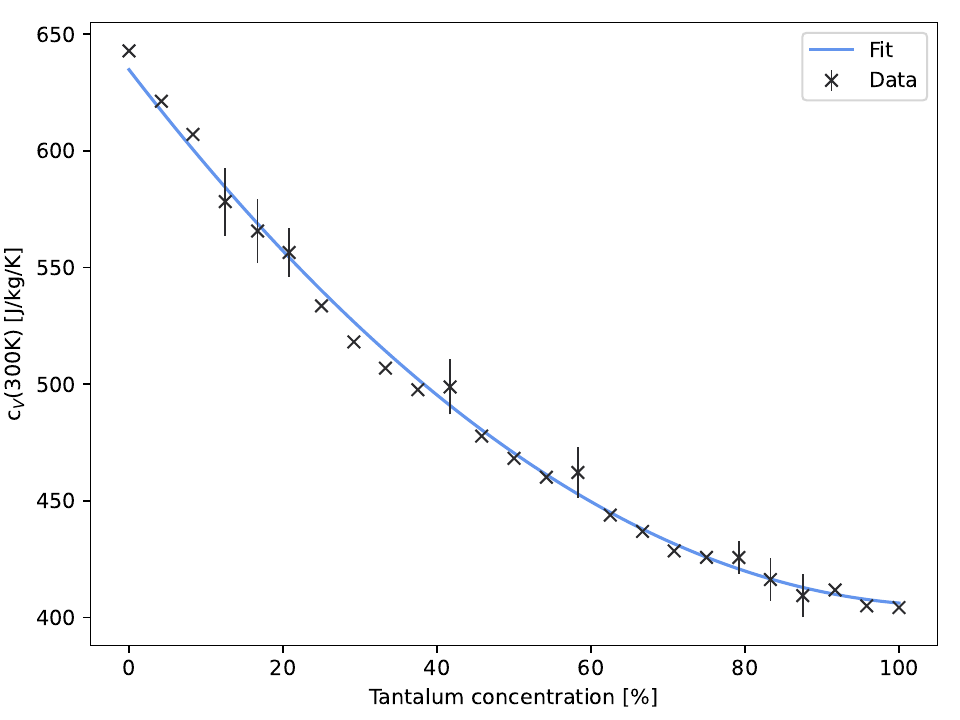}%
\caption{\label{fig:c_v} Left panel: DFT calculated heat capacity $c_V$ of the 
LiNb$_{1-x}$Ta$_x$O$_3$ solid solutions as a function of the temperature and of 
the composition. Right panel: DFT calculated heat capacity at 300\,K as a function
of the composition.} 
\end{figure}

The heat capacity continuously decreases with the Ta content from LiNbO$_3$ to 
LiTaO$_3$, as shown in figure \ref{fig:c_v} (lhs). Considering $c_V$ at a given 
temperature, e.g., 300\,K as shown in figure \ref{fig:c_v} (rhs), it becomes 
clear that the heat capacity decreases sub-linearly with the Ta content. The 
values calculated for the end compounds LiNbO$_3$ (0.64\,Jg$^{-1}$K$^{-1}$) and
LiTaO$_3$ (0.40\,Jg$^{-1}$K$^{-1}$) are in very good agreement with 
experimental (0.70\,Jg$^{-1}$K$^{-1}$ at 303\,K for nearly stoichiometric 
LiNbO$_3$ \cite{YAO2008501} and 0.42\,Jg$^{-1}$K$^{-1}$ for LiTaO$_3$
\cite{Korth}) as well as theoretical results \cite{Friedrich15} available in the 
literature for $c_P$. 

In order to extract the Debye temperature $T_D$ from the calculated heat capacity, 
the result is fitted to the Debye model:

\begin{equation}
\label{eq:debye}
c_V=9nk_B\left(\frac{T}{T_D}\right)^3\int^{T_D/T}_0 \frac{x^4e^x}{(e^x-1)^2} dx.
\end{equation}

Due to the model character of equation \ref{eq:debye}, the numerical results 
strongly depend on the temperature range considered for the fit. Performing
the fit between 0 and 20\,K, e.g., a Debye Temperature $T_D=572$\,K for LiNbO$_3$ 
is obtained, which perfectly matches a previously calculated value of 574\,K \cite{Friedrich15}
and is in good agreement with the experimentally determined value of 593\,K \cite{Villar86}.
Moreover, our calculations show that the Debye temperature $T_D$ of the solid 
solutions grows roughly linearly with the Ta content.

\subsection{Optical properties}

\begin{figure}
\includegraphics[width=0.48\linewidth]{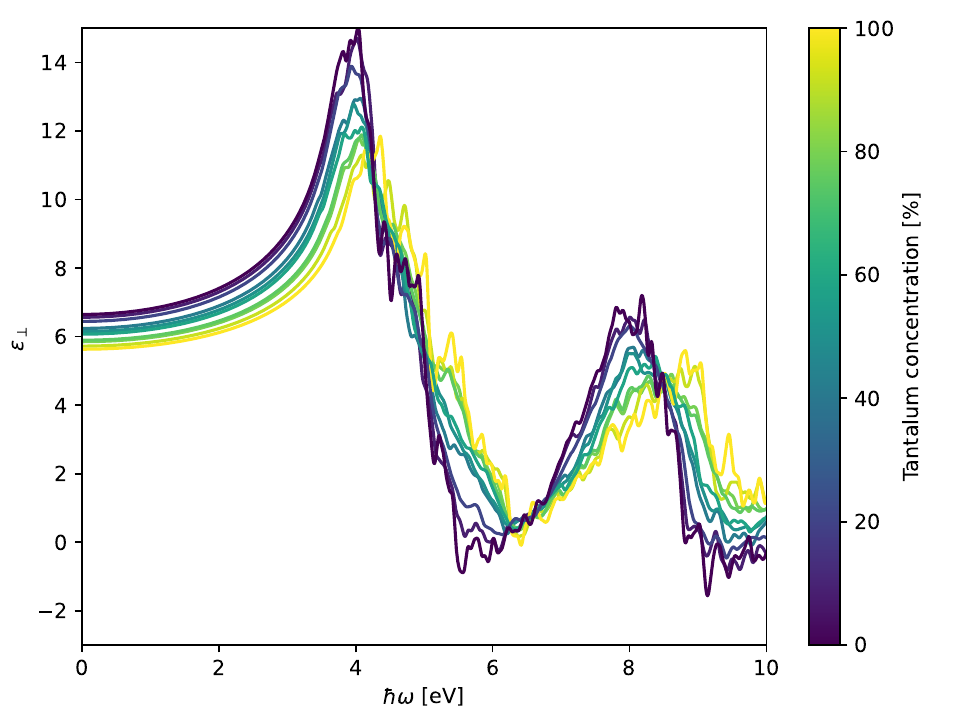}\includegraphics[width=0.48\linewidth]{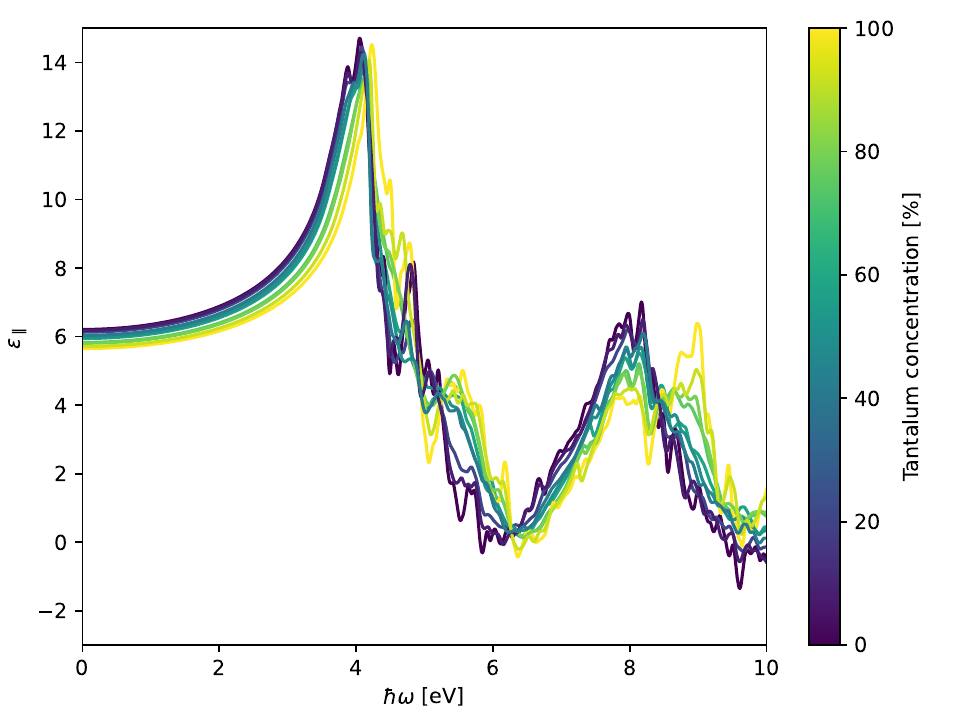}%%

\includegraphics[width=0.48\linewidth]{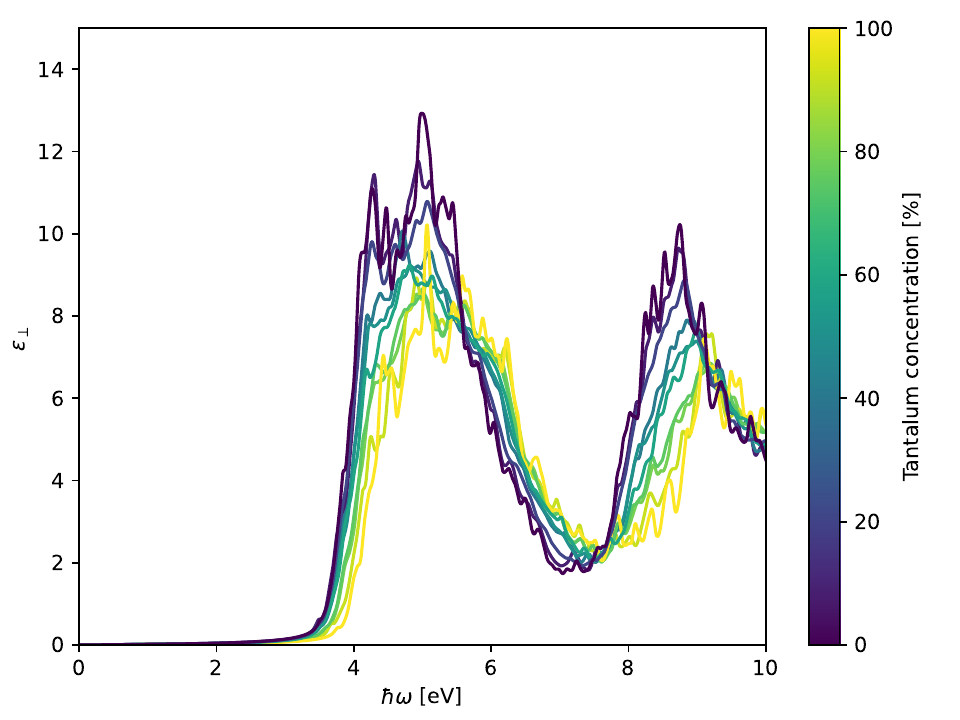}\includegraphics[width=0.48\linewidth]{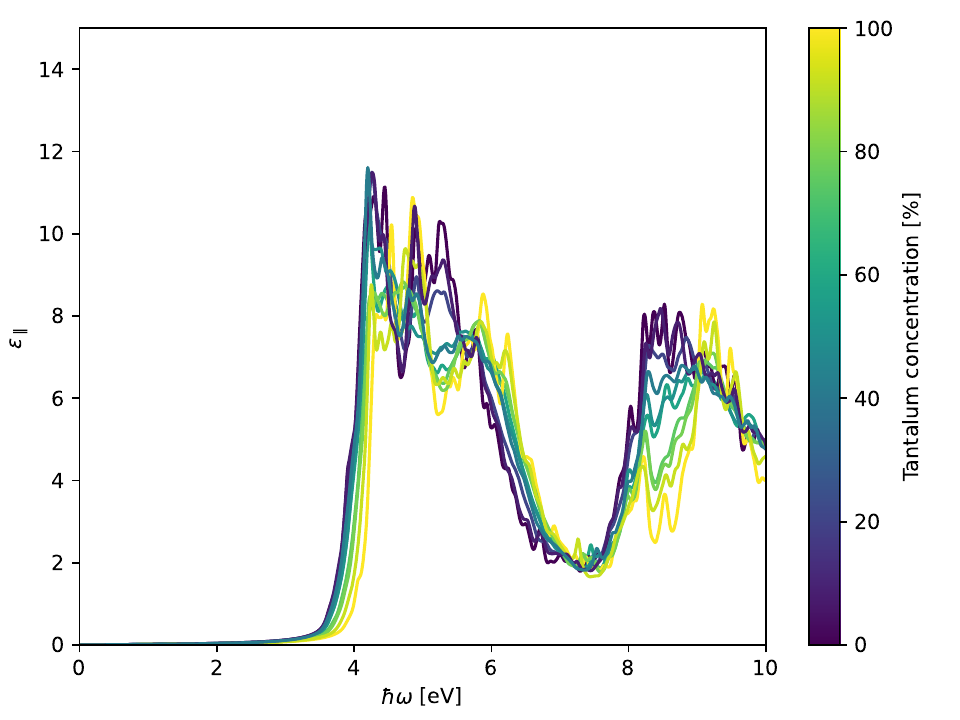}%%
\caption{\label{fig:dielectric} Real (upper row) and imaginary part (lower row) of the dielectric function
of ferroelectric LiNb$_{1-x}$Ta$_x$O$_3$ solid solutions calculated within DFT in the IPA. The panels at
the left and at the right hand side show the ordinary and extraordinary component of the dielectric 
function, respectively.}
\end{figure}

The linear optical response of the LNT solid solutions in their ferroelectric phase is 
discussed on the basis of the calculated dielectric function. The latter is displayed 
in figure \ref{fig:dielectric} as a function of the composition. The upper row of 
figure \ref{fig:dielectric} shows the real part and the bottom row the imaginary part 
of the IPA calculated dielectric function. As well, the panels on the left hand side
show the ordinary optical direction ($\varepsilon_{xx}$) and the panels on the right 
hand side show the extraordinary optical direction ($\varepsilon_{zz}$). The linear 
optical response of LiNbO$_3$ at the IPA level of accuracy (black curve) features the
two main absorption bands at about 5\,eV and 8.5\,eV known from previous investigations,
e.g., Ref. \cite{ArthurRiefer13}. Also the double peak structure of the main absorption
band, with a different relative intensity of the two peaks in the ordinary and 
extraordinary optical direction, is in close agreement with previous investigations. 
LiTaO$_3$ (yellow curves in figure \ref{fig:dielectric}) features an absorption spectrum 
that is rather similar to the absorption spectrum of LiNbO$_3$, however with all 
spectral signatures shifted at higher energies due to the somewhat larger electronic
bandgap. Between the two end compounds, the linear optical answer of the 
ferroelectric LiNb$_{1-x}$Ta$_x$O$_3$ mixed crystals substantially interpolates
the features of LiNbO$_3$ and LiTaO$_3$, however, the energetically lowest absorption 
edge is given by solid solution with about 40\% Ta content. This mirrors the 
deviations from the Vegard behavior of the fundamental bandgap previously discussed.
Moreover, the experimentally determined absorption edge is underestimated due to 
the DFT underestimation of the electronic bandgap.

A more refined calculation of the optical answer can be obtained, e.g., including 
self-energy effects within the GW approximation. Unfortunately, a calculation of the 
full optical spectra for all concentrations and the employed cell size is beyond 
todays computational power. Nonetheless, to estimate magnitude of the many-body
effects on the LNT mixed crystals we calculate the quasiparticle shifts of the 
IPA-calculated electronic eigenvalues. We discuss exemplarily as a spot
check the LiNb$_{0.42}$Ta$_{0.58}$O$_3$ composition. All other compositions,
including the end compounds LiNbO$_3$ and LiTaO$_3$ behave very similarly.

\begin{figure}
\includegraphics[width=0.46\linewidth]{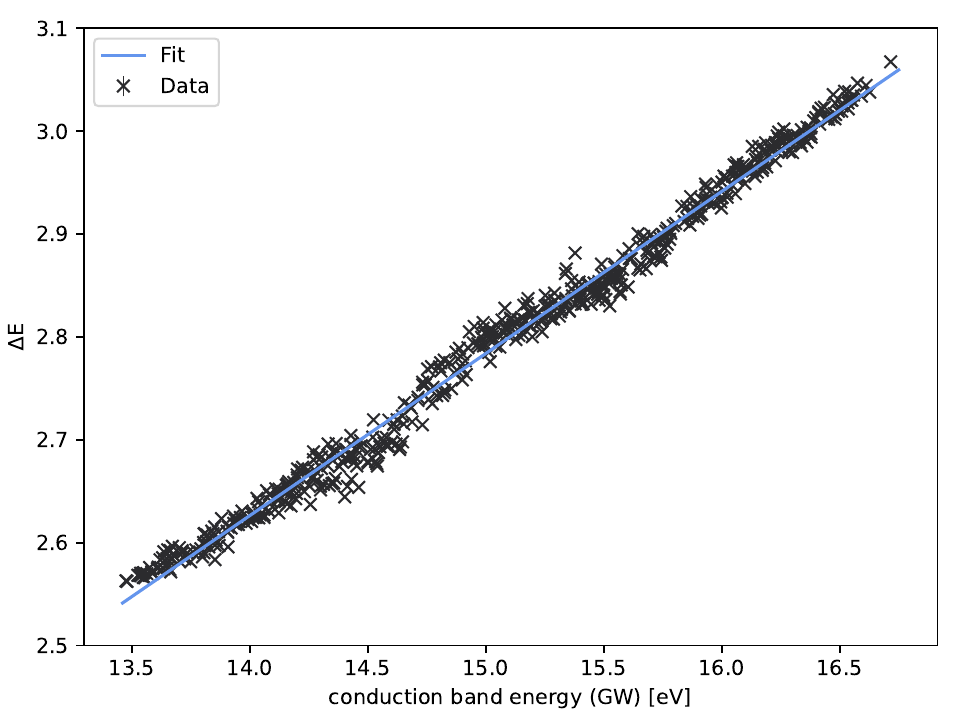}\includegraphics[width=0.46\linewidth]{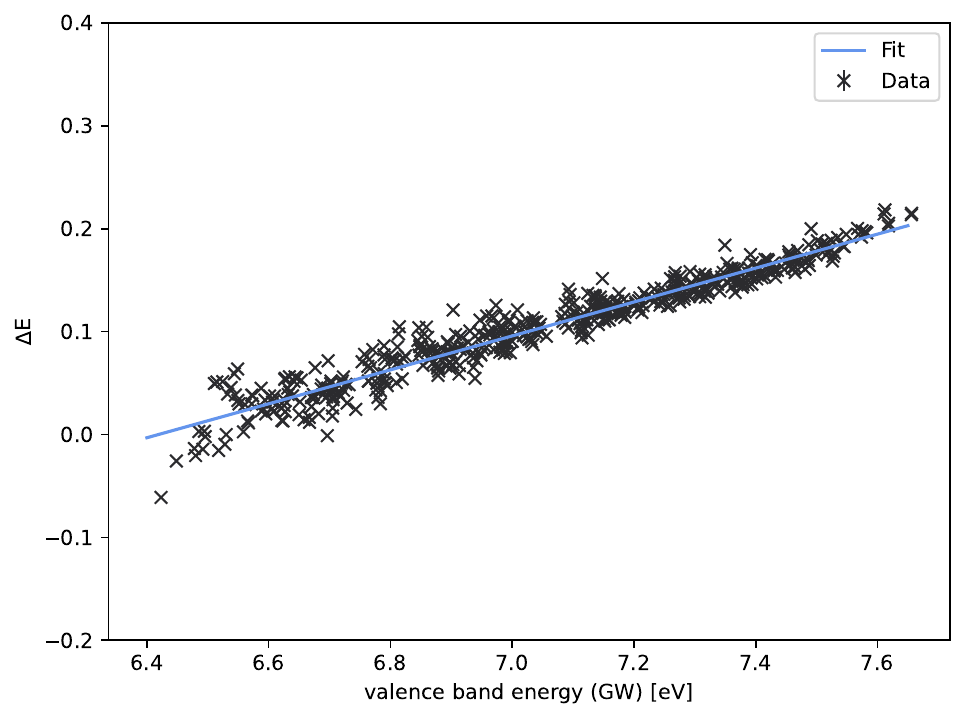}
\caption{\label{fig:IQA} Quasiparticle shifts calculated with respect to the DFT-IPA
band energy for the ferroelectric LiNb$_{1-x}$Ta$_x$O$_3$ solid solutions. The conduction
bands are more affected by many-body effects than the valence bands.}
\end{figure}

The energy shifts at each point of the Brillouin zone  
with respect to the band positions in the IPA are shown in
figure \ref{fig:IQA}. The quasiparticle shifts are nearly independent 
from the k-point, so that the bands are mainly rigidly translated.
Valence bands are less affected by many-body effects than conduction bands.
The magnitude of the translation is roughly proportional to the band energy, 
with the shift of the conduction band minimum of ca. 2.55\,eV in agreement with 
available literature data \cite{ArthurRiefer13}.

The rigid shift of the single bands is not surprising, as the valence band and the conduction
band edges have --witin a given composition-- the same orbital nature over the whole Brillouin
zone. As shown, e.g., in the DOS of figure \ref{fig:dos}, the valence band is dominated by
oxygen states and the conduction band by Nb/Ta-O hybrids. The present calculations are very 
similar to our previous results for LiNbO$_3$, where the self-energy 
was also evaluated within the frequency-dependent random-phase approximation employed here
\cite{ArthurRiefer13}. As the many-body effects mainly translate all spectral features to higher
energies (independently from the composition), their effect on the adsorption edge 
might be correctly simulated by a 
computationally convenient scissors shift in upcoming calculations. 

One of the peculiarities of the LNT solid solutions is their optical birefringence, 
defined as the difference between the extraordinary and ordinary refractive index:

\begin{equation}
\label{eq:birefringence}
\Delta n = (n_e-n_o).
\end{equation}

Indeed, LiNbO$_3$ shows negative birefringence, while LiTaO$_3$ has a positive 
birefringence. This suggests the existence of some concentration with vanishing 
birefringence, resulting in a highly unusual combination of ferroelectricity and 
optical isotropy \cite{Wood08}.

\begin{figure}
\includegraphics[width=0.48\linewidth]{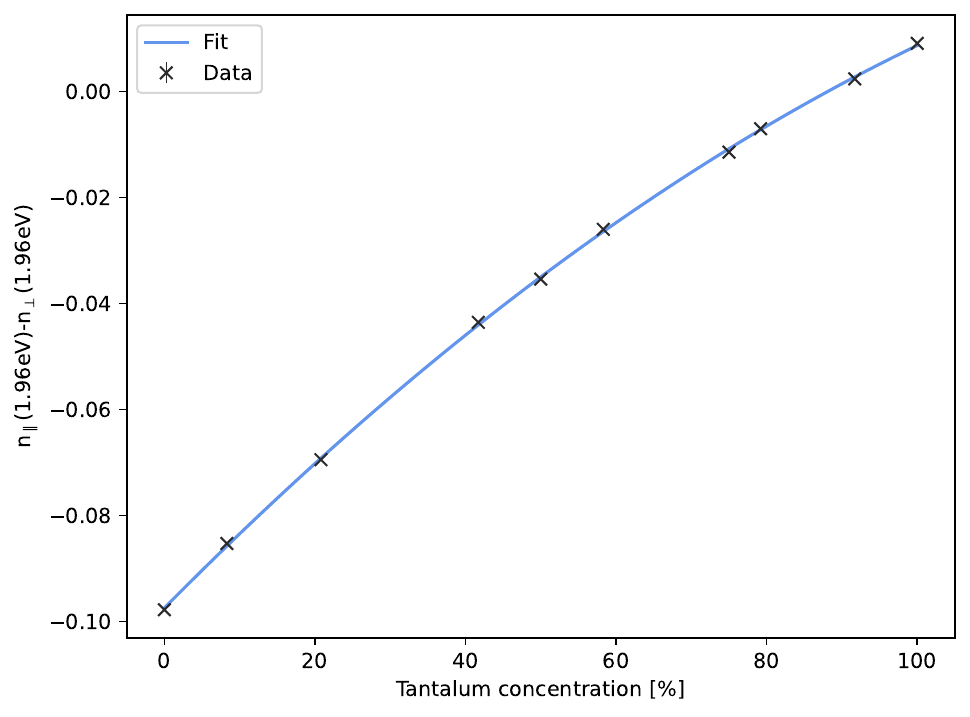}
\caption{\label{fig:birefringence} Birefringence of ferroelectric LiNb$_{1-x}$Ta$_x$O$_3$ 
solid solutions as a function of the composition calculated within DFT in the IPA for a 
laser wavelength of 633\,nm.}
\end{figure}

With the knowledge of the dielectric function as shown in figure \ref{fig:dielectric}
and equation \ref{eq:brechzahl}, we calculate the birefringence of the 
LiNb$_{1-x}$Ta$_x$O$_3$ solid solutions as a function of $x$ for a wavelength 
of 633\,nm (1.96\,eV) typical for the commonly diffused He:Ne lasers. The corresponding 
data is shown in figure \ref{fig:birefringence}. The birefringence is a slightly super 
linear function of the concentration, as experimentally found by Wood \textit{et al.} 
\cite{Wood08} and firstly predicted by Riefer \textit{et al.} \cite{ArthurRiefer13}.  
Our calculations reveal that zero birefringence occurs for a composition of $x=0.89$. 
This is in qualitative agreement both with the room temperature measurements of LNT 
bulk by Wood \textit{et al.} \cite{Wood08}, who determined the accidentally-isotropic 
point at $x = 0.94$ and with measurements on LiTa$_x$Nb$_{1-x}$O$_3$ thin films of 
by Kondo \textit{et al.} \cite{Kondo79}, who reported the accidentally-isotropic 
composition at about $x = 0.93$ (also for a laser wavelength of 633\,nm).
We remark that at this composition the crystal is electrically anisotropic (it still
has a permanent, spontaneous macroscopic polarization directed along the crystallographic
[0001] axis) and optically anisotropic, a seldom and unusual combination.

\section{\label{sec:conclusions}Conclusions}

In order to theoretically characterize LiNb$_{1-x}$Ta$_x$O$_3$ solid solutions as a function of the 
composition, a set of special quasirandom structures spanning the whole composition range is created.
With the help of the SQS, the structural parameters of the mixed crystals are calculated and deviations from
the Vegard behavior are demonstrated. Moreover, the composition dependent positions of the cations
within the oxygen octahedra qualitatively explain the differences in the spontaneous polarization.
An analysis of the electronic structures also reveal deviations from the Vegard law. These arise
from the fact that the conduction band shifts linearly with the Ta content, while valence band does not,
resulting in an overall sublinear behavior. The predicted composition dependence of the fundamental  
bandgap is experimentally verified by specifically performed optical spectroscopy measurements, which
are in qualitative agreement with the calculations. The determined value of the bowing parameter can 
be employed to design devices such as photodetectors based on the LiNb$_{1-x}$Ta$_x$O$_3$ solid 
solutions. 

The linear optical answer of the mixed crystals is quantified by the calculation of the dielectric 
function (ordinary and extraordinary component). The latter are calculated in the independent
particle  approximation. The dielectric functions of the end compounds LiNbO$_3$ and LiTaO$_3$ are in 
agreement with existing results, while the intermediate compositions show a moderate  dependence 
on the Ta content.
The inclusion of self-energy effects within the GWA opens the bandgap by more than 2\,eV.
The birefringence of the mixed crystals is shown to depend in first approximation 
linearly on the Ta concentration
and to increase from the negative values of the Nb rich samples to the positive values of the Ta rich 
samples. The particular composition with $x$=0.89 is characterized by an optical
isotropy that is quite peculiar for ferroelectrics.
Thermodynamical quantities such as the specific heat $c_V$ or the Debye temperature $T_D$ 
of the LiNb$_{1-x}$Ta$_x$O$_3$ mixed crystals are calculated as a function of the composition as 
well. While $T_D$ increases roughly linearly with $x$, $c_V$ decreases sub-linearly with 
the Ta content.

%%%%%%%---------------Ausblick
Although this work provides a rather comprehensive characterization of the 
LiNb$_{1-x}$Ta$_x$O$_3$ solid solutions, our knowledge of this appealing
material class is far from being complete. Both the end compounds LiNbO$_3$ and
LiTaO$_3$ are characterized by a strong Li deficiency, resulting in the so 
called congruent composition. The congruently grown crystals differ in many
aspects from the stoichiometric samples. Whether the LNT solid solutions
crystallize in a congruent composition remains to be settled. Similarly, 
little is known about the defect structure of the mixed crystals, both 
concerning intrinsic defects and doping, which are both known to massively
impact the materials properties of LN and LT. A somewhat related point regards
the homogeneity of the solid solutions. Within the SQS approach, we model by
definition a random distribution of the cations. However, clustering of Ta
or Nb in local regions with a composition similar to that of the end compounds is
also conceivable. This and other aspects, including in particular excitonic 
effects in the linear optical response or the 
temperature dependence of the materials properties, must be explored in 
future investigations.

\begin{acknowledgments}
We gratefully acknowledge financial support by the Deutsche Forschungsgemeinschaft (DFG) through
the research group FOR5044 (Grant No. 426703838 \cite{FOR5044}, SA1948/3-1, IM37/12-1).
Calculations for this research were
conducted on the Lichtenberg high-performance computer of the TU Darmstadt and at the
H\"ochstleistungrechenzentrum Stuttgart (HLRS). The authors furthermore acknowledge the
computational resources provided by the HPC Core Facility and the HRZ of the Justus-Liebig-Universit\"at
Gie{\ss}en.
\end{acknowledgments}

\bibliography{literature_LNT}

\begin{thebibliography}{69}%
\makeatletter
\providecommand \@ifxundefined [1]{%
 \@ifx{#1\undefined}
}%
\providecommand \@ifnum [1]{%
 \ifnum #1\expandafter \@firstoftwo
 \else \expandafter \@secondoftwo
 \fi
}%
\providecommand \@ifx [1]{%
 \ifx #1\expandafter \@firstoftwo
 \else \expandafter \@secondoftwo
 \fi
}%
\providecommand \natexlab [1]{#1}%
\providecommand \enquote  [1]{``#1''}%
\providecommand \bibnamefont  [1]{#1}%
\providecommand \bibfnamefont [1]{#1}%
\providecommand \citenamefont [1]{#1}%
\providecommand \href@noop [0]{\@secondoftwo}%
\providecommand \href [0]{\begingroup \@sanitize@url \@href}%
\providecommand \@href[1]{\@@startlink{#1}\@@href}%
\providecommand \@@href[1]{\endgroup#1\@@endlink}%
\providecommand \@sanitize@url [0]{\catcode `\\12\catcode `\$12\catcode `\&12\catcode `\#12\catcode `\^12\catcode `\_12\catcode `\%12\relax}%
\providecommand \@@startlink[1]{}%
\providecommand \@@endlink[0]{}%
\providecommand \url  [0]{\begingroup\@sanitize@url \@url }%
\providecommand \@url [1]{\endgroup\@href {#1}{\urlprefix }}%
\providecommand \urlprefix  [0]{URL }%
\providecommand \Eprint [0]{\href }%
\providecommand \doibase [0]{https://doi.org/}%
\providecommand \selectlanguage [0]{\@gobble}%
\providecommand \bibinfo  [0]{\@secondoftwo}%
\providecommand \bibfield  [0]{\@secondoftwo}%
\providecommand \translation [1]{[#1]}%
\providecommand \BibitemOpen [0]{}%
\providecommand \bibitemStop [0]{}%
\providecommand \bibitemNoStop [0]{.\EOS\space}%
\providecommand \EOS [0]{\spacefactor3000\relax}%
\providecommand \BibitemShut  [1]{\csname bibitem#1\endcsname}%
\let\auto@bib@innerbib\@empty
%</preamble>
\bibitem [{\citenamefont {Vert}\ \emph {et~al.}(2012)\citenamefont {Vert}, \citenamefont {Doi}, \citenamefont {Hellwich}, \citenamefont {Hess}, \citenamefont {Hodge}, \citenamefont {Kubisa}, \citenamefont {Rinaudo},\ and\ \citenamefont {Schu{\'e}}}]{IUPAC}%
  \BibitemOpen
  \bibfield  {author} {\bibinfo {author} {\bibfnamefont {M.}~\bibnamefont {Vert}}, \bibinfo {author} {\bibfnamefont {Y.}~\bibnamefont {Doi}}, \bibinfo {author} {\bibfnamefont {K.-H.}\ \bibnamefont {Hellwich}}, \bibinfo {author} {\bibfnamefont {M.}~\bibnamefont {Hess}}, \bibinfo {author} {\bibfnamefont {P.}~\bibnamefont {Hodge}}, \bibinfo {author} {\bibfnamefont {P.}~\bibnamefont {Kubisa}}, \bibinfo {author} {\bibfnamefont {M.}~\bibnamefont {Rinaudo}},\ and\ \bibinfo {author} {\bibfnamefont {F.}~\bibnamefont {Schu{\'e}}},\ }\bibfield  {title} {\bibinfo {title} {Terminology for biorelated polymers and applications (iupac recommendations 2012)},\ }\href {https://doi.org/doi:10.1351/PAC-REC-10-12-04} {\bibfield  {journal} {\bibinfo  {journal} {Pure and Applied Chemistry}\ }\textbf {\bibinfo {volume} {84}},\ \bibinfo {pages} {377} (\bibinfo {year} {2012})}\BibitemShut {NoStop}%
\bibitem [{\citenamefont {Hume-Rothery}\ and\ \citenamefont {Powell}(1935)}]{Rules1}%
  \BibitemOpen
  \bibfield  {author} {\bibinfo {author} {\bibfnamefont {W.}~\bibnamefont {Hume-Rothery}}\ and\ \bibinfo {author} {\bibfnamefont {H.~M.}\ \bibnamefont {Powell}},\ }\bibfield  {title} {\bibinfo {title} {On the theory of super-lattice structures in alloys},\ }\href {https://api.semanticscholar.org/CorpusID:100648687} {\bibfield  {journal} {\bibinfo  {journal} {Zeitschrift f{\"u}r Kristallographie - Crystalline Materials}\ }\textbf {\bibinfo {volume} {91}},\ \bibinfo {pages} {23 } (\bibinfo {year} {1935})}\BibitemShut {NoStop}%
\bibitem [{\citenamefont {Hume-Rothery}\ \emph {et~al.}(1969)\citenamefont {Hume-Rothery}, \citenamefont {Haworth},\ and\ \citenamefont {Smallman}}]{Rules2}%
  \BibitemOpen
  \bibfield  {author} {\bibinfo {author} {\bibfnamefont {W.}~\bibnamefont {Hume-Rothery}}, \bibinfo {author} {\bibfnamefont {C.~W.}\ \bibnamefont {Haworth}},\ and\ \bibinfo {author} {\bibfnamefont {R.~E.}\ \bibnamefont {Smallman}},\ }\href {https://worldcat.org/title/639778957} {\emph {\bibinfo {title} {The structure of metals and alloys [by] William Hume-Rothery, R.E. Smallman and C W. Haworth}}},\ \bibinfo {edition} {5th}\ ed.\ (\bibinfo  {publisher} {Institute of Metals and the Institution of Metallurgists London},\ \bibinfo {year} {1969})\BibitemShut {NoStop}%
\bibitem [{\citenamefont {Keloglu}\ and\ \citenamefont {Fedorko}(1972)}]{Keloglu1972}%
  \BibitemOpen
  \bibfield  {author} {\bibinfo {author} {\bibfnamefont {Y.~P.}\ \bibnamefont {Keloglu}}\ and\ \bibinfo {author} {\bibfnamefont {A.~S.}\ \bibnamefont {Fedorko}},\ }\bibinfo {title} {Vegard's law for some binary and pseudobinary semiconductor systems},\ in\ \href {https://doi.org/10.1007/978-1-4684-8682-7_22} {\emph {\bibinfo {booktitle} {Chemical Bonds in Solids: Volume 4: Semiconductor Crystals, Glasses, and Liquids}}},\ \bibinfo {editor} {edited by\ \bibinfo {editor} {\bibfnamefont {A.~N.~N.}\ \bibnamefont {Sirota}}}\ (\bibinfo  {publisher} {Springer US},\ \bibinfo {address} {New York, NY},\ \bibinfo {year} {1972})\ pp.\ \bibinfo {pages} {113--117}\BibitemShut {NoStop}%
\bibitem [{\citenamefont {Polimeni}\ \emph {et~al.}(2004)\citenamefont {Polimeni}, \citenamefont {Baldassarri H{\"o}ger~von H{\"o}gersthal}, \citenamefont {Masia}, \citenamefont {Frova}, \citenamefont {Capizzi}, \citenamefont {Sanna}, \citenamefont {Fiorentini}, \citenamefont {Klar},\ and\ \citenamefont {Stolz}}]{Polimeni2004}%
  \BibitemOpen
  \bibfield  {author} {\bibinfo {author} {\bibfnamefont {A.}~\bibnamefont {Polimeni}}, \bibinfo {author} {\bibfnamefont {G.}~\bibnamefont {Baldassarri H{\"o}ger~von H{\"o}gersthal}}, \bibinfo {author} {\bibfnamefont {F.}~\bibnamefont {Masia}}, \bibinfo {author} {\bibfnamefont {A.}~\bibnamefont {Frova}}, \bibinfo {author} {\bibfnamefont {M.}~\bibnamefont {Capizzi}}, \bibinfo {author} {\bibfnamefont {S.}~\bibnamefont {Sanna}}, \bibinfo {author} {\bibfnamefont {V.}~\bibnamefont {Fiorentini}}, \bibinfo {author} {\bibfnamefont {P.~J.}\ \bibnamefont {Klar}},\ and\ \bibinfo {author} {\bibfnamefont {W.}~\bibnamefont {Stolz}},\ }\bibfield  {title} {\bibinfo {title} {Tunable variation of the electron effective mass and exciton radius in hydrogenated {Ga}{As}$_{1-x}${N}$_x$},\ }\bibfield  {journal} {\bibinfo  {journal} {Physical Review B - Condensed Matter and Materials Physics}\ }\textbf {\bibinfo {volume} {69}},\ \href {https://doi.org/10.1103/PhysRevB.69.041201} {10.1103/PhysRevB.69.041201} (\bibinfo {year}
  {2004})\BibitemShut {NoStop}%
\bibitem [{\citenamefont {Sanna}\ and\ \citenamefont {Fiorentini}(2004)}]{Sanna2004}%
  \BibitemOpen
  \bibfield  {author} {\bibinfo {author} {\bibfnamefont {S.}~\bibnamefont {Sanna}}\ and\ \bibinfo {author} {\bibfnamefont {V.}~\bibnamefont {Fiorentini}},\ }\bibfield  {title} {\bibinfo {title} {Lattice constant, effective mass, and gap recovery in hydrogenated {Ga}{As}$_{1-x}${N}$_x$},\ }\bibfield  {journal} {\bibinfo  {journal} {Physical Review B - Condensed Matter and Materials Physics}\ }\textbf {\bibinfo {volume} {69}},\ \href {https://doi.org/10.1103/PhysRevB.69.125208} {10.1103/PhysRevB.69.125208} (\bibinfo {year} {2004})\BibitemShut {NoStop}%
\bibitem [{\citenamefont {Vegard}(1921)}]{Vegard1}%
  \BibitemOpen
  \bibfield  {author} {\bibinfo {author} {\bibfnamefont {L.}~\bibnamefont {Vegard}},\ }\bibfield  {title} {\bibinfo {title} {Die konstitution der mischkristalle und die raumf\"ullung der atome},\ }\href {https://doi.org/10.1007/BF01349680} {\bibfield  {journal} {\bibinfo  {journal} {Zeitschrift f\"ur Physik}\ }\textbf {\bibinfo {volume} {5}},\ \bibinfo {pages} {17} (\bibinfo {year} {1921})}\BibitemShut {NoStop}%
\bibitem [{\citenamefont {Denton}\ and\ \citenamefont {Ashcroft}(1991)}]{Vegard2}%
  \BibitemOpen
  \bibfield  {author} {\bibinfo {author} {\bibfnamefont {A.~R.}\ \bibnamefont {Denton}}\ and\ \bibinfo {author} {\bibfnamefont {N.~W.}\ \bibnamefont {Ashcroft}},\ }\bibfield  {title} {\bibinfo {title} {Vegard's law},\ }\href {https://doi.org/10.1103/PhysRevA.43.3161} {\bibfield  {journal} {\bibinfo  {journal} {Phys. Rev. A}\ }\textbf {\bibinfo {volume} {43}},\ \bibinfo {pages} {3161} (\bibinfo {year} {1991})}\BibitemShut {NoStop}%
\bibitem [{\citenamefont {Bombardi}\ \emph {et~al.}(2003)\citenamefont {Bombardi}, \citenamefont {d'Acapito}, \citenamefont {Mattenberger}, \citenamefont {Vogt},\ and\ \citenamefont {Lander}}]{Bombardi03}%
  \BibitemOpen
  \bibfield  {author} {\bibinfo {author} {\bibfnamefont {A.}~\bibnamefont {Bombardi}}, \bibinfo {author} {\bibfnamefont {F.}~\bibnamefont {d'Acapito}}, \bibinfo {author} {\bibfnamefont {K.}~\bibnamefont {Mattenberger}}, \bibinfo {author} {\bibfnamefont {O.}~\bibnamefont {Vogt}},\ and\ \bibinfo {author} {\bibfnamefont {G.~H.}\ \bibnamefont {Lander}},\ }\bibfield  {title} {\bibinfo {title} {{Non-Vegard behavior of the ${\mathrm{U}}_{x}{\mathrm{La}}_{1\ensuremath{-}x}\mathrm{S}$ system}},\ }\href {https://doi.org/10.1103/PhysRevB.68.104414} {\bibfield  {journal} {\bibinfo  {journal} {Phys. Rev. B}\ }\textbf {\bibinfo {volume} {68}},\ \bibinfo {pages} {104414} (\bibinfo {year} {2003})}\BibitemShut {NoStop}%
\bibitem [{\citenamefont {Baidya}\ \emph {et~al.}(2016)\citenamefont {Baidya}, \citenamefont {Bera}, \citenamefont {Kr\"ocher}, \citenamefont {Safonova}, \citenamefont {Abdala}, \citenamefont {Gerke}, \citenamefont {P\"ottgen}, \citenamefont {Priolkar},\ and\ \citenamefont {Mandal}}]{Baidya16}%
  \BibitemOpen
  \bibfield  {author} {\bibinfo {author} {\bibfnamefont {T.}~\bibnamefont {Baidya}}, \bibinfo {author} {\bibfnamefont {P.}~\bibnamefont {Bera}}, \bibinfo {author} {\bibfnamefont {O.}~\bibnamefont {Kr\"ocher}}, \bibinfo {author} {\bibfnamefont {O.}~\bibnamefont {Safonova}}, \bibinfo {author} {\bibfnamefont {P.~M.}\ \bibnamefont {Abdala}}, \bibinfo {author} {\bibfnamefont {B.}~\bibnamefont {Gerke}}, \bibinfo {author} {\bibfnamefont {R.}~\bibnamefont {P\"ottgen}}, \bibinfo {author} {\bibfnamefont {K.~R.}\ \bibnamefont {Priolkar}},\ and\ \bibinfo {author} {\bibfnamefont {T.~K.}\ \bibnamefont {Mandal}},\ }\bibfield  {title} {\bibinfo {title} {{Understanding the anomalous behavior of Vegard{'}s law in Ce$_{1-x}$M$_x$O$_2$ (M = Sn and Ti; 0 $<$ x $\leq$ 0.5) solid solutions}},\ }\href {https://doi.org/10.1039/C6CP01525E} {\bibfield  {journal} {\bibinfo  {journal} {Phys. Chem. Chem. Phys.}\ }\textbf {\bibinfo {volume} {18}},\ \bibinfo {pages} {13974} (\bibinfo {year} {2016})}\BibitemShut {NoStop}%
\bibitem [{\citenamefont {Wood}\ \emph {et~al.}(2008)\citenamefont {Wood}, \citenamefont {Daniels}, \citenamefont {Brown},\ and\ \citenamefont {Glazer}}]{Wood08}%
  \BibitemOpen
  \bibfield  {author} {\bibinfo {author} {\bibfnamefont {I.~G.}\ \bibnamefont {Wood}}, \bibinfo {author} {\bibfnamefont {P.}~\bibnamefont {Daniels}}, \bibinfo {author} {\bibfnamefont {R.~H.}\ \bibnamefont {Brown}},\ and\ \bibinfo {author} {\bibfnamefont {A.~M.}\ \bibnamefont {Glazer}},\ }\bibfield  {title} {\bibinfo {title} {Optical birefringence study of the ferroelectric phase transition in lithium niobate tantalate mixed crystals: $\text{LiNb}_{1-x}\text{Ta}_x\text{O}_3$},\ }\href {https://doi.org/10.1088/0953-8984/20/23/235237} {\bibfield  {journal} {\bibinfo  {journal} {Journal of Physics: Condensed Matter}\ }\textbf {\bibinfo {volume} {20}},\ \bibinfo {pages} {235237} (\bibinfo {year} {2008})}\BibitemShut {NoStop}%
\bibitem [{\citenamefont {Yakhnevych}\ \emph {et~al.}(2024)\citenamefont {Yakhnevych}, \citenamefont {El~Azzouzi}, \citenamefont {Bernhardt}, \citenamefont {Kofahl}, \citenamefont {Suhak}, \citenamefont {Sanna}, \citenamefont {Becker}, \citenamefont {Schmidt}, \citenamefont {Ganschow},\ and\ \citenamefont {Fritze}}]{UlianaDez23}%
  \BibitemOpen
  \bibfield  {author} {\bibinfo {author} {\bibfnamefont {U.}~\bibnamefont {Yakhnevych}}, \bibinfo {author} {\bibfnamefont {F.}~\bibnamefont {El~Azzouzi}}, \bibinfo {author} {\bibfnamefont {F.}~\bibnamefont {Bernhardt}}, \bibinfo {author} {\bibfnamefont {C.}~\bibnamefont {Kofahl}}, \bibinfo {author} {\bibfnamefont {Y.}~\bibnamefont {Suhak}}, \bibinfo {author} {\bibfnamefont {S.}~\bibnamefont {Sanna}}, \bibinfo {author} {\bibfnamefont {K.-D.}\ \bibnamefont {Becker}}, \bibinfo {author} {\bibfnamefont {H.}~\bibnamefont {Schmidt}}, \bibinfo {author} {\bibfnamefont {S.}~\bibnamefont {Ganschow}},\ and\ \bibinfo {author} {\bibfnamefont {H.}~\bibnamefont {Fritze}},\ }\bibfield  {title} {\bibinfo {title} {Oxygen partial pressure and temperature dependent electrical conductivity of lithium- niobate-tantalate solid solutions},\ }\href@noop {} {\bibfield  {journal} {\bibinfo  {journal} {Solid State Ionics}\ } (\bibinfo {year} {2024})},\ \bibinfo {note} {accepted}\BibitemShut {NoStop}%
\bibitem [{\citenamefont {Gureva}\ \emph {et~al.}(2023)\citenamefont {Gureva}, \citenamefont {Kulikov}, \citenamefont {Mololkin}, \citenamefont {Fakhrtdinov}, \citenamefont {Artemev}, \citenamefont {Demkiv}, \citenamefont {Pisarevsky},\ and\ \citenamefont {Marchenkov}}]{Gureva23}%
  \BibitemOpen
  \bibfield  {author} {\bibinfo {author} {\bibfnamefont {P.}~\bibnamefont {Gureva}}, \bibinfo {author} {\bibfnamefont {A.}~\bibnamefont {Kulikov}}, \bibinfo {author} {\bibfnamefont {A.}~\bibnamefont {Mololkin}}, \bibinfo {author} {\bibfnamefont {R.}~\bibnamefont {Fakhrtdinov}}, \bibinfo {author} {\bibfnamefont {A.}~\bibnamefont {Artemev}}, \bibinfo {author} {\bibfnamefont {A.}~\bibnamefont {Demkiv}}, \bibinfo {author} {\bibfnamefont {Y.}~\bibnamefont {Pisarevsky}},\ and\ \bibinfo {author} {\bibfnamefont {N.}~\bibnamefont {Marchenkov}},\ }\bibfield  {title} {\bibinfo {title} {{Local variations of the piezoelectric properties of an LiNb$_{1-x}$Ta${_x}$O$_3$ crystal}},\ }\href {https://doi.org/10.1107/S160057672300211X} {\bibfield  {journal} {\bibinfo  {journal} {Journal of Applied Crystallography}\ }\textbf {\bibinfo {volume} {56}},\ \bibinfo {pages} {539} (\bibinfo {year} {2023})}\BibitemShut {NoStop}%
\bibitem [{\citenamefont {Inbar}\ and\ \citenamefont {Cohen}(1996)}]{Inbar96}%
  \BibitemOpen
  \bibfield  {author} {\bibinfo {author} {\bibfnamefont {I.}~\bibnamefont {Inbar}}\ and\ \bibinfo {author} {\bibfnamefont {R.~E.}\ \bibnamefont {Cohen}},\ }\bibfield  {title} {\bibinfo {title} {Comparison of the electronic structures and energetics of ferroelectric $\text{LiNbO}_{3}$ and $\text{LiTaO}_{3}$},\ }\href {https://doi.org/10.1103/PhysRevB.53.1193} {\bibfield  {journal} {\bibinfo  {journal} {Phys. Rev. B}\ }\textbf {\bibinfo {volume} {53}},\ \bibinfo {pages} {1193} (\bibinfo {year} {1996})}\BibitemShut {NoStop}%
\bibitem [{\citenamefont {Weis}\ and\ \citenamefont {Gaylord}(1985)}]{Weis85}%
  \BibitemOpen
  \bibfield  {author} {\bibinfo {author} {\bibfnamefont {R.~S.}\ \bibnamefont {Weis}}\ and\ \bibinfo {author} {\bibfnamefont {T.~K.}\ \bibnamefont {Gaylord}},\ }\bibfield  {title} {\bibinfo {title} {Lithium niobate: Summary of physical properties and crystal structure},\ }\href {https://doi.org/10.1007/BF00614817} {\bibfield  {journal} {\bibinfo  {journal} {Applied Physics A}\ }\textbf {\bibinfo {volume} {37}},\ \bibinfo {pages} {191} (\bibinfo {year} {1985})}\BibitemShut {NoStop}%
\bibitem [{\citenamefont {of~Sheffield}(2023)}]{webelements}%
  \BibitemOpen
  \bibfield  {author} {\bibinfo {author} {\bibfnamefont {T.~U.}\ \bibnamefont {of~Sheffield}},\ }\href@noop {} {\bibinfo {title} {{WebElements}}},\ \bibinfo {howpublished} {\url{https://www.webelements.com}} (\bibinfo {year} {2023}),\ \bibinfo {note} {last accessed 10th January 2024}\BibitemShut {NoStop}%
\bibitem [{\citenamefont {Bartasyte}\ \emph {et~al.}(2012)\citenamefont {Bartasyte}, \citenamefont {Glazer}, \citenamefont {Wondre}, \citenamefont {Prabhakaran}, \citenamefont {Thomas}, \citenamefont {Huband}, \citenamefont {Keeble},\ and\ \citenamefont {Margueron}}]{Bartasyte12}%
  \BibitemOpen
  \bibfield  {author} {\bibinfo {author} {\bibfnamefont {A.}~\bibnamefont {Bartasyte}}, \bibinfo {author} {\bibfnamefont {A.}~\bibnamefont {Glazer}}, \bibinfo {author} {\bibfnamefont {F.}~\bibnamefont {Wondre}}, \bibinfo {author} {\bibfnamefont {D.}~\bibnamefont {Prabhakaran}}, \bibinfo {author} {\bibfnamefont {P.}~\bibnamefont {Thomas}}, \bibinfo {author} {\bibfnamefont {S.}~\bibnamefont {Huband}}, \bibinfo {author} {\bibfnamefont {D.}~\bibnamefont {Keeble}},\ and\ \bibinfo {author} {\bibfnamefont {S.}~\bibnamefont {Margueron}},\ }\bibfield  {title} {\bibinfo {title} {{Growth of LiNb$_{1-x}$Ta$_x$O$_3$ solid solution crystals}},\ }\href {https://doi.org/https://doi.org/10.1016/j.matchemphys.2012.03.060} {\bibfield  {journal} {\bibinfo  {journal} {Materials Chemistry and Physics}\ }\textbf {\bibinfo {volume} {134}},\ \bibinfo {pages} {728} (\bibinfo {year} {2012})}\BibitemShut {NoStop}%
\bibitem [{\citenamefont {Manzoor}\ \emph {et~al.}(2018)\citenamefont {Manzoor}, \citenamefont {Pandey}, \citenamefont {Chakraborty}, \citenamefont {Phillpot},\ and\ \citenamefont {Aidhy}}]{Manzoor18}%
  \BibitemOpen
  \bibfield  {author} {\bibinfo {author} {\bibfnamefont {A.}~\bibnamefont {Manzoor}}, \bibinfo {author} {\bibfnamefont {S.}~\bibnamefont {Pandey}}, \bibinfo {author} {\bibfnamefont {D.}~\bibnamefont {Chakraborty}}, \bibinfo {author} {\bibfnamefont {S.~R.}\ \bibnamefont {Phillpot}},\ and\ \bibinfo {author} {\bibfnamefont {D.~S.}\ \bibnamefont {Aidhy}},\ }\bibfield  {title} {\bibinfo {title} {Entropy contributions to phase stability in binary random solid solutions},\ }\href {https://doi.org/10.1038/s41524-018-0102-y} {\bibfield  {journal} {\bibinfo  {journal} {Computational Materials}\ }\textbf {\bibinfo {volume} {4}},\ \bibinfo {pages} {47} (\bibinfo {year} {2018})}\BibitemShut {NoStop}%
\bibitem [{\citenamefont {R\"using}\ \emph {et~al.}(2016)\citenamefont {R\"using}, \citenamefont {Sanna}, \citenamefont {Neufeld}, \citenamefont {Berth}, \citenamefont {Schmidt}, \citenamefont {Zrenner}, \citenamefont {Yu}, \citenamefont {Wang},\ and\ \citenamefont {Zhang}}]{Micha16}%
  \BibitemOpen
  \bibfield  {author} {\bibinfo {author} {\bibfnamefont {M.}~\bibnamefont {R\"using}}, \bibinfo {author} {\bibfnamefont {S.}~\bibnamefont {Sanna}}, \bibinfo {author} {\bibfnamefont {S.}~\bibnamefont {Neufeld}}, \bibinfo {author} {\bibfnamefont {G.}~\bibnamefont {Berth}}, \bibinfo {author} {\bibfnamefont {W.~G.}\ \bibnamefont {Schmidt}}, \bibinfo {author} {\bibfnamefont {A.}~\bibnamefont {Zrenner}}, \bibinfo {author} {\bibfnamefont {H.}~\bibnamefont {Yu}}, \bibinfo {author} {\bibfnamefont {Y.}~\bibnamefont {Wang}},\ and\ \bibinfo {author} {\bibfnamefont {H.}~\bibnamefont {Zhang}},\ }\bibfield  {title} {\bibinfo {title} {{Vibrational properties of ${\mathrm{LiNb}}_{1\ensuremath{-}x}{\mathrm{Ta}}_{x}{\mathrm{O}}_{3}$ mixed crystals}},\ }\href {https://doi.org/10.1103/PhysRevB.93.184305} {\bibfield  {journal} {\bibinfo  {journal} {Phys. Rev. B}\ }\textbf {\bibinfo {volume} {93}},\ \bibinfo {pages} {184305} (\bibinfo {year} {2016})}\BibitemShut {NoStop}%
\bibitem [{\citenamefont {Xue}\ \emph {et~al.}(2000)\citenamefont {Xue}, \citenamefont {Betzler},\ and\ \citenamefont {Hesse}}]{XUE2000581}%
  \BibitemOpen
  \bibfield  {author} {\bibinfo {author} {\bibfnamefont {D.}~\bibnamefont {Xue}}, \bibinfo {author} {\bibfnamefont {K.}~\bibnamefont {Betzler}},\ and\ \bibinfo {author} {\bibfnamefont {H.}~\bibnamefont {Hesse}},\ }\bibfield  {title} {\bibinfo {title} {Dielectric properties of lithium niobate–tantalate crystals},\ }\href {https://doi.org/https://doi.org/10.1016/S0038-1098(00)00243-X} {\bibfield  {journal} {\bibinfo  {journal} {Solid State Communications}\ }\textbf {\bibinfo {volume} {115}},\ \bibinfo {pages} {581} (\bibinfo {year} {2000})}\BibitemShut {NoStop}%
\bibitem [{\citenamefont {El~Azzouzi}\ \emph {et~al.}(2024)\citenamefont {El~Azzouzi}, \citenamefont {Klimm}, \citenamefont {Verhoff}, \citenamefont {Sch\"afer}, \citenamefont {Ganschow}, \citenamefont {Becker}, \citenamefont {Sanna},\ and\ \citenamefont {Fritze}}]{FatimaDez23}%
  \BibitemOpen
  \bibfield  {author} {\bibinfo {author} {\bibfnamefont {F.}~\bibnamefont {El~Azzouzi}}, \bibinfo {author} {\bibfnamefont {D.}~\bibnamefont {Klimm}}, \bibinfo {author} {\bibfnamefont {L.~M.}\ \bibnamefont {Verhoff}}, \bibinfo {author} {\bibfnamefont {N.~A.}\ \bibnamefont {Sch\"afer}}, \bibinfo {author} {\bibfnamefont {S.}~\bibnamefont {Ganschow}}, \bibinfo {author} {\bibfnamefont {K.-D.}\ \bibnamefont {Becker}}, \bibinfo {author} {\bibfnamefont {S.}~\bibnamefont {Sanna}},\ and\ \bibinfo {author} {\bibfnamefont {H.}~\bibnamefont {Fritze}},\ }\bibfield  {title} {\bibinfo {title} {{Phase Transformation in Lithium Niobate-Lithium Tantalate Solid Solutions (LiNb$_{1-x}$Ta$_x$O$_3$)}},\ }\href@noop {} {\bibfield  {journal} {\bibinfo  {journal} {Phys. Stat. Sol. (a)}\ } (\bibinfo {year} {2024})},\ \bibinfo {note} {submitted}\BibitemShut {NoStop}%
\bibitem [{\citenamefont {Riefer}\ \emph {et~al.}(2013{\natexlab{a}})\citenamefont {Riefer}, \citenamefont {Sanna},\ and\ \citenamefont {Schmidt}}]{Riefer13SS}%
  \BibitemOpen
  \bibfield  {author} {\bibinfo {author} {\bibfnamefont {A.}~\bibnamefont {Riefer}}, \bibinfo {author} {\bibfnamefont {S.}~\bibnamefont {Sanna}},\ and\ \bibinfo {author} {\bibfnamefont {W.~G.}\ \bibnamefont {Schmidt}},\ }\bibfield  {title} {\bibinfo {title} {{LiNb$_{1-x}$Ta$_x$O$_3$ Electronic Structure and Optical Response from First-Principles Calculations}},\ }\href {https://doi.org/10.1080/00150193.2013.821904} {\bibfield  {journal} {\bibinfo  {journal} {Ferroelectrics}\ }\textbf {\bibinfo {volume} {447}},\ \bibinfo {pages} {78} (\bibinfo {year} {2013}{\natexlab{a}})}\BibitemShut {NoStop}%
\bibitem [{\citenamefont {Sanna}\ \emph {et~al.}(2013)\citenamefont {Sanna}, \citenamefont {Riefer}, \citenamefont {Neufeld}, \citenamefont {Schmidt}, \citenamefont {Berth}, \citenamefont {R\"using}, \citenamefont {Widhalm},\ and\ \citenamefont {Zrenner}}]{Sanna13SS}%
  \BibitemOpen
  \bibfield  {author} {\bibinfo {author} {\bibfnamefont {S.}~\bibnamefont {Sanna}}, \bibinfo {author} {\bibfnamefont {A.}~\bibnamefont {Riefer}}, \bibinfo {author} {\bibfnamefont {S.}~\bibnamefont {Neufeld}}, \bibinfo {author} {\bibfnamefont {W.~G.}\ \bibnamefont {Schmidt}}, \bibinfo {author} {\bibfnamefont {G.}~\bibnamefont {Berth}}, \bibinfo {author} {\bibfnamefont {M.}~\bibnamefont {R\"using}}, \bibinfo {author} {\bibfnamefont {A.}~\bibnamefont {Widhalm}},\ and\ \bibinfo {author} {\bibfnamefont {A.}~\bibnamefont {Zrenner}},\ }\bibfield  {title} {\bibinfo {title} {{Vibrational Fingerprints of LiNbO$_3$-LiTaO$_3$ Mixed Crystals}},\ }\href {https://doi.org/10.1080/00150193.2013.821893} {\bibfield  {journal} {\bibinfo  {journal} {Ferroelectrics}\ }\textbf {\bibinfo {volume} {447}},\ \bibinfo {pages} {63} (\bibinfo {year} {2013})}\BibitemShut {NoStop}%
\bibitem [{\citenamefont {Wei}\ \emph {et~al.}(1990)\citenamefont {Wei}, \citenamefont {Ferreira}, \citenamefont {Bernard},\ and\ \citenamefont {Zunger}}]{Wei90}%
  \BibitemOpen
  \bibfield  {author} {\bibinfo {author} {\bibfnamefont {S.-H.}\ \bibnamefont {Wei}}, \bibinfo {author} {\bibfnamefont {L.~G.}\ \bibnamefont {Ferreira}}, \bibinfo {author} {\bibfnamefont {J.~E.}\ \bibnamefont {Bernard}},\ and\ \bibinfo {author} {\bibfnamefont {A.}~\bibnamefont {Zunger}},\ }\bibfield  {title} {\bibinfo {title} {Electronic properties of random alloys: Special quasirandom structures},\ }\href {https://doi.org/10.1103/PhysRevB.42.9622} {\bibfield  {journal} {\bibinfo  {journal} {Phys. Rev. B}\ }\textbf {\bibinfo {volume} {42}},\ \bibinfo {pages} {9622} (\bibinfo {year} {1990})}\BibitemShut {NoStop}%
\bibitem [{\citenamefont {von Pezold}\ \emph {et~al.}(2010)\citenamefont {von Pezold}, \citenamefont {Dick}, \citenamefont {Fri\'ak},\ and\ \citenamefont {Neugebauer}}]{Petzold10}%
  \BibitemOpen
  \bibfield  {author} {\bibinfo {author} {\bibfnamefont {J.}~\bibnamefont {von Pezold}}, \bibinfo {author} {\bibfnamefont {A.}~\bibnamefont {Dick}}, \bibinfo {author} {\bibfnamefont {M.}~\bibnamefont {Fri\'ak}},\ and\ \bibinfo {author} {\bibfnamefont {J.}~\bibnamefont {Neugebauer}},\ }\bibfield  {title} {\bibinfo {title} {{Generation and performance of special quasirandom structures for studying the elastic properties of random alloys: Application to Al-Ti}},\ }\href {https://doi.org/10.1103/PhysRevB.81.094203} {\bibfield  {journal} {\bibinfo  {journal} {Phys. Rev. B}\ }\textbf {\bibinfo {volume} {81}},\ \bibinfo {pages} {094203} (\bibinfo {year} {2010})}\BibitemShut {NoStop}%
\bibitem [{\citenamefont {Bellaiche}\ and\ \citenamefont {Vanderbilt}(2000)}]{VCA}%
  \BibitemOpen
  \bibfield  {author} {\bibinfo {author} {\bibfnamefont {L.}~\bibnamefont {Bellaiche}}\ and\ \bibinfo {author} {\bibfnamefont {D.}~\bibnamefont {Vanderbilt}},\ }\bibfield  {title} {\bibinfo {title} {Virtual crystal approximation revisited: Application to dielectric and piezoelectric properties of perovskites},\ }\href {https://doi.org/10.1103/PhysRevB.61.7877} {\bibfield  {journal} {\bibinfo  {journal} {Phys. Rev. B}\ }\textbf {\bibinfo {volume} {61}},\ \bibinfo {pages} {7877} (\bibinfo {year} {2000})}\BibitemShut {NoStop}%
\bibitem [{\citenamefont {Soven}(1967)}]{CPA1}%
  \BibitemOpen
  \bibfield  {author} {\bibinfo {author} {\bibfnamefont {P.}~\bibnamefont {Soven}},\ }\bibfield  {title} {\bibinfo {title} {Coherent-potential model of substitutional disordered alloys},\ }\href {https://doi.org/10.1103/PhysRev.156.809} {\bibfield  {journal} {\bibinfo  {journal} {Phys. Rev.}\ }\textbf {\bibinfo {volume} {156}},\ \bibinfo {pages} {809} (\bibinfo {year} {1967})}\BibitemShut {NoStop}%
\bibitem [{\citenamefont {Sanchez}\ \emph {et~al.}(1984)\citenamefont {Sanchez}, \citenamefont {Ducastelle},\ and\ \citenamefont {Gratias}}]{Sanchez84}%
  \BibitemOpen
  \bibfield  {author} {\bibinfo {author} {\bibfnamefont {J.}~\bibnamefont {Sanchez}}, \bibinfo {author} {\bibfnamefont {F.}~\bibnamefont {Ducastelle}},\ and\ \bibinfo {author} {\bibfnamefont {D.}~\bibnamefont {Gratias}},\ }\bibfield  {title} {\bibinfo {title} {Generalized cluster description of multicomponent systems},\ }\href {https://doi.org/10.1016/0378-4371(84)90096-7} {\bibfield  {journal} {\bibinfo  {journal} {Physica A: Statistical Mechanics and its Applications}\ }\textbf {\bibinfo {volume} {128}},\ \bibinfo {pages} {334} (\bibinfo {year} {1984})}\BibitemShut {NoStop}%
\bibitem [{\citenamefont {Laks}\ \emph {et~al.}(1992)\citenamefont {Laks}, \citenamefont {Ferreira}, \citenamefont {Froyen},\ and\ \citenamefont {Zunger}}]{Laks92}%
  \BibitemOpen
  \bibfield  {author} {\bibinfo {author} {\bibfnamefont {D.~B.}\ \bibnamefont {Laks}}, \bibinfo {author} {\bibfnamefont {L.~G.}\ \bibnamefont {Ferreira}}, \bibinfo {author} {\bibfnamefont {S.}~\bibnamefont {Froyen}},\ and\ \bibinfo {author} {\bibfnamefont {A.}~\bibnamefont {Zunger}},\ }\bibfield  {title} {\bibinfo {title} {Efficient cluster expansion for substitutional systems},\ }\href {https://doi.org/10.1103/PhysRevB.46.12587} {\bibfield  {journal} {\bibinfo  {journal} {Phys. Rev. B}\ }\textbf {\bibinfo {volume} {46}},\ \bibinfo {pages} {12587} (\bibinfo {year} {1992})}\BibitemShut {NoStop}%
\bibitem [{\citenamefont {van~de Walle}(2008)}]{CExp}%
  \BibitemOpen
  \bibfield  {author} {\bibinfo {author} {\bibfnamefont {A.}~\bibnamefont {van~de Walle}},\ }\bibfield  {title} {\bibinfo {title} {A complete representation of structure–property relationships in crystals},\ }\href {https://doi.org/10.1038/nmat2200} {\bibfield  {journal} {\bibinfo  {journal} {Nature materials}\ }\textbf {\bibinfo {volume} {7}},\ \bibinfo {pages} {455} (\bibinfo {year} {2008})}\BibitemShut {NoStop}%
\bibitem [{\citenamefont {Ghosh}\ \emph {et~al.}(2008)\citenamefont {Ghosh}, \citenamefont {{van de Walle}},\ and\ \citenamefont {Asta}}]{Ghosh08}%
  \BibitemOpen
  \bibfield  {author} {\bibinfo {author} {\bibfnamefont {G.}~\bibnamefont {Ghosh}}, \bibinfo {author} {\bibfnamefont {A.}~\bibnamefont {{van de Walle}}},\ and\ \bibinfo {author} {\bibfnamefont {M.}~\bibnamefont {Asta}},\ }\bibfield  {title} {\bibinfo {title} {First-principles calculations of the structural and thermodynamic properties of bcc, fcc and hcp solid solutions in the al–tm (tm=ti, zr and hf) systems: A comparison of cluster expansion and supercell methods},\ }\href {https://doi.org/https://doi.org/10.1016/j.actamat.2008.03.006} {\bibfield  {journal} {\bibinfo  {journal} {Acta Materialia}\ }\textbf {\bibinfo {volume} {56}},\ \bibinfo {pages} {3202} (\bibinfo {year} {2008})}\BibitemShut {NoStop}%
\bibitem [{\citenamefont {Kresse}\ and\ \citenamefont {Hafner}(1993)}]{Kresse1993}%
  \BibitemOpen
  \bibfield  {author} {\bibinfo {author} {\bibfnamefont {G.}~\bibnamefont {Kresse}}\ and\ \bibinfo {author} {\bibfnamefont {J.}~\bibnamefont {Hafner}},\ }\bibfield  {title} {\bibinfo {title} {{Ab initio molecular dynamics for liquid metals}},\ }\href {https://doi.org/10.1103/PhysRevB.47.558} {\bibfield  {journal} {\bibinfo  {journal} {Physical Review B}\ }\textbf {\bibinfo {volume} {47}},\ \bibinfo {pages} {558} (\bibinfo {year} {1993})}\BibitemShut {NoStop}%
\bibitem [{\citenamefont {Kresse}\ and\ \citenamefont {Furthm{\"{u}}ller}(1996{\natexlab{a}})}]{Kresse1996}%
  \BibitemOpen
  \bibfield  {author} {\bibinfo {author} {\bibfnamefont {G.}~\bibnamefont {Kresse}}\ and\ \bibinfo {author} {\bibfnamefont {J.}~\bibnamefont {Furthm{\"{u}}ller}},\ }\bibfield  {title} {\bibinfo {title} {{Efficient iterative schemes for ab initio total-energy calculations using a plane-wave basis set}},\ }\href {https://doi.org/10.1103/PhysRevB.54.11169} {\bibfield  {journal} {\bibinfo  {journal} {Physical Review B - Condensed Matter and Materials Physics}\ }\textbf {\bibinfo {volume} {54}},\ \bibinfo {pages} {11169} (\bibinfo {year} {1996}{\natexlab{a}})}\BibitemShut {NoStop}%
\bibitem [{\citenamefont {Kresse}\ and\ \citenamefont {Furthm{\"{u}}ller}(1996{\natexlab{b}})}]{Kresse1996_2}%
  \BibitemOpen
  \bibfield  {author} {\bibinfo {author} {\bibfnamefont {G.}~\bibnamefont {Kresse}}\ and\ \bibinfo {author} {\bibfnamefont {J.}~\bibnamefont {Furthm{\"{u}}ller}},\ }\bibfield  {title} {\bibinfo {title} {{Efficiency of ab-initio total energy calculations for metals and semiconductors using a plane-wave basis set}},\ }\href {https://doi.org/10.1016/0927-0256(96)00008-0} {\bibfield  {journal} {\bibinfo  {journal} {Computational Materials Science}\ }\textbf {\bibinfo {volume} {6}},\ \bibinfo {pages} {15} (\bibinfo {year} {1996}{\natexlab{b}})}\BibitemShut {NoStop}%
\bibitem [{\citenamefont {Bl\"ochl}(1994)}]{Bloechl94}%
  \BibitemOpen
  \bibfield  {author} {\bibinfo {author} {\bibfnamefont {P.~E.}\ \bibnamefont {Bl\"ochl}},\ }\bibfield  {title} {\bibinfo {title} {Projector augmented-wave method},\ }\href {https://doi.org/10.1103/PhysRevB.50.17953} {\bibfield  {journal} {\bibinfo  {journal} {Phys. Rev. B}\ }\textbf {\bibinfo {volume} {50}},\ \bibinfo {pages} {17953} (\bibinfo {year} {1994})}\BibitemShut {NoStop}%
\bibitem [{\citenamefont {Joubert}(1999)}]{Joubert1999}%
  \BibitemOpen
  \bibfield  {author} {\bibinfo {author} {\bibfnamefont {D.}~\bibnamefont {Joubert}},\ }\bibfield  {title} {\bibinfo {title} {{From ultrasoft pseudopotentials to the projector augmented-wave method}},\ }\href {https://doi.org/10.1103/PhysRevB.59.1758} {\bibfield  {journal} {\bibinfo  {journal} {Physical Review B - Condensed Matter and Materials Physics}\ }\textbf {\bibinfo {volume} {59}},\ \bibinfo {pages} {1758} (\bibinfo {year} {1999})}\BibitemShut {NoStop}%
\bibitem [{\citenamefont {Perdew}\ \emph {et~al.}(2008)\citenamefont {Perdew}, \citenamefont {Ruzsinszky}, \citenamefont {Csonka}, \citenamefont {Vydrov}, \citenamefont {Scuseria}, \citenamefont {Constantin}, \citenamefont {Zhou},\ and\ \citenamefont {Burke}}]{Perdew2008}%
  \BibitemOpen
  \bibfield  {author} {\bibinfo {author} {\bibfnamefont {J.~P.}\ \bibnamefont {Perdew}}, \bibinfo {author} {\bibfnamefont {A.}~\bibnamefont {Ruzsinszky}}, \bibinfo {author} {\bibfnamefont {G.~I.}\ \bibnamefont {Csonka}}, \bibinfo {author} {\bibfnamefont {O.~A.}\ \bibnamefont {Vydrov}}, \bibinfo {author} {\bibfnamefont {G.~E.}\ \bibnamefont {Scuseria}}, \bibinfo {author} {\bibfnamefont {L.~A.}\ \bibnamefont {Constantin}}, \bibinfo {author} {\bibfnamefont {X.}~\bibnamefont {Zhou}},\ and\ \bibinfo {author} {\bibfnamefont {K.}~\bibnamefont {Burke}},\ }\bibfield  {title} {\bibinfo {title} {{Restoring the density-gradient expansion for exchange in solids and surfaces}},\ }\href {https://doi.org/10.1103/PhysRevLett.100.136406} {\bibfield  {journal} {\bibinfo  {journal} {Physical Review Letters}\ }\textbf {\bibinfo {volume} {100}},\ \bibinfo {pages} {1} (\bibinfo {year} {2008})}\BibitemShut {NoStop}%
\bibitem [{\citenamefont {Murnaghan}(1944)}]{Murnaghan44}%
  \BibitemOpen
  \bibfield  {author} {\bibinfo {author} {\bibfnamefont {F.~D.}\ \bibnamefont {Murnaghan}},\ }\bibfield  {title} {\bibinfo {title} {The compressibility of media under extreme pressures},\ }\href {https://doi.org/10.1073/pnas.30.9.244} {\bibfield  {journal} {\bibinfo  {journal} {Proceedings of the National Academy of Sciences}\ }\textbf {\bibinfo {volume} {30}},\ \bibinfo {pages} {244} (\bibinfo {year} {1944})}\BibitemShut {NoStop}%
\bibitem [{\citenamefont {Pack}\ and\ \citenamefont {Monkhorst}(1977)}]{Pack1977}%
  \BibitemOpen
  \bibfield  {author} {\bibinfo {author} {\bibfnamefont {J.~D.}\ \bibnamefont {Pack}}\ and\ \bibinfo {author} {\bibfnamefont {H.~J.}\ \bibnamefont {Monkhorst}},\ }\bibfield  {title} {\bibinfo {title} {{''Special points for Brillouin-zone integrations'' - a reply}},\ }\href {https://doi.org/10.1103/PhysRevB.16.1748} {\bibfield  {journal} {\bibinfo  {journal} {Physical Review B}\ }\textbf {\bibinfo {volume} {16}},\ \bibinfo {pages} {1748} (\bibinfo {year} {1977})}\BibitemShut {NoStop}%
\bibitem [{\citenamefont {Feynman}(1939)}]{Forces}%
  \BibitemOpen
  \bibfield  {author} {\bibinfo {author} {\bibfnamefont {R.~P.}\ \bibnamefont {Feynman}},\ }\bibfield  {title} {\bibinfo {title} {Forces in molecules},\ }\href {https://doi.org/10.1103/PhysRev.56.340} {\bibfield  {journal} {\bibinfo  {journal} {Phys. Rev.}\ }\textbf {\bibinfo {volume} {56}},\ \bibinfo {pages} {340} (\bibinfo {year} {1939})}\BibitemShut {NoStop}%
\bibitem [{\citenamefont {Togo}\ \emph {et~al.}(2023)\citenamefont {Togo}, \citenamefont {Chaput}, \citenamefont {Tadano},\ and\ \citenamefont {Tanaka}}]{phonopy-phono3py-JPCM}%
  \BibitemOpen
  \bibfield  {author} {\bibinfo {author} {\bibfnamefont {A.}~\bibnamefont {Togo}}, \bibinfo {author} {\bibfnamefont {L.}~\bibnamefont {Chaput}}, \bibinfo {author} {\bibfnamefont {T.}~\bibnamefont {Tadano}},\ and\ \bibinfo {author} {\bibfnamefont {I.}~\bibnamefont {Tanaka}},\ }\bibfield  {title} {\bibinfo {title} {Implementation strategies in phonopy and phono3py},\ }\href {https://doi.org/10.1088/1361-648X/acd831} {\bibfield  {journal} {\bibinfo  {journal} {J. Phys. Condens. Matter}\ }\textbf {\bibinfo {volume} {35}},\ \bibinfo {pages} {353001} (\bibinfo {year} {2023})}\BibitemShut {NoStop}%
\bibitem [{\citenamefont {Togo}(2023)}]{phonopy-phono3py-JPSJ}%
  \BibitemOpen
  \bibfield  {author} {\bibinfo {author} {\bibfnamefont {A.}~\bibnamefont {Togo}},\ }\bibfield  {title} {\bibinfo {title} {First-principles phonon calculations with phonopy and phono3py},\ }\href {https://doi.org/10.7566/JPSJ.92.012001} {\bibfield  {journal} {\bibinfo  {journal} {J. Phys. Soc. Jpn.}\ }\textbf {\bibinfo {volume} {92}},\ \bibinfo {pages} {012001} (\bibinfo {year} {2023})}\BibitemShut {NoStop}%
\bibitem [{\citenamefont {Parlinski}\ \emph {et~al.}(1997)\citenamefont {Parlinski}, \citenamefont {Li},\ and\ \citenamefont {Kawazoe}}]{Parlinski1997}%
  \BibitemOpen
  \bibfield  {author} {\bibinfo {author} {\bibfnamefont {K.}~\bibnamefont {Parlinski}}, \bibinfo {author} {\bibfnamefont {Z.~Q.}\ \bibnamefont {Li}},\ and\ \bibinfo {author} {\bibfnamefont {Y.}~\bibnamefont {Kawazoe}},\ }\bibfield  {title} {\bibinfo {title} {{First-principles determination of the soft mode in cubic ZrO$_2$}},\ }\href {https://doi.org/10.1103/PhysRevLett.78.4063} {\bibfield  {journal} {\bibinfo  {journal} {Physical Review Letters}\ }\textbf {\bibinfo {volume} {78}},\ \bibinfo {pages} {4063} (\bibinfo {year} {1997})}\BibitemShut {NoStop}%
\bibitem [{\citenamefont {Giannozzi}\ \emph {et~al.}(2017)\citenamefont {Giannozzi}, \citenamefont {Andreussi}, \citenamefont {Brumme}, \citenamefont {Bunau}, \citenamefont {Nardelli}, \citenamefont {Calandra}, \citenamefont {Car}, \citenamefont {Cavazzoni}, \citenamefont {Ceresoli}, \citenamefont {Cococcioni}, \citenamefont {Colonna}, \citenamefont {Carnimeo}, \citenamefont {Corso}, \citenamefont {de~Gironcoli}, \citenamefont {Delugas}, \citenamefont {Jr}, \citenamefont {Ferretti}, \citenamefont {Floris}, \citenamefont {Fratesi}, \citenamefont {Fugallo}, \citenamefont {Gebauer}, \citenamefont {Gerstmann}, \citenamefont {Giustino}, \citenamefont {Gorni}, \citenamefont {Jia}, \citenamefont {Kawamura}, \citenamefont {Ko}, \citenamefont {Kokalj}, \citenamefont {Küçükbenli}, \citenamefont {Lazzeri}, \citenamefont {Marsili}, \citenamefont {Marzari}, \citenamefont {Mauri}, \citenamefont {Nguyen}, \citenamefont {Nguyen}, \citenamefont {de-la Roza}, \citenamefont {Paulatto}, \citenamefont {Poncé}, \citenamefont
  {Rocca}, \citenamefont {Sabatini}, \citenamefont {Santra}, \citenamefont {Schlipf}, \citenamefont {Seitsonen}, \citenamefont {Smogunov}, \citenamefont {Timrov}, \citenamefont {Thonhauser}, \citenamefont {Umari}, \citenamefont {Vast}, \citenamefont {Wu},\ and\ \citenamefont {Baroni}}]{QE-2017}%
  \BibitemOpen
  \bibfield  {author} {\bibinfo {author} {\bibfnamefont {P.}~\bibnamefont {Giannozzi}}, \bibinfo {author} {\bibfnamefont {O.}~\bibnamefont {Andreussi}}, \bibinfo {author} {\bibfnamefont {T.}~\bibnamefont {Brumme}}, \bibinfo {author} {\bibfnamefont {O.}~\bibnamefont {Bunau}}, \bibinfo {author} {\bibfnamefont {M.~B.}\ \bibnamefont {Nardelli}}, \bibinfo {author} {\bibfnamefont {M.}~\bibnamefont {Calandra}}, \bibinfo {author} {\bibfnamefont {R.}~\bibnamefont {Car}}, \bibinfo {author} {\bibfnamefont {C.}~\bibnamefont {Cavazzoni}}, \bibinfo {author} {\bibfnamefont {D.}~\bibnamefont {Ceresoli}}, \bibinfo {author} {\bibfnamefont {M.}~\bibnamefont {Cococcioni}}, \bibinfo {author} {\bibfnamefont {N.}~\bibnamefont {Colonna}}, \bibinfo {author} {\bibfnamefont {I.}~\bibnamefont {Carnimeo}}, \bibinfo {author} {\bibfnamefont {A.~D.}\ \bibnamefont {Corso}}, \bibinfo {author} {\bibfnamefont {S.}~\bibnamefont {de~Gironcoli}}, \bibinfo {author} {\bibfnamefont {P.}~\bibnamefont {Delugas}}, \bibinfo {author} {\bibfnamefont
  {R.~A.~D.}\ \bibnamefont {Jr}}, \bibinfo {author} {\bibfnamefont {A.}~\bibnamefont {Ferretti}}, \bibinfo {author} {\bibfnamefont {A.}~\bibnamefont {Floris}}, \bibinfo {author} {\bibfnamefont {G.}~\bibnamefont {Fratesi}}, \bibinfo {author} {\bibfnamefont {G.}~\bibnamefont {Fugallo}}, \bibinfo {author} {\bibfnamefont {R.}~\bibnamefont {Gebauer}}, \bibinfo {author} {\bibfnamefont {U.}~\bibnamefont {Gerstmann}}, \bibinfo {author} {\bibfnamefont {F.}~\bibnamefont {Giustino}}, \bibinfo {author} {\bibfnamefont {T.}~\bibnamefont {Gorni}}, \bibinfo {author} {\bibfnamefont {J.}~\bibnamefont {Jia}}, \bibinfo {author} {\bibfnamefont {M.}~\bibnamefont {Kawamura}}, \bibinfo {author} {\bibfnamefont {H.-Y.}\ \bibnamefont {Ko}}, \bibinfo {author} {\bibfnamefont {A.}~\bibnamefont {Kokalj}}, \bibinfo {author} {\bibfnamefont {E.}~\bibnamefont {Küçükbenli}}, \bibinfo {author} {\bibfnamefont {M.}~\bibnamefont {Lazzeri}}, \bibinfo {author} {\bibfnamefont {M.}~\bibnamefont {Marsili}}, \bibinfo {author} {\bibfnamefont
  {N.}~\bibnamefont {Marzari}}, \bibinfo {author} {\bibfnamefont {F.}~\bibnamefont {Mauri}}, \bibinfo {author} {\bibfnamefont {N.~L.}\ \bibnamefont {Nguyen}}, \bibinfo {author} {\bibfnamefont {H.-V.}\ \bibnamefont {Nguyen}}, \bibinfo {author} {\bibfnamefont {A.~O.}\ \bibnamefont {de-la Roza}}, \bibinfo {author} {\bibfnamefont {L.}~\bibnamefont {Paulatto}}, \bibinfo {author} {\bibfnamefont {S.}~\bibnamefont {Poncé}}, \bibinfo {author} {\bibfnamefont {D.}~\bibnamefont {Rocca}}, \bibinfo {author} {\bibfnamefont {R.}~\bibnamefont {Sabatini}}, \bibinfo {author} {\bibfnamefont {B.}~\bibnamefont {Santra}}, \bibinfo {author} {\bibfnamefont {M.}~\bibnamefont {Schlipf}}, \bibinfo {author} {\bibfnamefont {A.~P.}\ \bibnamefont {Seitsonen}}, \bibinfo {author} {\bibfnamefont {A.}~\bibnamefont {Smogunov}}, \bibinfo {author} {\bibfnamefont {I.}~\bibnamefont {Timrov}}, \bibinfo {author} {\bibfnamefont {T.}~\bibnamefont {Thonhauser}}, \bibinfo {author} {\bibfnamefont {P.}~\bibnamefont {Umari}}, \bibinfo {author}
  {\bibfnamefont {N.}~\bibnamefont {Vast}}, \bibinfo {author} {\bibfnamefont {X.}~\bibnamefont {Wu}},\ and\ \bibinfo {author} {\bibfnamefont {S.}~\bibnamefont {Baroni}},\ }\bibfield  {title} {\bibinfo {title} {Advanced capabilities for materials modelling with quantum espresso},\ }\href {http://stacks.iop.org/0953-8984/29/i=46/a=465901} {\bibfield  {journal} {\bibinfo  {journal} {Journal of Physics: Condensed Matter}\ }\textbf {\bibinfo {volume} {29}},\ \bibinfo {pages} {465901} (\bibinfo {year} {2017})}\BibitemShut {NoStop}%
\bibitem [{\citenamefont {Hamann}(2013)}]{Hamann13}%
  \BibitemOpen
  \bibfield  {author} {\bibinfo {author} {\bibfnamefont {D.~R.}\ \bibnamefont {Hamann}},\ }\bibfield  {title} {\bibinfo {title} {Optimized norm-conserving vanderbilt pseudopotentials},\ }\href {https://doi.org/10.1103/PhysRevB.88.085117} {\bibfield  {journal} {\bibinfo  {journal} {Phys. Rev. B}\ }\textbf {\bibinfo {volume} {88}},\ \bibinfo {pages} {085117} (\bibinfo {year} {2013})}\BibitemShut {NoStop}%
\bibitem [{\citenamefont {Deslippe}\ \emph {et~al.}(2012)\citenamefont {Deslippe}, \citenamefont {Samsonidze}, \citenamefont {Strubbe}, \citenamefont {Jain}, \citenamefont {Cohen},\ and\ \citenamefont {Louie}}]{Deslippe2012}%
  \BibitemOpen
  \bibfield  {author} {\bibinfo {author} {\bibfnamefont {J.}~\bibnamefont {Deslippe}}, \bibinfo {author} {\bibfnamefont {G.}~\bibnamefont {Samsonidze}}, \bibinfo {author} {\bibfnamefont {D.~A.}\ \bibnamefont {Strubbe}}, \bibinfo {author} {\bibfnamefont {M.}~\bibnamefont {Jain}}, \bibinfo {author} {\bibfnamefont {M.~L.}\ \bibnamefont {Cohen}},\ and\ \bibinfo {author} {\bibfnamefont {S.~G.}\ \bibnamefont {Louie}},\ }\bibfield  {title} {\bibinfo {title} {{BerkeleyGW: A massively parallel computer package for the calculation of the quasiparticle and optical properties of materials and nanostructures}},\ }\href {https://doi.org/10.1016/j.cpc.2011.12.006} {\bibfield  {journal} {\bibinfo  {journal} {Computer Physics Communications}\ }\textbf {\bibinfo {volume} {183}},\ \bibinfo {pages} {1269} (\bibinfo {year} {2012})}\BibitemShut {NoStop}%
\bibitem [{\citenamefont {Hybertsen}\ and\ \citenamefont {Louie}(1986)}]{Hybertson1986}%
  \BibitemOpen
  \bibfield  {author} {\bibinfo {author} {\bibfnamefont {M.~S.}\ \bibnamefont {Hybertsen}}\ and\ \bibinfo {author} {\bibfnamefont {S.~G.}\ \bibnamefont {Louie}},\ }\bibfield  {title} {\bibinfo {title} {Electron correlation in semiconductors and insulators: Band gaps and quasiparticle energies},\ }\href {https://doi.org/10.1103/PhysRevB.34.5390} {\bibfield  {journal} {\bibinfo  {journal} {Phys. Rev. B}\ }\textbf {\bibinfo {volume} {34}},\ \bibinfo {pages} {5390} (\bibinfo {year} {1986})}\BibitemShut {NoStop}%
\bibitem [{\citenamefont {Tauc}\ \emph {et~al.}(1966)\citenamefont {Tauc}, \citenamefont {Grigorovici},\ and\ \citenamefont {Vancu}}]{Tauc66}%
  \BibitemOpen
  \bibfield  {author} {\bibinfo {author} {\bibfnamefont {J.}~\bibnamefont {Tauc}}, \bibinfo {author} {\bibfnamefont {R.}~\bibnamefont {Grigorovici}},\ and\ \bibinfo {author} {\bibfnamefont {A.}~\bibnamefont {Vancu}},\ }\bibfield  {title} {\bibinfo {title} {Optical properties and electronic structure of amorphous germanium},\ }\href {https://doi.org/10.1002/pssb.19660150224} {\bibfield  {journal} {\bibinfo  {journal} {physica status solidi (b)}\ }\textbf {\bibinfo {volume} {15}},\ \bibinfo {pages} {627} (\bibinfo {year} {1966})}\BibitemShut {NoStop}%
\bibitem [{\citenamefont {Zanatta}(2019)}]{Zanatta19}%
  \BibitemOpen
  \bibfield  {author} {\bibinfo {author} {\bibfnamefont {A.~R.}\ \bibnamefont {Zanatta}},\ }\bibfield  {title} {\bibinfo {title} {Revisiting the optical bandgap of semiconductors and the proposal of a unified methodology to its determination},\ }\bibfield  {journal} {\bibinfo  {journal} {Scientific Reports}\ }\textbf {\bibinfo {volume} {9}},\ \href {https://doi.org/10.1038/s41598-019-47670-y} {10.1038/s41598-019-47670-y} (\bibinfo {year} {2019})\BibitemShut {NoStop}%
\bibitem [{\citenamefont {Bock}\ \emph {et~al.}(2019)\citenamefont {Bock}, \citenamefont {Kijatkin}, \citenamefont {Berben},\ and\ \citenamefont {Imlau}}]{Bock19}%
  \BibitemOpen
  \bibfield  {author} {\bibinfo {author} {\bibfnamefont {S.}~\bibnamefont {Bock}}, \bibinfo {author} {\bibfnamefont {C.}~\bibnamefont {Kijatkin}}, \bibinfo {author} {\bibfnamefont {D.}~\bibnamefont {Berben}},\ and\ \bibinfo {author} {\bibfnamefont {M.}~\bibnamefont {Imlau}},\ }\bibfield  {title} {\bibinfo {title} {Absorption and remission characterization of pure, dielectric (nano-)powders using diffuse reflectance spectroscopy: An end-to-end instruction},\ }\bibfield  {journal} {\bibinfo  {journal} {Applied Sciences}\ }\textbf {\bibinfo {volume} {9}},\ \href {https://doi.org/10.3390/app9224933} {10.3390/app9224933} (\bibinfo {year} {2019})\BibitemShut {NoStop}%
\bibitem [{\citenamefont {Bhatt}\ \emph {et~al.}(2012)\citenamefont {Bhatt}, \citenamefont {Bhaumik}, \citenamefont {Ganesamoorthy}, \citenamefont {Karnal}, \citenamefont {Swami}, \citenamefont {Patel},\ and\ \citenamefont {Gupta}}]{Bhatt12}%
  \BibitemOpen
  \bibfield  {author} {\bibinfo {author} {\bibfnamefont {R.}~\bibnamefont {Bhatt}}, \bibinfo {author} {\bibfnamefont {I.}~\bibnamefont {Bhaumik}}, \bibinfo {author} {\bibfnamefont {S.}~\bibnamefont {Ganesamoorthy}}, \bibinfo {author} {\bibfnamefont {A.~K.}\ \bibnamefont {Karnal}}, \bibinfo {author} {\bibfnamefont {M.~K.}\ \bibnamefont {Swami}}, \bibinfo {author} {\bibfnamefont {H.~S.}\ \bibnamefont {Patel}},\ and\ \bibinfo {author} {\bibfnamefont {P.~K.}\ \bibnamefont {Gupta}},\ }\bibfield  {title} {\bibinfo {title} {Urbach tail and bandgap analysis in near stoichiometric {LiNbO}$_3$ crystals},\ }\href {https://doi.org/https://doi.org/10.1002/pssa.201127361} {\bibfield  {journal} {\bibinfo  {journal} {physica status solidi (a)}\ }\textbf {\bibinfo {volume} {209}},\ \bibinfo {pages} {176} (\bibinfo {year} {2012})}\BibitemShut {NoStop}%
\bibitem [{\citenamefont {O’Donnell}\ and\ \citenamefont {Chen}(1991)}]{ODonnel91}%
  \BibitemOpen
  \bibfield  {author} {\bibinfo {author} {\bibfnamefont {K.~P.}\ \bibnamefont {O’Donnell}}\ and\ \bibinfo {author} {\bibfnamefont {X.}~\bibnamefont {Chen}},\ }\bibfield  {title} {\bibinfo {title} {{Temperature dependence of semiconductor band gaps}},\ }\href {https://doi.org/10.1063/1.104723} {\bibfield  {journal} {\bibinfo  {journal} {Applied Physics Letters}\ }\textbf {\bibinfo {volume} {58}},\ \bibinfo {pages} {2924} (\bibinfo {year} {1991})}\BibitemShut {NoStop}%
\bibitem [{\citenamefont {{Huang}}\ and\ \citenamefont {{Rhys}}(1950)}]{Huang50}%
  \BibitemOpen
  \bibfield  {author} {\bibinfo {author} {\bibfnamefont {K.}~\bibnamefont {{Huang}}}\ and\ \bibinfo {author} {\bibfnamefont {A.}~\bibnamefont {{Rhys}}},\ }\bibfield  {title} {\bibinfo {title} {{Theory of Light Absorption and Non-Radiative Transitions in F-Centres}},\ }\href {https://doi.org/10.1098/rspa.1950.0184} {\bibfield  {journal} {\bibinfo  {journal} {Proceedings of the Royal Society of London Series A}\ }\textbf {\bibinfo {volume} {204}},\ \bibinfo {pages} {406} (\bibinfo {year} {1950})}\BibitemShut {NoStop}%
\bibitem [{\citenamefont {Sanna}\ and\ \citenamefont {Schmidt}(2017)}]{Sanna17surf}%
  \BibitemOpen
  \bibfield  {author} {\bibinfo {author} {\bibfnamefont {S.}~\bibnamefont {Sanna}}\ and\ \bibinfo {author} {\bibfnamefont {W.~G.}\ \bibnamefont {Schmidt}},\ }\bibfield  {title} {\bibinfo {title} {$\text{LiNbO}_3$ surfaces from a microscopic perspective},\ }\href {https://doi.org/10.1088/1361-648X/aa818d} {\bibfield  {journal} {\bibinfo  {journal} {Journal of Physics: Condensed Matter}\ }\textbf {\bibinfo {volume} {29}},\ \bibinfo {pages} {413001} (\bibinfo {year} {2017})}\BibitemShut {NoStop}%
\bibitem [{\citenamefont {Gaillac}\ \emph {et~al.}(2016)\citenamefont {Gaillac}, \citenamefont {Pullumbi},\ and\ \citenamefont {Coudert}}]{Gaillac16}%
  \BibitemOpen
  \bibfield  {author} {\bibinfo {author} {\bibfnamefont {R.}~\bibnamefont {Gaillac}}, \bibinfo {author} {\bibfnamefont {P.}~\bibnamefont {Pullumbi}},\ and\ \bibinfo {author} {\bibfnamefont {F.-X.}\ \bibnamefont {Coudert}},\ }\bibfield  {title} {\bibinfo {title} {Elate: an open-source online application for analysis and visualization of elastic tensors},\ }\href {https://doi.org/10.1088/0953-8984/28/27/275201} {\bibfield  {journal} {\bibinfo  {journal} {Journal of Physics: Condensed Matter}\ }\textbf {\bibinfo {volume} {28}},\ \bibinfo {pages} {275201} (\bibinfo {year} {2016})}\BibitemShut {NoStop}%
\bibitem [{\citenamefont {Jain}\ \emph {et~al.}(2013)\citenamefont {Jain}, \citenamefont {Ong}, \citenamefont {Hautier}, \citenamefont {Chen}, \citenamefont {Richards}, \citenamefont {Dacek}, \citenamefont {Cholia}, \citenamefont {Gunter}, \citenamefont {Skinner}, \citenamefont {Ceder},\ and\ \citenamefont {Persson}}]{Jain2013}%
  \BibitemOpen
  \bibfield  {author} {\bibinfo {author} {\bibfnamefont {A.}~\bibnamefont {Jain}}, \bibinfo {author} {\bibfnamefont {S.~P.}\ \bibnamefont {Ong}}, \bibinfo {author} {\bibfnamefont {G.}~\bibnamefont {Hautier}}, \bibinfo {author} {\bibfnamefont {W.}~\bibnamefont {Chen}}, \bibinfo {author} {\bibfnamefont {W.~D.}\ \bibnamefont {Richards}}, \bibinfo {author} {\bibfnamefont {S.}~\bibnamefont {Dacek}}, \bibinfo {author} {\bibfnamefont {S.}~\bibnamefont {Cholia}}, \bibinfo {author} {\bibfnamefont {D.}~\bibnamefont {Gunter}}, \bibinfo {author} {\bibfnamefont {D.}~\bibnamefont {Skinner}}, \bibinfo {author} {\bibfnamefont {G.}~\bibnamefont {Ceder}},\ and\ \bibinfo {author} {\bibfnamefont {K.~a.}\ \bibnamefont {Persson}},\ }\bibfield  {title} {\bibinfo {title} {{Commentary: The Materials Project: A materials genome approach to accelerating materials innovation}},\ }\href {https://doi.org/10.1063/1.4812323} {\bibfield  {journal} {\bibinfo  {journal} {APL Materials}\ }\textbf {\bibinfo {volume} {1}},\ \bibinfo {pages}
  {011002} (\bibinfo {year} {2013})}\BibitemShut {NoStop}%
\bibitem [{\citenamefont {Chen}\ \emph {et~al.}(2001)\citenamefont {Chen}, \citenamefont {Xu}, \citenamefont {Chen}, \citenamefont {Kong},\ and\ \citenamefont {Zhang}}]{Chen01}%
  \BibitemOpen
  \bibfield  {author} {\bibinfo {author} {\bibfnamefont {Y.-L.}\ \bibnamefont {Chen}}, \bibinfo {author} {\bibfnamefont {J.-J.}\ \bibnamefont {Xu}}, \bibinfo {author} {\bibfnamefont {X.-J.}\ \bibnamefont {Chen}}, \bibinfo {author} {\bibfnamefont {Y.-F.}\ \bibnamefont {Kong}},\ and\ \bibinfo {author} {\bibfnamefont {G.-Y.}\ \bibnamefont {Zhang}},\ }\bibfield  {title} {\bibinfo {title} {Domain reversion process in near-stoichiometric {L}i{N}b{O}$_3$ crystals},\ }\href {https://doi.org/https://doi.org/10.1016/S0030-4018(00)01137-8} {\bibfield  {journal} {\bibinfo  {journal} {Optics Communications}\ }\textbf {\bibinfo {volume} {188}},\ \bibinfo {pages} {359} (\bibinfo {year} {2001})}\BibitemShut {NoStop}%
\bibitem [{\citenamefont {Kitamura}\ \emph {et~al.}(1998)\citenamefont {Kitamura}, \citenamefont {Furukawa}, \citenamefont {Niwa}, \citenamefont {Gopalan},\ and\ \citenamefont {Mitchell}}]{Kitamura98}%
  \BibitemOpen
  \bibfield  {author} {\bibinfo {author} {\bibfnamefont {K.}~\bibnamefont {Kitamura}}, \bibinfo {author} {\bibfnamefont {Y.}~\bibnamefont {Furukawa}}, \bibinfo {author} {\bibfnamefont {K.}~\bibnamefont {Niwa}}, \bibinfo {author} {\bibfnamefont {V.}~\bibnamefont {Gopalan}},\ and\ \bibinfo {author} {\bibfnamefont {T.~E.}\ \bibnamefont {Mitchell}},\ }\bibfield  {title} {\bibinfo {title} {{Crystal growth and low coercive field 180° domain switching characteristics of stoichiometric {L}i{T}a{O}$_3$}},\ }\href {https://doi.org/10.1063/1.122676} {\bibfield  {journal} {\bibinfo  {journal} {Applied Physics Letters}\ }\textbf {\bibinfo {volume} {73}},\ \bibinfo {pages} {3073} (\bibinfo {year} {1998})}\BibitemShut {NoStop}%
\bibitem [{\citenamefont {Krampf}\ \emph {et~al.}(2021)\citenamefont {Krampf}, \citenamefont {Imlau}, \citenamefont {Suhak}, \citenamefont {Fritze},\ and\ \citenamefont {Sanna}}]{Krampf_2021}%
  \BibitemOpen
  \bibfield  {author} {\bibinfo {author} {\bibfnamefont {A.}~\bibnamefont {Krampf}}, \bibinfo {author} {\bibfnamefont {M.}~\bibnamefont {Imlau}}, \bibinfo {author} {\bibfnamefont {Y.}~\bibnamefont {Suhak}}, \bibinfo {author} {\bibfnamefont {H.}~\bibnamefont {Fritze}},\ and\ \bibinfo {author} {\bibfnamefont {S.}~\bibnamefont {Sanna}},\ }\bibfield  {title} {\bibinfo {title} {Evaluation of similarities and differences of {LiTaO}$_3$ and {LiNbO}$_3$ based on high-t-conductivity, nonlinear optical fs-spectroscopy and ab initio modeling of polaronic structures},\ }\href {https://doi.org/10.1088/1367-2630/abe3ac} {\bibfield  {journal} {\bibinfo  {journal} {New Journal of Physics}\ }\textbf {\bibinfo {volume} {23}},\ \bibinfo {pages} {033016} (\bibinfo {year} {2021})}\BibitemShut {NoStop}%
\bibitem [{\citenamefont {Gaczynski}\ \emph {et~al.}(2024)\citenamefont {Gaczynski}, \citenamefont {Suhak}, \citenamefont {Ganschow}, \citenamefont {Sanna}, \citenamefont {Fritze},\ and\ \citenamefont {Becker}}]{Becker24}%
  \BibitemOpen
  \bibfield  {author} {\bibinfo {author} {\bibfnamefont {P.}~\bibnamefont {Gaczynski}}, \bibinfo {author} {\bibfnamefont {Y.}~\bibnamefont {Suhak}}, \bibinfo {author} {\bibfnamefont {S.}~\bibnamefont {Ganschow}}, \bibinfo {author} {\bibfnamefont {S.}~\bibnamefont {Sanna}}, \bibinfo {author} {\bibfnamefont {H.}~\bibnamefont {Fritze}},\ and\ \bibinfo {author} {\bibfnamefont {K.-D.}\ \bibnamefont {Becker}},\ }\bibfield  {title} {\bibinfo {title} {{A High-Temperature Optical Spectroscopy Study of the Fundamental Absorption Edge in LiNbO$_3$ - LiTaO$_3$ Solid Solutions}},\ }\href@noop {} {\bibfield  {journal} {\bibinfo  {journal} {Physica Status Solidi a}\ }\textbf {\bibinfo {volume} {xx}} (\bibinfo {year} {2024})}\BibitemShut {NoStop}%
\bibitem [{\citenamefont {Klenen}\ \emph {et~al.}(2024)\citenamefont {Klenen}, \citenamefont {Sauerwein}, \citenamefont {Vittadello}, \citenamefont {Koempe}, \citenamefont {Hreb}, \citenamefont {Sydorchuk}, \citenamefont {Yakhnevych}, \citenamefont {Sugak}, \citenamefont {Vasylechko},\ and\ \citenamefont {Imlau}}]{MircoNLO24}%
  \BibitemOpen
  \bibfield  {author} {\bibinfo {author} {\bibfnamefont {J.}~\bibnamefont {Klenen}}, \bibinfo {author} {\bibfnamefont {F.}~\bibnamefont {Sauerwein}}, \bibinfo {author} {\bibfnamefont {L.}~\bibnamefont {Vittadello}}, \bibinfo {author} {\bibfnamefont {K.}~\bibnamefont {Koempe}}, \bibinfo {author} {\bibfnamefont {V.}~\bibnamefont {Hreb}}, \bibinfo {author} {\bibfnamefont {V.}~\bibnamefont {Sydorchuk}}, \bibinfo {author} {\bibfnamefont {U.}~\bibnamefont {Yakhnevych}}, \bibinfo {author} {\bibfnamefont {D.}~\bibnamefont {Sugak}}, \bibinfo {author} {\bibfnamefont {L.}~\bibnamefont {Vasylechko}},\ and\ \bibinfo {author} {\bibfnamefont {M.}~\bibnamefont {Imlau}},\ }\bibfield  {title} {\bibinfo {title} {Gap-free tuning of second and third harmonic generation in mechanochemically synthesized nanocrystalline {LiNb$_{1-x}$Ta$_x$O$_3$} ($0\leq x\leq 1$) studied by nonlinear diffuse femtosecond-pulse reflectometry},\ }\href@noop {} {\bibfield  {journal} {\bibinfo  {journal} {Nanomaterials}\ }\textbf {\bibinfo {volume}
  {xxx}},\ \bibinfo {pages} {xxx} (\bibinfo {year} {2024})}\BibitemShut {NoStop}%
\bibitem [{\citenamefont {Bernhardt}\ \emph {et~al.}(2024)\citenamefont {Bernhardt}, \citenamefont {Fink}, \citenamefont {Verhoff}, \citenamefont {Sch\"afer}, \citenamefont {Kapp}, \citenamefont {Nachwati}, \citenamefont {Bashir}, \citenamefont {Azzouzi}, \citenamefont {Yakhnevych}, \citenamefont {Suhak}, \citenamefont {Becker}, \citenamefont {Klimm}, \citenamefont {Ganschow}, \citenamefont {Schmidt}, \citenamefont {Fritze},\ and\ \citenamefont {Sanna}}]{SimoMD23}%
  \BibitemOpen
  \bibfield  {author} {\bibinfo {author} {\bibfnamefont {F.}~\bibnamefont {Bernhardt}}, \bibinfo {author} {\bibfnamefont {C.}~\bibnamefont {Fink}}, \bibinfo {author} {\bibfnamefont {L.~M.}\ \bibnamefont {Verhoff}}, \bibinfo {author} {\bibfnamefont {N.~A.}\ \bibnamefont {Sch\"afer}}, \bibinfo {author} {\bibfnamefont {A.}~\bibnamefont {Kapp}}, \bibinfo {author} {\bibfnamefont {W.~A.}\ \bibnamefont {Nachwati}}, \bibinfo {author} {\bibfnamefont {U.}~\bibnamefont {Bashir}}, \bibinfo {author} {\bibfnamefont {F.~E.}\ \bibnamefont {Azzouzi}}, \bibinfo {author} {\bibfnamefont {U.}~\bibnamefont {Yakhnevych}}, \bibinfo {author} {\bibfnamefont {Y.}~\bibnamefont {Suhak}}, \bibinfo {author} {\bibfnamefont {K.-D.}\ \bibnamefont {Becker}}, \bibinfo {author} {\bibfnamefont {D.}~\bibnamefont {Klimm}}, \bibinfo {author} {\bibfnamefont {S.}~\bibnamefont {Ganschow}}, \bibinfo {author} {\bibfnamefont {H.}~\bibnamefont {Schmidt}}, \bibinfo {author} {\bibfnamefont {H.}~\bibnamefont {Fritze}},\ and\ \bibinfo {author} {\bibfnamefont
  {S.}~\bibnamefont {Sanna}},\ }\href@noop {} {\bibfield  {journal} {\bibinfo  {journal} {Phys. Rev. M}\ } (\bibinfo {year} {2024})},\ \bibinfo {note} {submitted}\BibitemShut {NoStop}%
\bibitem [{\citenamefont {Friedrich}\ \emph {et~al.}(2015)\citenamefont {Friedrich}, \citenamefont {Riefer}, \citenamefont {Sanna}, \citenamefont {Schmidt},\ and\ \citenamefont {Schindlmayr}}]{Friedrich15}%
  \BibitemOpen
  \bibfield  {author} {\bibinfo {author} {\bibfnamefont {M.}~\bibnamefont {Friedrich}}, \bibinfo {author} {\bibfnamefont {A.}~\bibnamefont {Riefer}}, \bibinfo {author} {\bibfnamefont {S.}~\bibnamefont {Sanna}}, \bibinfo {author} {\bibfnamefont {W.~G.}\ \bibnamefont {Schmidt}},\ and\ \bibinfo {author} {\bibfnamefont {A.}~\bibnamefont {Schindlmayr}},\ }\bibfield  {title} {\bibinfo {title} {Phonon dispersion and zero-point renormalization of {LiNbO}$_3$ from density-functional perturbation theory},\ }\href {https://doi.org/10.1088/0953-8984/27/38/385402} {\bibfield  {journal} {\bibinfo  {journal} {Journal of Physics: Condensed Matter}\ }\textbf {\bibinfo {volume} {27}},\ \bibinfo {pages} {385402} (\bibinfo {year} {2015})}\BibitemShut {NoStop}%
\bibitem [{\citenamefont {Yao}\ \emph {et~al.}(2008)\citenamefont {Yao}, \citenamefont {Wang}, \citenamefont {Liu}, \citenamefont {Hu}, \citenamefont {Zhang}, \citenamefont {Cheng},\ and\ \citenamefont {Ling}}]{YAO2008501}%
  \BibitemOpen
  \bibfield  {author} {\bibinfo {author} {\bibfnamefont {S.}~\bibnamefont {Yao}}, \bibinfo {author} {\bibfnamefont {J.}~\bibnamefont {Wang}}, \bibinfo {author} {\bibfnamefont {H.}~\bibnamefont {Liu}}, \bibinfo {author} {\bibfnamefont {X.}~\bibnamefont {Hu}}, \bibinfo {author} {\bibfnamefont {H.}~\bibnamefont {Zhang}}, \bibinfo {author} {\bibfnamefont {X.}~\bibnamefont {Cheng}},\ and\ \bibinfo {author} {\bibfnamefont {Z.}~\bibnamefont {Ling}},\ }\bibfield  {title} {\bibinfo {title} {{Growth, optical and thermal properties of near-stoichiometric LiNbO$_3$ single crystal}},\ }\href {https://doi.org/https://doi.org/10.1016/j.jallcom.2007.02.001} {\bibfield  {journal} {\bibinfo  {journal} {Journal of Alloys and Compounds}\ }\textbf {\bibinfo {volume} {455}},\ \bibinfo {pages} {501} (\bibinfo {year} {2008})}\BibitemShut {NoStop}%
\bibitem [{Kor(2024)}]{Korth}%
  \BibitemOpen
  \href@noop {} {\bibinfo {title} {{Korth Kristalle GmbH}}},\ \bibinfo {howpublished} {\url{https://www.korth.de/en/materials/detail/Lithium\%20Tantalate}} (\bibinfo {year} {2024}),\ \bibinfo {note} {accessed: 2024-02-05}\BibitemShut {NoStop}%
\bibitem [{\citenamefont {Villar}\ \emph {et~al.}(1986)\citenamefont {Villar}, \citenamefont {Gmelin},\ and\ \citenamefont {Grimm}}]{Villar86}%
  \BibitemOpen
  \bibfield  {author} {\bibinfo {author} {\bibfnamefont {R.}~\bibnamefont {Villar}}, \bibinfo {author} {\bibfnamefont {E.}~\bibnamefont {Gmelin}},\ and\ \bibinfo {author} {\bibfnamefont {H.}~\bibnamefont {Grimm}},\ }\bibfield  {title} {\bibinfo {title} {Specific heat of crystalline ferroelectrics at low temperatures},\ }\href {https://doi.org/10.1080/00150198608008190} {\bibfield  {journal} {\bibinfo  {journal} {Ferroelectrics}\ }\textbf {\bibinfo {volume} {69}},\ \bibinfo {pages} {165} (\bibinfo {year} {1986})}\BibitemShut {NoStop}%
\bibitem [{\citenamefont {Riefer}\ \emph {et~al.}(2013{\natexlab{b}})\citenamefont {Riefer}, \citenamefont {Sanna}, \citenamefont {Schindlmayr},\ and\ \citenamefont {Schmidt}}]{ArthurRiefer13}%
  \BibitemOpen
  \bibfield  {author} {\bibinfo {author} {\bibfnamefont {A.}~\bibnamefont {Riefer}}, \bibinfo {author} {\bibfnamefont {S.}~\bibnamefont {Sanna}}, \bibinfo {author} {\bibfnamefont {A.}~\bibnamefont {Schindlmayr}},\ and\ \bibinfo {author} {\bibfnamefont {W.~G.}\ \bibnamefont {Schmidt}},\ }\bibfield  {title} {\bibinfo {title} {Optical response of stoichiometric and congruent lithium niobate from first-principles calculations},\ }\href {https://doi.org/10.1103/PhysRevB.87.195208} {\bibfield  {journal} {\bibinfo  {journal} {Phys. Rev. B}\ }\textbf {\bibinfo {volume} {87}},\ \bibinfo {pages} {195208} (\bibinfo {year} {2013}{\natexlab{b}})}\BibitemShut {NoStop}%
\bibitem [{\citenamefont {Kondo}\ \emph {et~al.}(1979)\citenamefont {Kondo}, \citenamefont {Sugii}, \citenamefont {Miyazawa},\ and\ \citenamefont {Uehara}}]{Kondo79}%
  \BibitemOpen
  \bibfield  {author} {\bibinfo {author} {\bibfnamefont {S.}~\bibnamefont {Kondo}}, \bibinfo {author} {\bibfnamefont {K.}~\bibnamefont {Sugii}}, \bibinfo {author} {\bibfnamefont {S.}~\bibnamefont {Miyazawa}},\ and\ \bibinfo {author} {\bibfnamefont {S.}~\bibnamefont {Uehara}},\ }\bibfield  {title} {\bibinfo {title} {{LPE growth of Li(Nb,Ta)O$_3$ solid-solution thin film waveguides on LiTaO$_3$ substrates}},\ }\href {https://doi.org/10.1016/0022-0248(79)90079-4} {\bibfield  {journal} {\bibinfo  {journal} {Journal of Crystal Growth}\ }\textbf {\bibinfo {volume} {46}},\ \bibinfo {pages} {314} (\bibinfo {year} {1979})}\BibitemShut {NoStop}%
\bibitem [{\citenamefont {Ganschow}\ \emph {et~al.}(2021)\citenamefont {Ganschow}, \citenamefont {Schmidt}, \citenamefont {Suhak}, \citenamefont {Imlau}, \citenamefont {R\"using}, \citenamefont {Eng}, \citenamefont {Fritze},\ and\ \citenamefont {Sanna}}]{FOR5044}%
  \BibitemOpen
  \bibfield  {author} {\bibinfo {author} {\bibfnamefont {S.}~\bibnamefont {Ganschow}}, \bibinfo {author} {\bibfnamefont {H.}~\bibnamefont {Schmidt}}, \bibinfo {author} {\bibfnamefont {Y.}~\bibnamefont {Suhak}}, \bibinfo {author} {\bibfnamefont {M.~K.}\ \bibnamefont {Imlau}}, \bibinfo {author} {\bibfnamefont {M.}~\bibnamefont {R\"using}}, \bibinfo {author} {\bibfnamefont {L.~M.}\ \bibnamefont {Eng}}, \bibinfo {author} {\bibfnamefont {H.}~\bibnamefont {Fritze}},\ and\ \bibinfo {author} {\bibfnamefont {S.}~\bibnamefont {Sanna}},\ }\href {https://www.for5044.de} {\bibinfo {title} {Periodic low-dimensional defect structures in polar oxides}} (\bibinfo {year} {2021})\BibitemShut {NoStop}%
\end{thebibliography}%

\end{document}